\documentclass[10pt,aps,prd,superscriptaddress,nofootinbib,nobibnotes,longbibliography,floatfix,twocolumn]{revtex4-2}

\usepackage{bm}
\usepackage{mathtools,
amsmath,
amssymb,
amsfonts,
mathrsfs,
chngcntr,
multirow}

\let\cc\corresponds
\let\corresponds\relax
\usepackage{mathabx}
\let\corresponds\cc

\usepackage{booktabs}
\usepackage[utf8]{inputenc}
\usepackage[T1]{fontenc}
\usepackage[dvipsnames]{xcolor}
\usepackage[unicode]{hyperref}
\hypersetup{colorlinks=true, citecolor=MidnightBlue,
            linkcolor=MidnightBlue, urlcolor=MidnightBlue, linktocpage=true}
\usepackage[normalem]{ulem}
\usepackage{orcidlink}
\usepackage[capitalize]{cleveref}

\newcommand{\orcid}[1]{\href{https://orcid.org/#1}{\includegraphics[width=10pt]{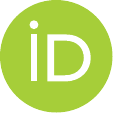}}}

\begin{document}
\title{Quantifying Systematic Biases in Black Hole Spectroscopy}

\author{Sebastian H.\,V\"olkel\,\orcid{0000-0002-9432-7690}}
\email{sebastian.voelkel@aei.mpg.de}
\affiliation{Max Planck Institute for Gravitational Physics (Albert Einstein Institute), \\ Am M\"uhlenberg 1, 
D-14476 Potsdam, Germany}

\author{Arnab Dhani\,\orcid{0000-0001-9930-9101}}
\email{arnab.dhani@aei.mpg.de}
\affiliation{Max Planck Institute for Gravitational Physics (Albert Einstein Institute), \\ Am M\"uhlenberg 1,
D-14476 Potsdam, Germany}

\date{\today}

\begin{abstract}
How long after the merger of two black holes can one rely on linear perturbation theory, and how many quasinormal modes are in the ringdown? 
Such questions suggest that black hole spectroscopy suffers from systematic uncertainties that potentially spoil ringdown analyses, both from high-accuracy simulations and in data from gravitational wave detectors. 
In this work, we demonstrate that linear-signal analysis is a powerful tool for quantifying biases, allowing for detailed explorations that are computationally too expensive for traditional Bayesian injection and recovery approaches. 
We quantify the validity of the Fisher information matrix and bias formula by comparing it to robust but slow Bayesian sampling. 
Working with flat noise in the time domain, statistical errors and systematic biases can mostly be detected analytically. 
Due to its efficiency, we provide detailed parameter space analyses for potentially unmodeled small contributions from overtones, quadratic modes, and tails. 
We find linear signal analysis well suited for predicting biases in simple ringdown models at intermediate signal-to-noise ratios (SNRs) of order 50 when unmodeled effects are small. 
It is also valuable in explaining ongoing issues in extracting quasinormal modes from high-precision simulations, as one can understand them as high-SNR signals. 
Therefore, this approach offers promising prospects for improving ringdown models by efficiently identifying and incorporating systematic uncertainties, ultimately enhancing the accuracy and robustness of black hole spectroscopy. 
\end{abstract}

\maketitle

\section{Introduction}

The intriguing concept of studying the properties of a black hole through its gravitational wave spectrum is known as \textit{black hole spectroscopy}~\cite{Detweiler:1980gk,Dreyer:2003bv,Berti:2005ys}. 
It is a fundamental prediction of black hole perturbation theory~\cite{Regge:1957td,Zerilli:1970se,Teukolsky:1973ha}. 
A perturbed black hole, such as during a binary black hole merger, emits characteristic gravitational waves that can be well approximated by a sum of damped sinusoid modes, known as quasinormal modes (QNMs). 
For a comprehensive discussion of this topic, we refer to the literature, including Refs.~\cite{Kokkotas:1999bd,Nollert:1999ji,Berti:2009kk,Konoplya:2011qq,Berti:2025hly}. 
According to general relativity (GR) and within the framework of no-hair theorems, the spectrum of QNMs is uniquely determined by the mass and spin of the resulting black hole, which is described by the Kerr metric~\cite{Israel:1967wq,Hawking:1971vc,Carter:1971zc,Robinson:1975bv}. 
As a result, measurements of multiple QNMs can be used to test the assumptions of the Kerr hypothesis and even the validity of GR itself.

Despite its elegance and similarities with other spectroscopy methods in physics, black hole spectroscopy faces several complex challenges.
One significant question is how long after the merger of two black holes one can rely on linear perturbation theory. 
Even after sufficient time has passed, quantifying the number of QNMs in a given signal remains challenging. 
This is in part because QNMs do not form a complete basis~\cite{1993AIHPA..59....3B,Beyer:1998nu}. 
On the one hand, QNMs are only part of the solution of the perturbed Einstein's equations, which also include the transient early part of the signal, known as the prompt response, and power-law tails which dominate the signal at very late times~\cite{Vishveshwara:1970zz,Price:1971fb,Leaver:1986gd,Gundlach:1993tp,Gundlach:1993tn,Barack:1998bw,Rosato:2025rtr}. 
On the other hand, the set of QNMs is overcomplete in the sense that any QNM can be expressed as a linear combination of the other QNMs, raising questions about their physical significance~\cite{Nollert:1998ys}. 
Standard black hole spectroscopy approaches, which analyze the intermediate part of the ringdown as a sum of damped sinusoids, encounter what is known as the short blanket problem. 
To minimize the effects of prompt responses and nonlinear contributions, the analysis must begin late enough but conclude before the signal's tails take over~\cite{Price:1971fb,Leaver:1986gd,Gundlach:1993tp,Barack:1998bw,Hod:1999ci,Hod:2000fh}. 
However, this reduced signal length requires high SNR signals. 
Unfortunately, increasing the SNR amplifies both the prompt response and nonlinear effects during the intermediate times. 
Pioneering work for the nonlinear case can be found in Refs.~\cite{Buonanno:2006ui,Berti:2007fi,London:2014cma}. 
In recent years, several studies have assessed how effectively QNMs can be inferred from high-accuracy linear, nonlinear simulations, and actual data, with varying conclusions regarding the questions previously mentioned~\cite{Giesler:2019uxc,Finch:2022ynt,Isi:2022mhy,Ma:2022wpv,Ma:2023cwe,Baibhav:2023clw,Nee:2023osy,Isi:2023nif,Carullo:2023gtf,Cheung:2023vki,Maselli:2023khq,Giesler:2024hcr,Mitman:2025hgy,Nobili:2025ydt,Thomopoulos:2025nuf,Gao:2025zvl}. 
While earlier studies used a heuristic set of overtones and retrograde modes to fit the complex amplitudes to the data, more recent works use refined fitting approaches to identify multiple QNMs, including overtones, retrograde modes, and quadratic modes, demonstrating the importance of careful extraction techniques from NR simulations~\cite{Cheung:2023vki,Giesler:2024hcr,Mitman:2025hgy,Gao:2025zvl}. 
However, once the QNMs are identified, the complex amplitudes are extracted by fixing the identified QNM frequencies, similar to earlier studies. 
A ringdown surrogate model based on Gaussian processes has been introduced in Ref.~\cite{MaganaZertuche:2024ajz}. 

One central aspect of ringdown analyses is controlling the number of free parameters. 
In a theory-agnostic approach, the goal is to use minimal information and treat all complex frequencies and amplitudes as independent parameters.
This method enables a flexible extraction of signal features, thereby capturing possible modifications from GR. 
In contrast, a theory-specific approach requires choosing the underlying theory, such as GR, and modeling all possible QNMs as a function of a few parameters, like the mass and spin. 
Additionally, QNM excitation coefficients could be obtained from numerical relativity simulations or perturbative calculations to further reduce the number of free parameters~\cite{Oshita:2021iyn,Motohashi:2024fwt,Oshita:2024wgt,Oshita:2025ibu,Lo:2025njp}. 
While a smaller number of independent parameters reduces the complexity when considering a large number of QNMs, it is challenging to extend this to rotating black holes in beyond-GR theories. 

Although each approach has its own advantages and disadvantages, both are prone to systematic errors in the extracted black hole parameters or agnostic QNMs. 
This is because, while almost all QNMs get excited in a binary merger, except for the ones excluded by the symmetries of the initial binary, it is challenging to calculate their amplitudes, except for simple cases~\cite{Berti:2006wq,Zhang:2013ksa}. 
Therefore, in both approaches, the number of QNMs needed to sufficiently describe a realistic ringdown signal is not known a priori~\cite{Giesler:2019uxc,Dhani:2020nik,Dhani:2021vac,Forteza:2021wfq,Giesler:2024hcr,Mitman:2022qdl,Cook:2020otn,Nee:2023osy,Baibhav:2023clw}. 
Relying on metrics like mismatches does not address the risk of overfitting. 
Moreover, in both approaches, little attention has been given to systematic errors due to the presence of tails. 
Tails are usually not modeled in parameter estimation studies because they are considered a weak late-time effect that can be avoided by stopping the analysis early enough. 
However, recent studies suggest that tails might be more relevant than anticipated~\cite{Carullo:2023tff,Cardoso:2024jme,DeAmicis:2024not,Islam:2024vro,Ma:2024bed,DeAmicis:2024eoy,Ma:2024hzq}. 
Other sources of systematics are the unmodeled prompt response, mode mixing, spin-precession from the progenitor system~\cite{OShaughnessy:2012iol,Zhu:2023fnf}, and detector calibration accuracy~\cite{Sinha:2025snr}. 
Nonlinear QNMs, so-called quadratic QNMs (which are sourced at quadratic order from linear perturbations), have also received significant attention recently and could be an important part of ringdown analyses, see Refs.~\cite{Cheung:2022rbm,Mitman:2022qdl,Redondo-Yuste:2023seq} for more precise details, along with other nonlinear effects~\cite{Sberna:2021eui}. 
They might also be detectable by future detectors~\cite{Yi:2024elj,Khera:2024bjs}. 
The importance of overtones in extracting the correct fundamental mode of collapsing neutron stars and binary-neutron star mergers forming a final black hole surrounded by leftover matter has recently been studied in Ref.~\cite{Steppohn:2025kbh}.

Quantifying the impact of unmodeled contributions to realistic signals in terms of systematic errors on the underlying parameters is a very complex and vast problem. 
Using traditional Bayesian methods like Markov-chain-Monte-Carlo (MCMC) sampling for injection and recovery studies to tackle this question is computationally expensive and thus limited. 
Simulation-based inference methods are promising alternatives and have recently been applied to the ringdown~\cite{Crisostomi:2023tle,Pacilio:2024qcq}. 

In this work, we show that linear signal approximation (LSA) methods~\cite{Finn:1992xs,Krolak:1993zy,Kokkotas:1994ef,Flanagan:1997kp,Cutler:2007mi}, based on the Fisher information matrix and the bias formula, are powerful tools for quantifying systematic errors from unmodeled effects to a simple ringdown model. 
For high-SNR signals from future detectors or high-accuracy simulations, LSA allows one to predict the biases in a model's parameters as long as the signal being analyzed has small deviations from the model. 
Therefore, the method is well suited for detailed explorations of the vast zoo of possibly overlooked effects and for quantifying at which SNR systematic errors due to a specific unmodeled effect become important. 
This further allows one to improve ringdown models and shows great potential to incorporate systematic uncertainties on top of the statistical ones. 
As detectors become more sensitive, black hole spectroscopy needs to put more attention to dealing with such systematic uncertainties to prevent false violations from GR~\cite{Gupta:2024gun,Chandramouli:2024vhw}, or properly quantify new physics that could be waiting out there. 

In Sec.~\ref{methods}, we introduce the ringdown models and the statistical framework and provide a comprehensive application of our framework in Sec.~\ref{applications}.   
We compare results obtained by LSA with those by Bayesian analysis for selected cases, and then explore a variety of effects, including unmodeled overtones, quadratic modes, and power-law tails. 
Our findings are discussed in Ref.~\ref{discussions}, and our conclusions can be found in Sec.~\ref{conclusions}. 
Complementary results can be found in Ref.~\ref{app_1}. 
In this work, we work in units with $G=c=1$.

\section{Methods}\label{methods}

In the following, we review ringdown models in Ref.~\ref{methods_a}, describe the Bayesian analysis framework and statistical methods Ref.~\ref{methods_b}, and outline LSA methods in Ref.~\ref{methods_c}.

\subsection{Modeling the Ringdown}\label{methods_a}

We model the ringdown signal as a sum of damped sinusoids, similar to the current ringdown analysis by the LVK Collaboration. 
The GW strain is then given by
\begin{align}\label{ringdown}
h(t; \bm{\theta}) = \sum_{\ell m n}^{N} A_{\ell m n} \exp\left[\mathrm{i} \left(\omega^{\mathrm{re}}_{\ell m n} t + \phi_{\ell m n} \right) - \omega^{\mathrm{im}}_{\ell m n} t\right]\,,
\end{align}
with time $t$ (chosen such that the signal starts at $t=t_0=0$), QNMs $\omega_{\ell m n} = \omega_{\ell m n}^{\mathrm{re}} + \mathrm{i} \omega_{\ell m n}^{\mathrm{im}}$, amplitudes $A_{\ell m n}$, and phases $\phi_{\ell m n}$. 
Note that $t_0$ should not be confused with the peak of a full inspiral-merger-ringdown waveform, as our analysis only employs a ringdown model. 
Instead, it sets a reference point with respect to amplitude ratios and phase shifts. 

Throughout this work, we will distinguish between two different ways of employing Eq.~\eqref{ringdown}: theory-agnostic and theory-specific. 
For the theory-agnostic case, all QNMs $\omega_{\ell m n }$, amplitudes $A_{\ell m n}$, and phases $\phi_{\ell m n}$ are treated as independent free parameters. 
Therefore, a ringdown waveform with $N$ modes corresponds to $4N$ free parameters.  
On the other hand, for the theory-specific case, the QNMs $\omega_{\ell m n } = \omega_{\ell m n }(M,a)$ are determined in terms of the stationary black hole mass $M$ and spin $a$, while the amplitudes $A_{\ell m n}$ and phases $\phi_{\ell m n}$ are still free parameters. 
In this case, a ringdown waveform with $N$ modes has $2N + 2$ free parameters. 
The numerical data for the QNMs is taken from Refs.~\cite{Berti:2009kk,Berti:2005ys}, which has been implemented in the parametrized QNM framework in Ref.~\cite{Cano:2024jkd} and is available on \texttt{github}~\cite{sebastian_volkel_2024_14001739}. 

To study the impact of unmodeled contributions to the ringdown signal, the signal, $h_\text{signal}(t; \bm{\theta}_\text{signal})$, is given by the sum of the model and the unmodeled effect, $\Delta h(t;\bm{\theta}_\text{effect})$, of interest. 
Therefore, it is described by
\begin{equation}
h_\text{signal}(t; \bm{\theta}_\text{signal}) = h(t; \bm{\theta}) + \Delta h(t;\bm{\theta}_\text{effect})\,,
\end{equation}
where $\bm{\theta}_\text{signal} = \left[\bm{\theta}, \bm{\theta}_\text{effect}\right]$. 
In Sec.~\ref{applications}, we study the effect of an overtone, a quadratic mode, and a power-law tail as the unmodeled feature. 
Whether its presence introduces biases at intermediate times, despite of not being a visible feature, is one of the question we will explore. 
The power-law tail is prescribed as 
\begin{equation}
h_\text{tail}(t) = \frac{A}{\left(t-t_\text{pole} \right)^{7}}\,,
\end{equation}
where the exponent $7$ is used to describe the $\ell=2$ Price tail~\cite{Price:1971fb,Leaver:1986gd,Gundlach:1993tp,Barack:1998bw,Hod:1999ci,Hod:2000fh}, and $A$ and $t_\text{pole}$ will be used as free parameters to control the contribution of the tail to the signal. 
These parameters, as well as the QNM amplitudes and phases, depend on the initial data, i.e., the properties of the binary. 
However, in this work, we restrict ourselves to canonical values for these parameters and do not identify them with any binary configuration.

\subsection{Bayesian Framework}\label{methods_b}

Below, we describe the standard time-domain statistical framework used in GW ringdown data analysis~\cite{Carullo:2019flw,Isi:2021iql}. 
The log-likelihood describing stationary Gaussian noise is given by 
\begin{equation}
\label{eq:likelihood}
\ln p(d |\bm{\theta}) = -\frac{1}{2}\left< d-h(\bm{\theta}) | d-h(\bm{\theta}) \right>\,,
\end{equation}
where $h(\bm{\theta})$ is the ringdown model and $d$ is the observed signal. 
We assume a flat noise power-spectral density (PSD), setting $S_h$ to be a constant, which allows for simple analytic calculations. 
While this does not accurately model the frequency-dependent noise of a realistic GW detector, it allows us to isolate the effect of just the unmodeled physics. 
The inner product used in Eq.~\eqref{eq:likelihood} then reduces to
\begin{align}\label{innerproduct_timedomain}
\left< h_1 | h_2 \right> = \frac{4}{S_h} \Re \left[ \int_{t_{\text{start}}}^{t_{\text{end}}} h_1(t) \times h_2^{*}(t) \text{d} t \right]\,,
\end{align}
which allows us to easily vary the start time $t_\text{start}$ and the end time $t_\text{end}$. 
Using a realistic power-spectral density would not directly impact the validity of LSA, but rather assign different weights and might favor a frequency-domain analysis. 
Moreover, using Eq.~\eqref{innerproduct_timedomain} is common when extracting QNM information from numerical relativity simulations, e.g., to calibrate excitation coefficients by minimum chi-square estimation. 
In Sec.~\ref{applications}, we will investigate different SNRs $\rho^2 \equiv \left< h | h\right>$ by simply adjusting the value of $S_h$, instead of changing the signal itself. 
This is equivalent to adjusting the distance between the detector and the source. 

In order to infer the parameters of a model, one can use Bayes' theorem
\begin{align}
p(\bm{\theta}|d) = \frac{p(d |\bm{\theta}) p(\bm{\theta})}{p(d)}\,,
\end{align}
which relates the probability distribution of the parameters given the data $p(\bm{\theta}|d)$, with the likelihood of the data given the parameters $p(d |\bm{\theta})$. 
The posterior further depends on our prior knowledge of the parameters $p(\bm{\theta})$, before analyzing the data, and the probability of the data itself $p(d)$, the evidence, which is just a normalization term. 
In this work, we assume uniform prior distributions for all parameters of our model. 

A direct computation of the posterior distribution from Bayes' theorem is only possible in special cases. 
However, there is a wide range of statistical methods that allow one to draw direct samples from the posterior by providing the likelihood and prior, independent from the evidence. 
The posterior distribution can then be approximated by obtaining many samples. 
Among these methods are MCMC techniques. 
In this work, we sample the likelihood Eq.~\eqref{eq:likelihood} using the \texttt{EMCEE} sampler~\cite{Foreman-Mackey:2012any}, which is a widely used MCMC code. 
Due to our simplifying assumptions of Gaussian noise in the time domain and using analytic functions (damped sinusoids), the computational cost of sampling is moderate. 
We use it to benchmark the results obtained from LSA, which is introduced next.

\subsection{Linear Signal Analysis}\label{methods_c}

Although MCMC methods are capable of providing posterior distributions, they are computationally expensive and thus quickly become a bottleneck for vast parameter space explorations. 
A well-established approach in gravitational wave data analysis, circumventing posterior sampling, is LSA. 
It is applied in the high-SNR limit, when an expansion of the likelihood around its maximum can be well approximated in terms of a linear expansion of the waveform model with respect to its parameters. 

We start with the prescription of obtaining statistical errors and correlations in LSA. 
To estimate the statistical errors of the ringdown parameters $\bm{\theta}$, we utilize the Fisher information matrix. 
It represents a quadratic approximation of the log-likelihood around its maximum, depending on the parameters of interest, and is valid for large SNR. 
Its inverse is the covariance matrix $C=\Gamma^{-1}$ whose entries $\sigma_{ij}$ contain the statistical errors $\Delta \bm{\theta}_i\equiv\sqrt{\sigma_{ii}}$ and correlations.  
The elements of the Fisher information matrix are given by 
\begin{align}
\Gamma_{ij} = \left< \partial_{i} h | \partial_{j} h \right> \,,
\end{align}
where the partial derivatives are with respect to the components of $\bm{\theta}$. 
Due to the simple structure of the ringdown model Eq.~\eqref{ringdown} and inner product, all of its elements can be computed analytically, both for the theory-specific and theory-agnostic case. 

A second application of LSA is the prediction of systematic errors for $\bm\theta$, also known as biases $\delta \bm{\theta}$. 
Biases are expected when the model used to analyze a signal is incomplete or inaccurate. 
Assuming that the difference between signal $h_\mathrm{signal}$ and model $h$ is small but finite, one can use the so-called bias formula
\begin{align}\label{eq_bias}
\delta \bm{\theta}_i = \left(\Gamma_{ij} \right)^{-1} \left< \partial_{j} h | h_\mathrm{signal} -  h\right>\,,
\end{align}
to predict the ``shifts'' in the parameters that one would obtain when performing inference (for large SNR)~\cite{Flanagan:1997kp,Cutler:2007mi}. 
Note that when applied to bias computations across different waveform families, e.g., for precessing binary black hole systems, an important alignment procedure has been pointed out in Ref.~\cite{Dhani:2024jja} and adopted in Ref.~\cite{Kapil:2024zdn}. 
Since our case allows for full analytic results for the elements of the Fisher information matrix $\Gamma_{ij}$, we only need to invert it numerically to obtain its inverse. 
Note that the bias formula does not require one to have an analytic representation of the signal. 
Because we also want to consider signals for which there is no analytic expression and be flexible for future applications, we compute the inner product for the bias formula numerically. 
Note that typical computation of the Fisher information matrix and bias formula is very fast for the ringdown model, i.e., many orders of magnitude faster than MCMC sampling.  

Finally, to quantify the relevance of biases in a compact form, we introduce the bias ratio $\delta \bm{\theta}_i/\Delta \bm{\theta}_i$. 
This quantity is smaller than one when parameters are dominated by statistical errors, and larger than one if dominated by systematic errors.

\section{Applications}\label{applications}

In this section, we apply our framework to different types of scenarios that are interesting for black hole spectroscopy. 
To set the stage for comprehensive parameter space explorations for which MCMC analyses become too expensive, we first quantify the agreement between LSA and MCMC analysis for selected cases in Sec.~\ref{applications_a}. 
We then proceed with different types of detailed parameter space explorations that are only feasible in a reasonable time with LSA in Sec.~\ref{applications_b}. 
The latter includes how biases from unmodeled QNMs depend on their phase difference, the impact of varying the start time of the ringdown analyses, and how the large zoo of possibly unmodeled QNMs introduces biases when being modeled one at a time, see Refs.~\cite{Yi:2025pxe,Capuano:2025kkl} for recent examples using LSA for LISA-related applications.

\subsection{LSA versus Bayesian Analysis}\label{applications_a}

\begin{figure}
\includegraphics[width=1.0\linewidth]{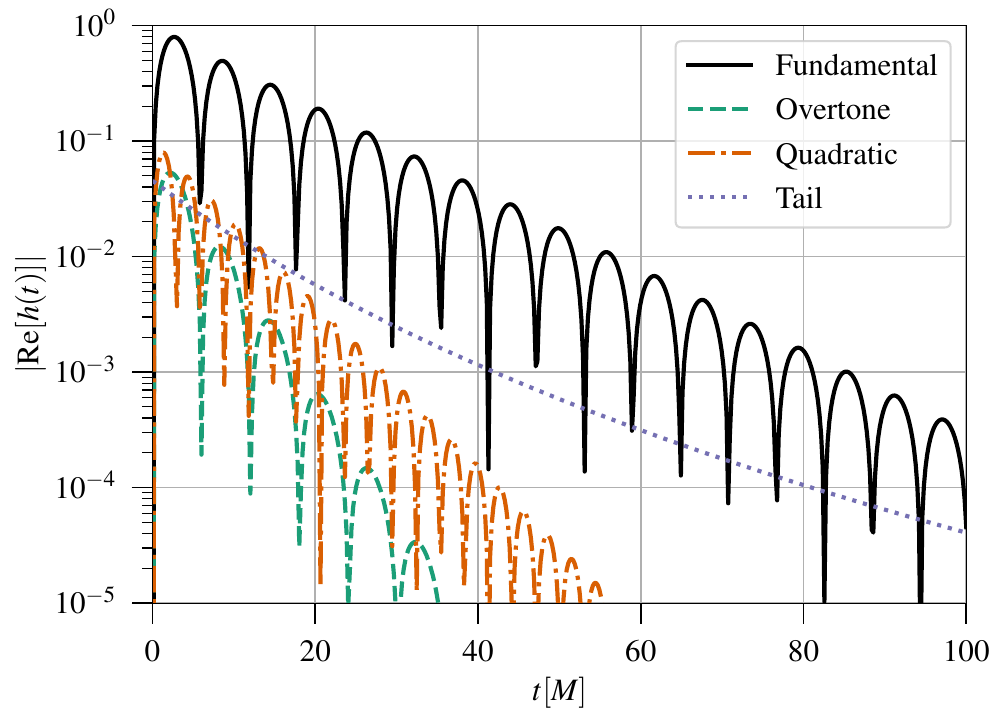}
\caption{\textit{Ringdown signal and model:} 
We show the fundamental mode $\omega_{220}$ along with the contribution of an unmodeled overtone $\omega_{221}$, unmodeled quadratic QNM $\omega_{220 \times 220}$, and unmodeled power-law tail. 
Numerical values of the parameters are reported in Table~\ref{table1}. 
\label{fig_waveform}}
\end{figure}

We consider specific examples to demonstrate that the predictions from LSA can be accurate when compared to the full Bayesian MCMC analysis. 
We consider a signal composed of a fundamental mode QNM along with either an overtone, a quadratic mode, or a power-law tail. 
The complex frequencies correspond to the prograde quadrupolar mode of a BH with unit mass and dimensionless spin of 0.7. 
The amplitude of the additional mode is fixed at 10\,\% of the fundamental mode at the beginning of the signal. 
Note that the beginning of the signal is not intended to represent the actual peak amplitude of a binary merger ringdown. 
In this case, the amplitudes would typically be much larger, and it would be important to incorporate additional linear and quadratic QNMs simultaneously. 
Instead, our example rather refers to a situation where the analysis is started when one has either subtracted other modes (but not completely) or assumes they are negligible. 
Our values are chosen as qualitative examples and do not correspond to any particular binary configuration. 
We summarize the different contributions in Sec.~\ref{fig_waveform}, which shows the fundamental mode ($\omega_{220}$), overtone ($\omega_{221}$), quadratic QNM ($\omega_{220 \times 220}$), and the power-law tail. 
Note that the quadratic mode oscillates and decays twice as fast as the fundamental mode but is still longer lived than the overtone. 
All numerical values are reported in Table~\ref{table1}.

\begin{table}
\centering
\caption{\textit{Parameters of the ringdown signals}: 
Values of each case are in terms of the theory-specific parametrization. 
Theory-agnostic real and imaginary parts of the QNMs are completely determined by mass and spin, as shown in the table. 
Their numerical values are $\omega_{220} = 0.53260024 +\mathrm{i} 0.08079287$, $\omega_{221} = 0.52116077 +\mathrm{i} 0.24423832$, $\omega_{220\times 220} = 1.06520049 +\mathrm{i} 0.16158575$. 
\label{table1}
}
\setlength{\tabcolsep}{6pt} 
\renewcommand{\arraystretch}{1.2} 

\bigskip
\begin{tabular}{@{} l c c c @{}}
\toprule
\textbf{Parameters}         & \textbf{Case 1}        & \textbf{Case 2}          & \textbf{Case 3} \\
\midrule
$M$                        & 1                        & 1                          & 1 \\
$a$                        & 0.7                      & 0.7                        & 0.7 \\
$A_{220}$                  & 1.0                      & 1.0                        & 1.0 \\
$\phi_{220}$               & $\pi/2$                  & $\pi/2$                    & $\pi/2$ \\
$A_{221}$                  & 0.1                      & ---                        & --- \\
$\phi_{221}$               & $\pi/2$                  & ---                        & --- \\
$A_{220 \times 220}$       & ---                      & 0.1                        & --- \\
$\phi_{220 \times 220}$    & ---                      & $\pi/2$                    & --- \\
$A_{\mathrm{tail}}$         & ---                      & ---                        & $1 \times 10^{11}$ \\
$t_{\mathrm{pole}} [M]$         & ---                      & ---                        & $-58$ \\
\bottomrule
\end{tabular}
\end{table}

\begin{figure}[h]
\includegraphics[width=1.0\linewidth]{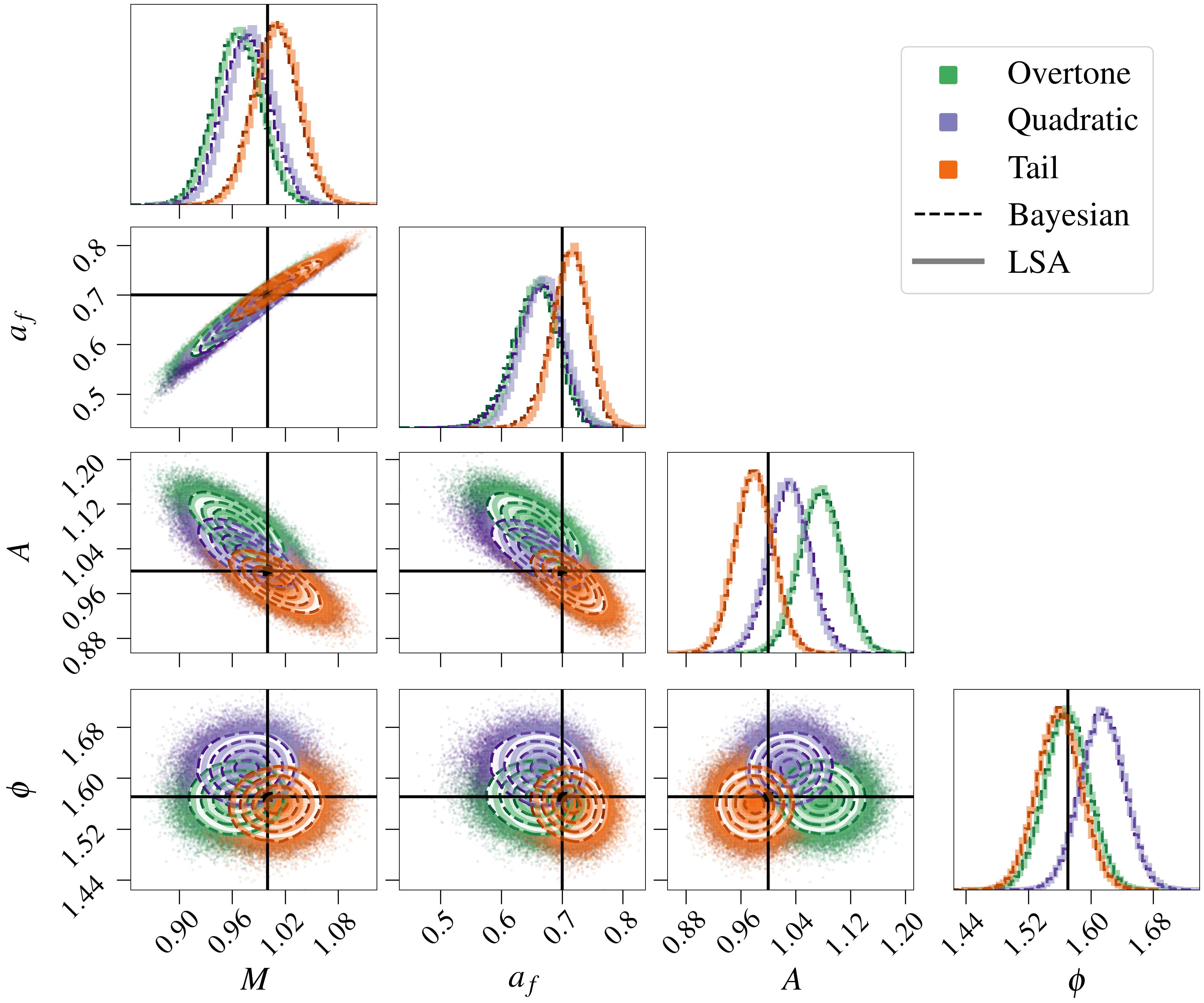}
\hfill
\vspace{0.3cm}
\includegraphics[width=1.0\linewidth]{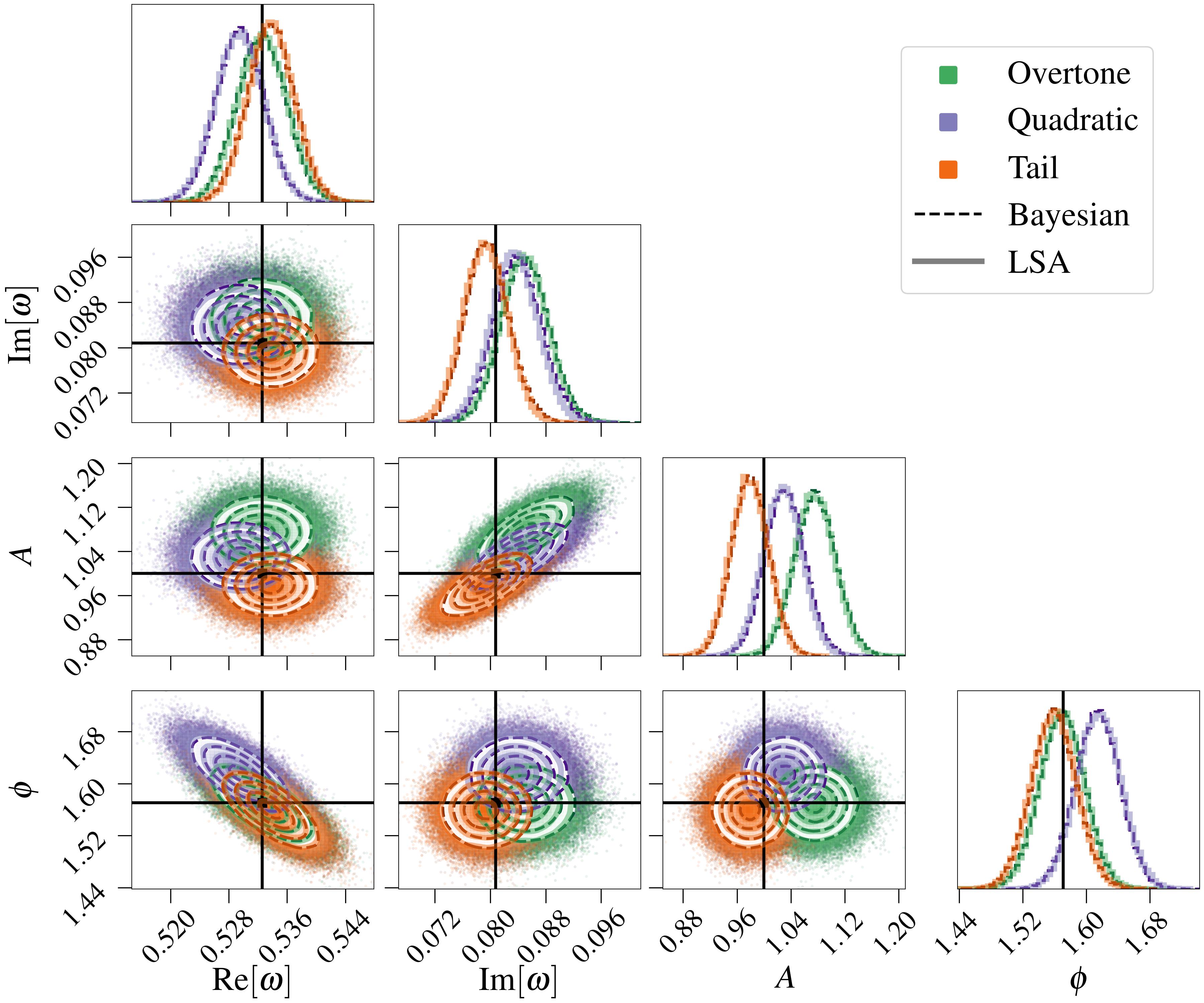}
\caption{\textit{Comparison of LSA versus Bayesian analysis:} 
Here we compare the theory-specific (top panel) and theory-agnostic (bottom panel) biases for three unmodeled effects (different colors) described in the main text. 
In each panel, we compare posteriors obtained using the Bayesian (thin dashed lines) and LSA (thick solid lines) methods, respectively. 
The values of the injected parameters are shown for reference (black crosshair). 
In this example, the signal has SNR=50. \label{fig_mcmc_fisher_50}}
\end{figure}

We inject a signal with SNR=50 containing the fundamental model and an additional effect, which could be either an overtone, a quadratic QNM, or a power-law tail. 
We analyze this signal using a single QNM model corresponding to the fundamental mode. 
We report the comparison between the Bayesian analysis and the Fisher estimates in Sec.~\ref{fig_mcmc_fisher_50}, with the top panel displaying the theory-specific case and the theory-agnostic case in the bottom panel. 
The crosshairs depict the true values of the parameters, while the different cases are shown in different colors.

As evident from Fig.~\ref{fig_mcmc_fisher_50}, the overall agreement between the full Bayesian posteriors and LSA predictions is excellent. 
In the theory-specific case, we observe that mass $M$ and spin $a$ posteriors follow a similar correlation in all the cases. 
In the theory-agnostic case, the posteriors of the real and imaginary part of $\omega$ do not follow such a correlation. 
Instead, the real part is less sensitive to the unmodeled effects, while the imaginary part is similarly biased as mass and spin in the theory-specific case. 
Although the biases in the examples presented here are not dramatic, they would already be problematic for tests of GR. 
Finally, we note that the biases in amplitude and phase are very similar in the theory-specific and theory-agnostic cases. 
In fact, the largest bias appears for the amplitude in the case of an unmodeled overtone. 
This could potentially be interesting in the context of amplitude-phase consistency tests of GR as proposed in Ref.~\cite{Forteza:2022tgq}. 
We provide additional results for the same unmodeled effects, but SNR=20 and SNR=100 in the appendix in Fig.~\ref{fig_mcmc_fisher_20_100}. 
We find that the accuracy of LSA degrades at SNR=20, with a small but notable better performance in the theory agnostic case. 
At SNR=100, the systematic errors typically exceed the statistical ones, which would be highly problematic for tests of GR. 

In Fig.~\ref{fig_snr_grid} of the appendix, we provide a comprehensive comparison of bias ratios as a function of selected SNRs starting from 20 to 300. 
They correspond to the same effects as studied previously, but condense the median and 68\,\% of the Bayesian results for the 1d marginalized posteriors of the parameters of the theory-specific and theory-agnostic model, respectively. 
In summary, as expected, the agreement between LSA and Bayesian results improves with increasing SNR. 
Remarkably, the quantitative details, i.e., which of the parameters is better approximated by LSA or experiences the largest biases, is very case dependent. 

\subsection{Parameter Space Explorations}\label{applications_b}

\subsubsection{Exploring QNM Phase Differences}\label{app2}

\begin{figure}[h]
\includegraphics[width=1.0\linewidth]{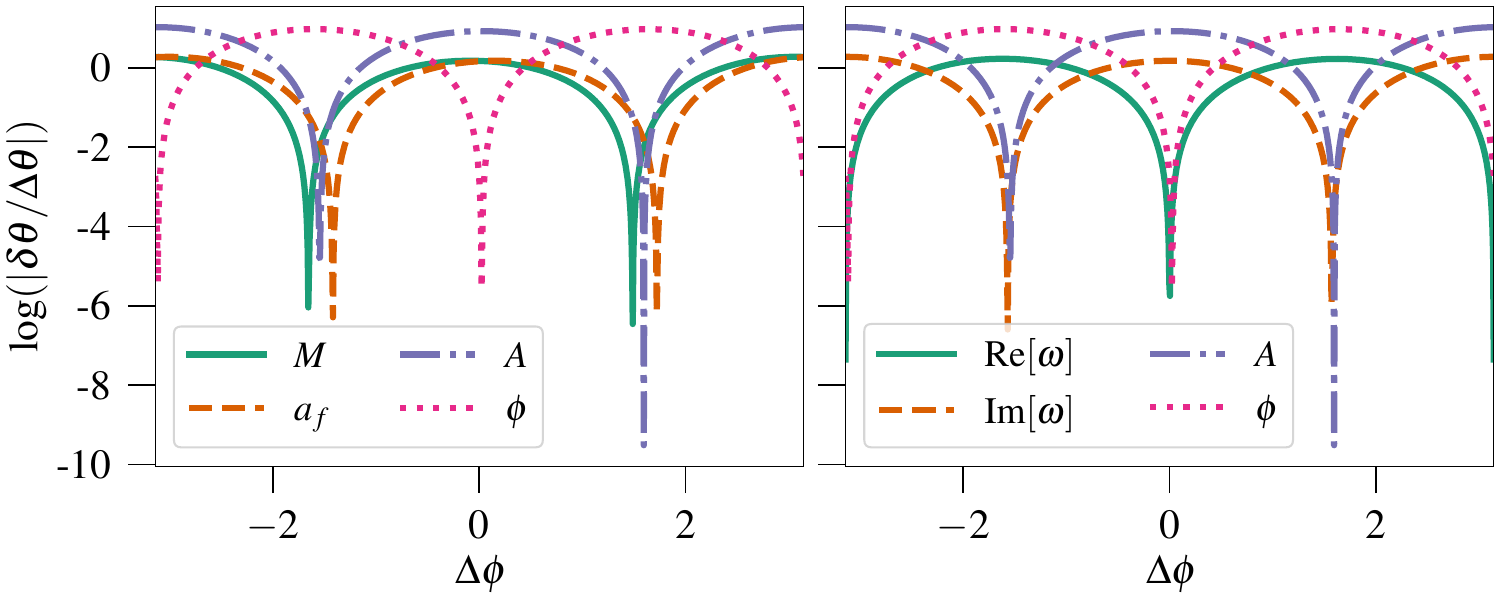}
\includegraphics[width=1.0\linewidth]{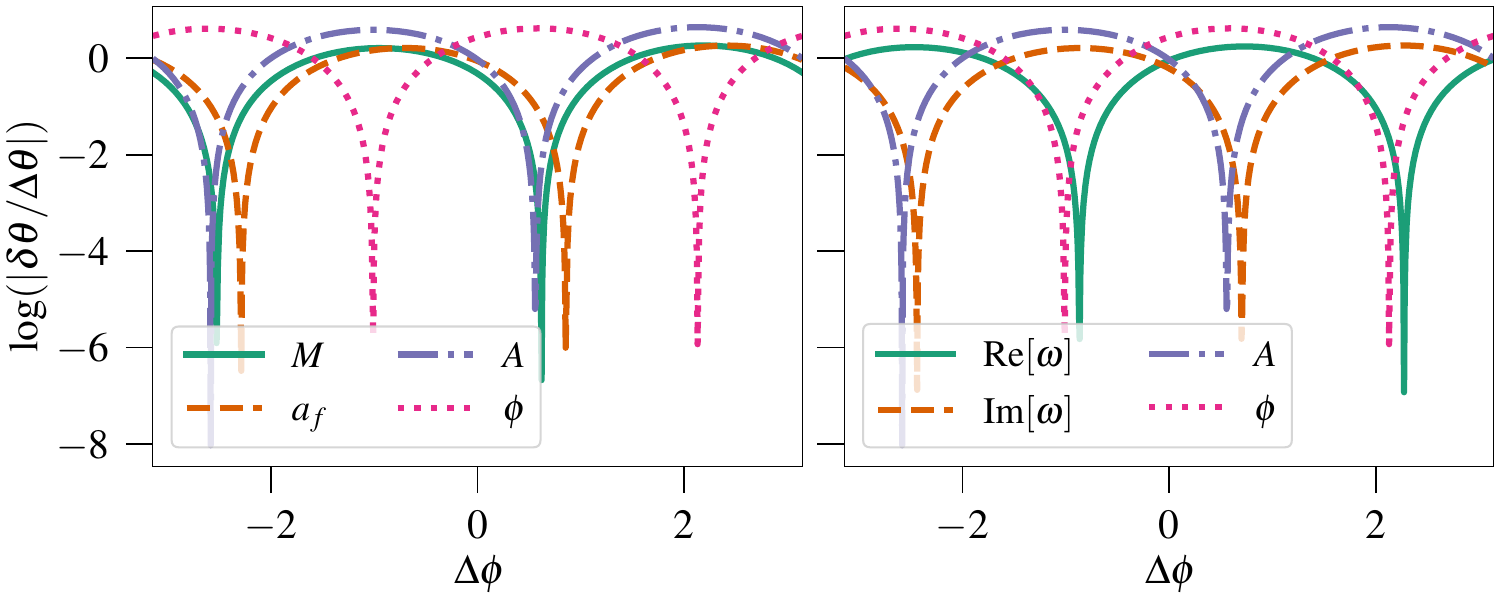}
\caption{\textit{Bias ratios as a function of QNM phase differences:} In all panels, we show the bias ratio originating from an unmodeled QNM with a relative amplitude of $0.1$ to the $\omega_{220}$ fundamental mode as a function of their phase difference $\Delta \phi$ for SNR=50. 
In the top row, we consider $\omega_{221}$, and in the bottom row, we consider the quadratic $\omega_{220\times220}$ as an unmodeled QNM. 
Left and right columns correspond to theory-specific and theory-agnostic models, respectively. 
\label{fig_phase_difference_1}
}
\end{figure}

Having established the validity of the LSA, we explore the impact of the phase difference between the fundamental mode and an unmodeled QNM on systematic biases. 
In Fig.~\ref{fig_phase_difference_1}, we show the bias ratios of the theory-specific and theory-agnostic case when using only the fundamental mode $\omega_{220}$ as model, but allowing either an unmodeled overtone $\omega_{221}$ or an unmodeled quadratic mode $\omega_{220 \times 220}$ with an initial amplitude  of $10\,\%$ of the fundamental, respectively. 

The results in Fig.~\ref{fig_phase_difference_1} show nontrivial features that might easily remain unnoticed in a less detailed analysis. 
We find that the phase difference has a huge impact on bias ratios, which vary by many orders of magnitude.  
Moreover, it also significantly changes which of the parameters is most or least biased. 
For instance, let us first compare results for the phase with those in Fig.~\ref{fig_mcmc_fisher_50}. 
We observe that the phase bias is almost vanishing for the unmodeled overtone $\omega_{221}$ (shown in the top row) only due to the specific choice of injecting it with zero phase difference. 
In contrast, the phase bias from the unmodeled overtone $\omega_{221}$ would be negligible around other phase differences. 
Moreover, we find that the bias ratios for amplitude and phase are very similar across the theory-specific and theory-agnostic models, as long as the unmodeled effect is the same. 
This generalizes the specific finding from Fig.~\ref{fig_mcmc_fisher_50} that amplitude and phase posteriors are very similar for the same unmodeled effect. 
Finally, the dependency of the bias ratios for the mass $M$ and spin $a$ or real and imaginary part of $\omega$ is quite different from each other. 
The value of phase differences at which one of its bias ratios vanishes seems quite unrelated. 

\subsubsection{Exploring Start Times of Ringdown Analysis}\label{app3}

In our next example, we quantify how starting the ringdown analysis at later times changes the bias ratios. 
Similar to varying the phase shifts in Sec.~\ref{applications_b}, each choice for the starting time corresponds to an independent analysis. 
One would expect that any effect decaying faster than the fundamental mode should result in smaller biases, while the statistical errors become larger. 
In the presence of tails, the same scaling reasoning does not apply to the biases. 
The power-law tail will take over the signal at very late times, but depending on when it starts in the ringdown, and what value one chooses for $t_\text{pole}$, it may also be relevant at early times.

\begin{figure}[h]
\includegraphics[width=1.0\linewidth]{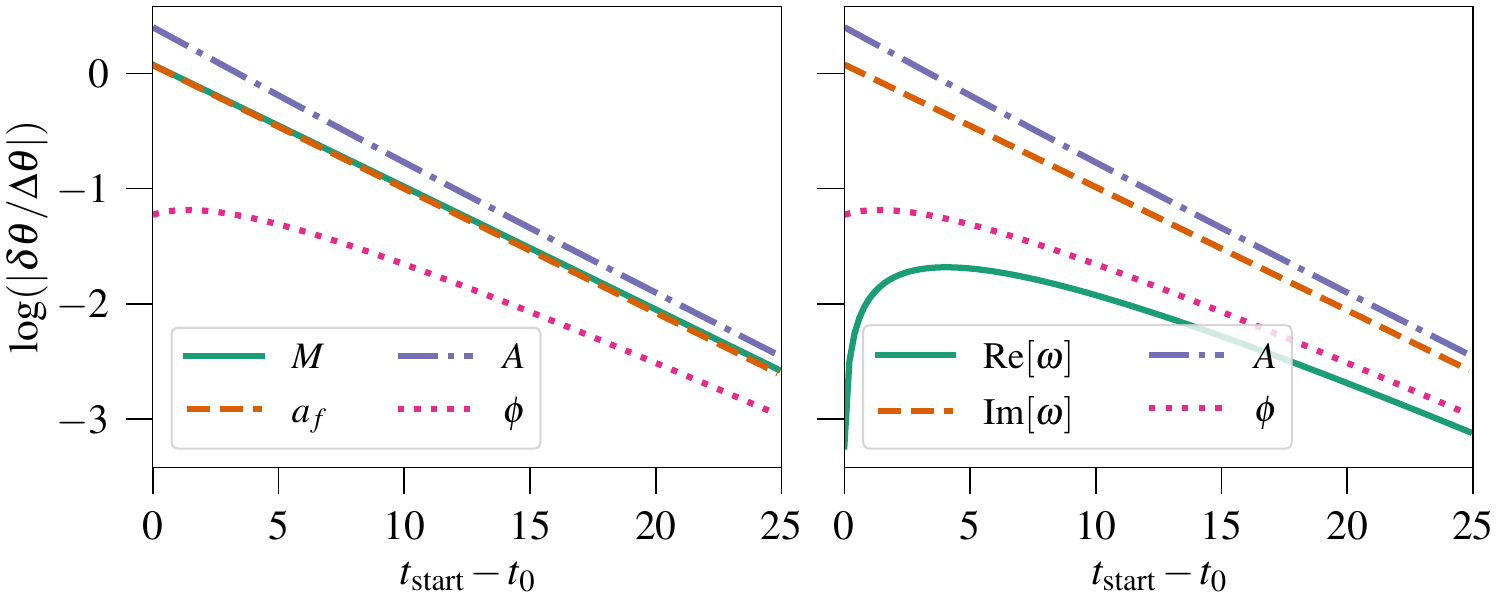}
\includegraphics[width=1.0\linewidth]{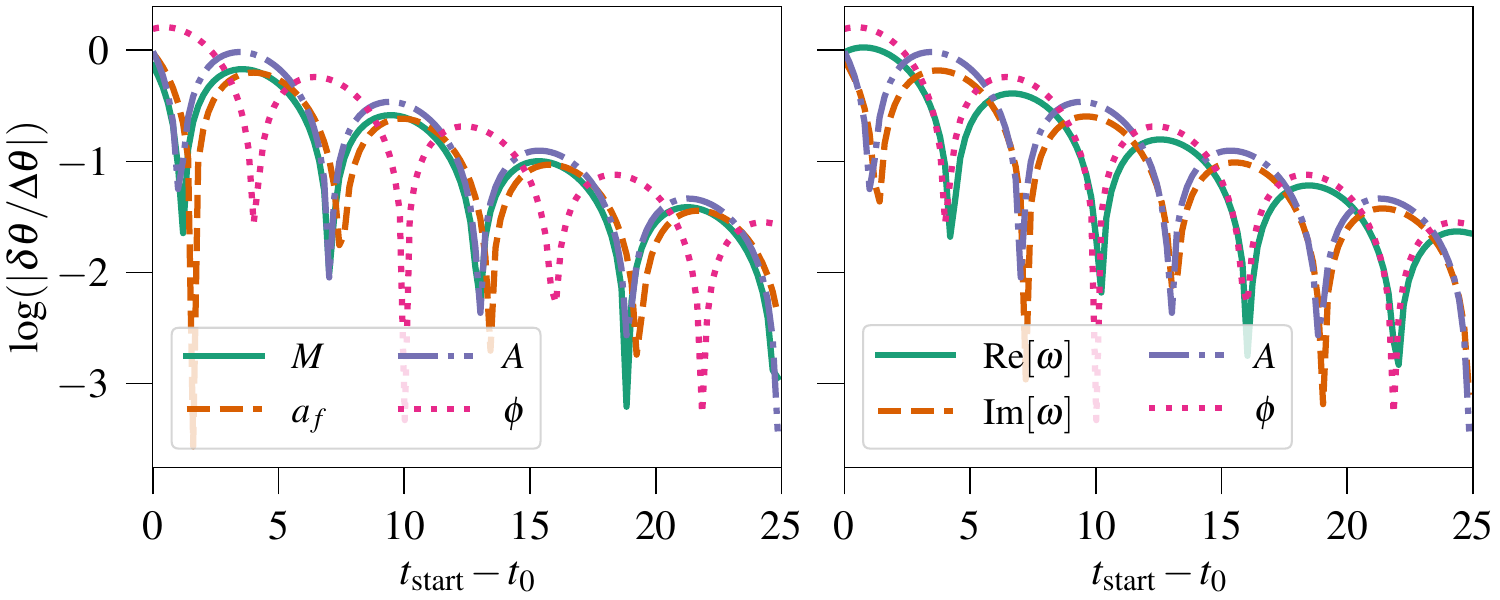}
\includegraphics[width=1.0\linewidth]{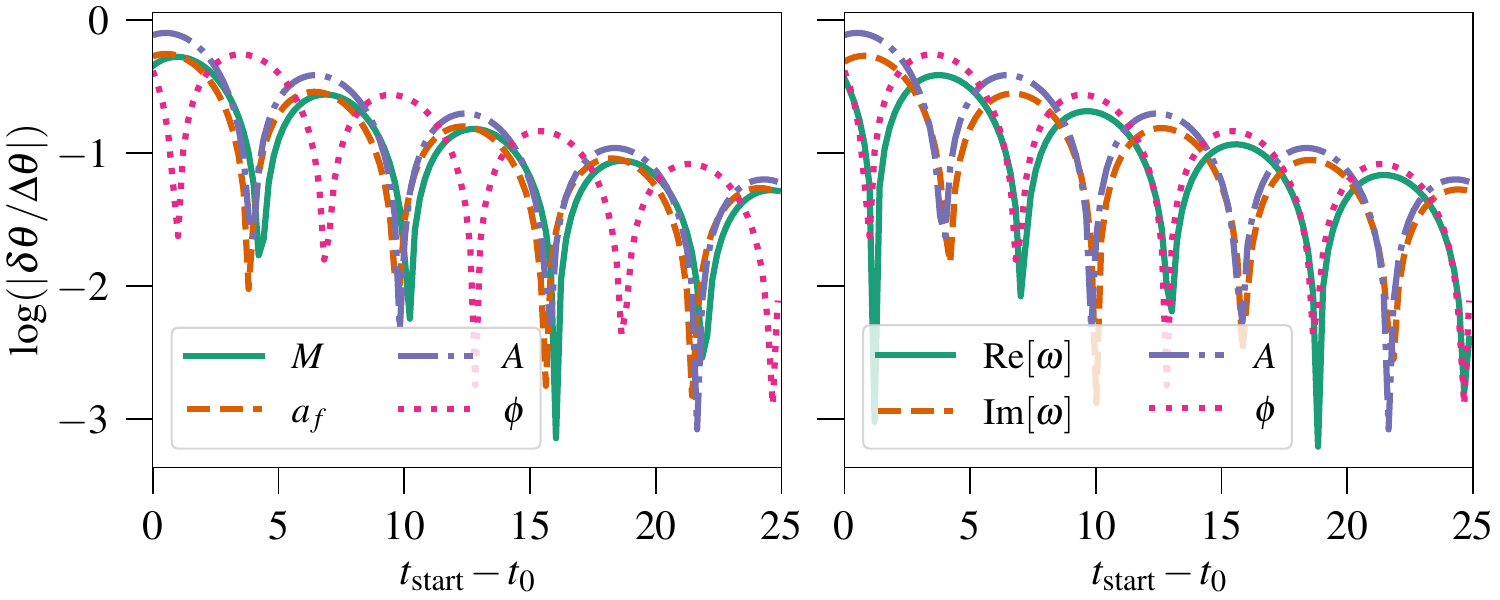}
\caption{\textit{Bias ratios as a function of starting time:} In all panels, we show the bias ratio originating from an unmodeled QNM with a relative amplitude of $0.1$ to the $\omega_{220}$ fundamental mode as a function of the starting time $t_\text{start}-t_\text{peak}$ for SNR=50. 
The top, middle, and bottom rows show biases from unmodeled $\omega_{221}$, quadratic $\omega_{220\times220}$, and power-law tail contributions, respectively. 
Left and right columns correspond to theory-specific and theory-agnostic models, respectively.
\label{fig_starting_time_1}}
\end{figure}

In Fig.~\ref{fig_starting_time_1}, we show the bias ratio as a function of the starting time for the same cases of unmodeled overtone, quadratic mode, and power-law tail as in Sec.~\ref{applications_a}. 
Starting with the overtone and quadratic mode, we find striking differences in the qualitative behavior of the bias ratios. 
Although all bias ratios tend to decrease toward later times, which is expected due to the faster decay of the effects compared to the fundamental mode, the oscillatory patterns are very different. 
In the case of the unmodeled overtone, the bias ratios do not seem to oscillate at all, while the ones from the unmodeled quadratic oscillate significantly. 
This is qualitatively in agreement with the observation that the difference in the real part of the overtone with respect to the fundamental mode is much smaller than the one between the quadratic mode and the fundamental mode. 
A quick look at the bias formula~\eqref{eq_bias} in connection with the simple model~\eqref{ringdown} (using only two modes) reveals that the modulation is either very small (overtone case), or of the same frequency as the fundamental mode (quadratic case). 
The bias ratios of the power-law tail shown in the bottom row admit qualitatively similar oscillations as the quadratic case, but decays slower. 
Also, this can be understood if one realizes that the tail is a ``low-frequency'' effect and the only oscillation comes from the fundamental mode. 
A careful look reveals that the bias ratios do not exactly decay exponentially but show indications of a slower, power-law behavior. 

We provide complementary results for SNR=100 in the appendix in Fig.~\ref{fig_starting_time_2}, which, due to the validity of LSA, is just a simple rescaling. 

\subsubsection{Exploring the Zoo of Possibly Unmodeled QNMs}\label{app4}

\begin{figure}[h]
\includegraphics[width=1.0\linewidth]{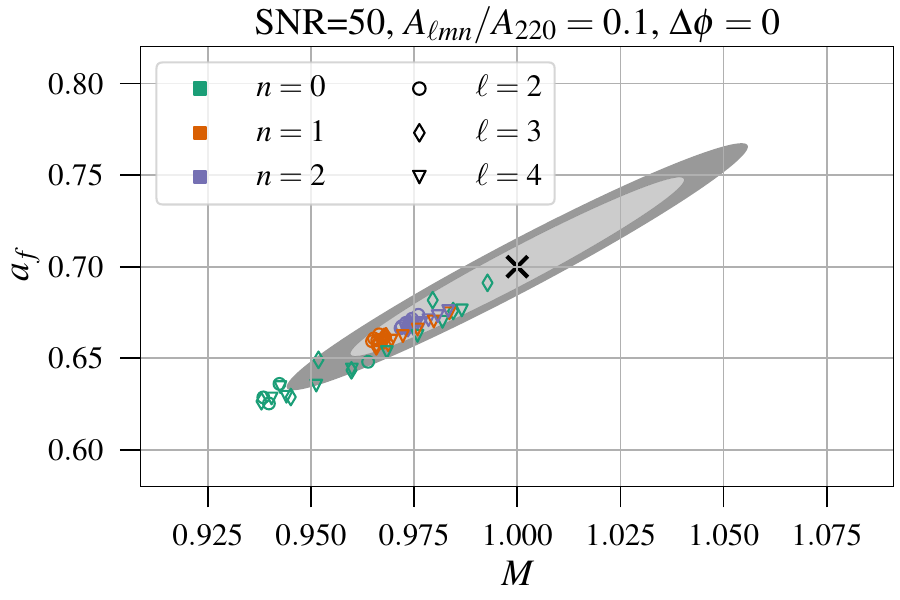}
\includegraphics[width=1.0\linewidth]{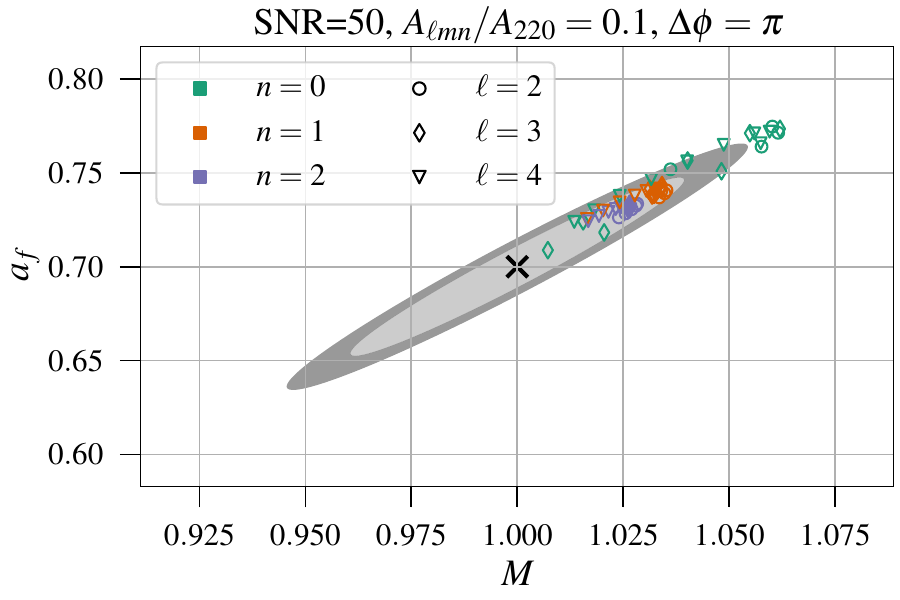}
\caption{\textit{Exploring biases from unmodeled QNMs:} 
Here we show biases of various unmodeled linear QNMs described in the main text using LSA. 
The contours indicate 70\,\% (light gray) and 90\,\% (dark gray) probability levels using the LSA prediction (Fisher) of the theory-specific fundamental mode, i.e., the marginalized errors on the mass and spin. 
The true value is shown in the center (black cross). 
In all cases, biases are computed one at a time with a relative mode amplitude of $A_{\ell m n}/A_{220}=0.1$ at SNR=50. 
The top panel shows modes with phase difference $\Delta \phi=0$, and the bottom panel with $\Delta \phi=\pi$. 
\label{fig_scatter_v1}}
\end{figure}

In Fig.~\ref{fig_scatter_v1}, we show the predicted biases for a variety of possibly unmodeled linear QNMs. 
Here we assume that the signal contains only one unmodeled QNM at a time, and it has an amplitude ratio of $A_{\ell m n}/A_{220} = 0.1$ and the SNR is 50. 
The uncertainty region corresponds to 70\,\% and 90\,\% confidence intervals. 
We show cases with phase difference between the 220 fundamental mode and any other QNM of $\Delta \phi = 0$ and $\Delta \phi = \pi$. 
We include $\ell=\left[2,3,4\right]$, $m\in[-\ell, \dots, \ell]$ and $n\in\left[0,1,2 \right]$. 
Complementary results for SNR=100 can be found in the appendix in Fig.~\ref{fig_scatter_v2}. 
To ease the presentation and avoid cluttering, we now focus on the theory-specific case and only show the bias of the final mass and spin. 

Overall, we find that biases are correlated and underestimate the mass and spin when being in phase ($\Delta \phi=0$), and overestimate both when out of phase ($\Delta \phi=\pi$), which is similar to the findings reported in Ref.~\cite{Berti:2007zu}. 
Moreover, likely due to the strong increase in damping time, biases from unmodeled overtones $n>0$ are less important overall compared to most, but not all, fundamental modes. 
Within the assumptions of LSA, one could, in principle, add biases from each QNM independently. 
Although this strategy might, in practice, break the assumption that the total effect is small, it is quite evident that biases from unmodeled QNMs of this amplitude are a significant challenge for tests of GR. 

\section{Discussion}\label{discussions}

We discuss the validity of LSA in Sec.~\ref{disc_lsa}, the extension to realistic detectors in Sec.~\ref{disc_psds}, and future application of the LSA framework in Sec.~\ref{disc_future}.

\subsection{Validity of LSA}\label{disc_lsa}

Within the range of applications considered in this work, we find LSA to be in very good agreement with accurate Bayesian MCMC results when considering SNRs of around 50. 
However, we have only considered cases in which the unmodeled effects are small. 
The bias formula is derived under the assumption that differences between signal and model are small and that the likelihood can be locally approximated by a multivariate Gaussian. 
In the following, we want to describe two cases for which LSA should be used with caution. 

First, if one has a model with several QNMs, and some of them have low SNR, then the posterior distribution of a subset of parameters may not be well approximated. 
This may cause inaccurate predictions when evaluating the bias formula, due to the contraction with the covariance matrix. 
In such a case, the bias calculation may be limited to those parameters well captured by the Fisher information matrix. 

Second, we comment on a subtle aspect regarding the high-SNR limit of the bias formula. 
In contrast to the Fisher information matrix, which should become more accurate with increasing SNR, the accuracy of the bias formula also depends on how similar the signal and model are. 
For very high SNRs ($\mathcal{O}(10^4) - \mathcal{O}(10^5)$), even a small but finite difference becomes relevant, i.e. the prediction of the bias itself has a finite error. 
Simply speaking, the predicted biases are a direct measure for the displacement of the means of the multivariate Gaussian describing the posterior, and the statistical errors simply encode its widths. 
Thus, for sufficiently high SNR, the true posterior and the one predicted from LSA would not overlap anymore, due to the finite error in predicting the systematic errors while statistical errors decrease. 
One may call this a systematic error of the systematic error itself, see also Ref.~\cite{Chandramouli:2024vhw}.
In our cases, it is much smaller than the statistical one and ellipses for the posteriors would only stop overlapping at much higher SNR (decreasing statistical errors) or if the strength of the unmodeled effect would be increased (increasing errors of the LSA prediction).

\subsection{Extension to Realistic Detectors}\label{disc_psds}

As outlined in Sec.~\ref{methods_b}, we assume flat noise in the time domain to allow for the analytic computation of the Fisher information matrix. 
This is well suited for the purpose of analyzing high-accuracy simulations to obtain best-fit parameters and their systematic errors. 
However, this simplification is not adequate to make quantitative predictions for realistic gravitational wave detectors, as those are characterized by a frequency-dependent power-spectral density. 
Since our work does not focus on specific sources and detectors but scales the detector directly by the signal SNRs, it rather demonstrates the methodology. 
Note that this aspect is not important when comparing Bayesian analysis and LSA, as we use the same approximation in both analyses. 

Future work focusing on detector-specific questions could either work with a more complicated detector response in the time domain, e.g., as described here~\cite{Carullo:2019flw,Isi:2021iql}, or transform the model and signal into the frequency domain. 
Although both approaches are, in principle, straightforward, one should not underestimate subtle complications due to the finite length of the signal, especially if the start time or end time are being varied~\cite{Isi:2019aib,Cotesta:2022pci,Isi:2022mhy,Wang:2023xsy}. 
A main complication would be that the Fisher information matrix cannot, in general, be computed analytically. 
This would reduce the overall efficiency of computations for large parameter space explorations and calls for a careful numerical treatment when exploring different regions of the parameter space. 
Nevertheless, LSA still provides an efficient way to complement traditional Bayesian MCMC injection and recovery campaigns.

\subsection{Future Applications of LSA Framework}\label{disc_future}

Given the computational advantages and robustness of LSA in our applications, there is a range of possible extensions that could be considered in future work. 
Since our study demonstrates the performance of LSA for several examples, one obvious extension could be to consider amplitudes and phases as extracted from NR simulations. 
This would allow one to connect the relevance of the various harmonics $\ell m n $ with the properties of BBH systems before merger. 
Especially in light of large mass ratios, eccentricity, and precession, it is far from trivial to rank subdominant excitations. 
The systematic errors originating from not including or mismodeling the amplitudes and phases themselves, e.g., how accurate are their predictions, have not been explored in depth. 
We expect such challenging BBH systems with high SNR to be particularly important for next-generation detectors like LISA. 
In this context, LSA was recently applied to study ringdown systematics for the LISA detector in Refs.~\cite{Yi:2025pxe,Capuano:2025kkl}, which demonstrates that future black hole spectroscopy relies crucially on identifying how many QNMs need to be included, see also Ref.~\cite{Toubiana:2023cwr}. 
Finally, it would be intriguing to extend our analysis to the more general scenario in which multiple QNMs are being modeled and multiple QNMs are treated as unmodeled effects. Such an analysis using LSA requires added care because the SNRs in the subleading order modes can be of $\mathcal{O}(few)$ even when the leading fundamental mode has large SNR (${\sim}50$), as can be seen in the recent event GW250114~\cite{LIGOScientific:2025obp}.

\section{Conclusions}\label{conclusions}

Black hole spectroscopy is a cornerstone of gravitational wave astronomy and opens up precision tests of strong gravity and dynamic spacetimes of black holes. 
Extracting multiple QNMs from the ringdown of two merging compact objects allows one to extract the source's properties and test GR's predictions. 
However, conceptually difficult questions like how many QNMs need to be included in a ringdown model make it clear that theoretical uncertainties need to be taken seriously. 
A large volume of studies, partially with conflicting results and differences in working principles, support this further~\cite{Isi:2019aib,Cotesta:2022pci,Correia:2023bfn,Wang:2023ljx,Giesler:2019uxc,Dhani:2020nik,Dhani:2021vac,Forteza:2021wfq,Giesler:2024hcr,Mitman:2022qdl,Cook:2020otn,Nee:2023osy,Baibhav:2023clw}. 
Although improvements in fitting routines allow one to extract more QNMs, the problem remains challenging, even in ideal conditions~\cite{Cheung:2023vki,Giesler:2024hcr,Mitman:2025hgy,Gao:2025zvl}. 
Thus, black hole spectroscopy must improve in quantifying theoretical uncertainties to provide robust results. 
Such steps are essential for unbiased parameter estimation and crucial for accurate tests of GR with the ringdown. 

In this work, we demonstrated that the Fisher information matrix and bias formula, both primary tools from LSA, can yield fast and accurate estimates of theoretical uncertainties. 
In particular, we have focused on theory-specific and theory-agnostic ringdown tests in which a small unmodeled contribution contaminates the analyzed signal. 
To quantify the validity of LSA, we first studied representative examples with traditional Bayesian MCMC sampling in Sec.~\ref{applications_a}. 
LSA can, in most cases, accurately describe biases from an unmodeled overtone, quadratic QNM, or power-law tail at an SNR of 50. 
While such ringdown SNRs are not common with current GW detectors, the LVK Collaboration has already observed one such event---GW250114~\cite{LIGOScientific:2025obp}---which has a ringdown SNR of ${\sim}40$. 
With increasing detector sensitivity in upcoming observing runs, such events---expected to be 1\,\% of the total detections~\cite{Others:2025nbi}---will have ringdown SNRs ${>}50$. 
Furthermore, future observatories, such as the Einstein Telescope and Cosmic Explorer, will routinely observe BBH mergers with large ringdown SNRs~\cite{ET:2019dnz,Bhagwat:2019dtm}.

In Sec.~\ref{applications_b}, we move to vast parameter space explorations for which traditional Bayesian MCMC sampling is computationally prohibitive. 
In Sec.~\ref{app2} and Sec.~\ref{app3}, we provide a detailed overview of how phase differences of an unmodeled QNM and different choices of starting time impact biases, respectively. 
LSA explains typical fitting instabilities/inaccuracies in the presence of ringdown data with a tail contribution. 
Thus, even if tails are not visible in the analyzed signal, they can cause significant biases. 
This observation may explain why QNM fitting in actual black hole signals, including a tail (even if not visible in the fitting window), is more difficult than for signals without a tail (either by construction or due to the underlying wave equation). 

In Sec.~\ref{app4}, we explore the zoo of possibly unmodeled QNMs of various $\ell m n $ and how they bias our estimates for mass and spin. 
We find that unmodeled QNMs with a relative amplitude ratio of only 0.1 are typically sufficient to obtain theoretical errors that exceed the statistical ones. 
Such contributions are clearly relevant if one starts the ringdown analysis later, when such QNMs may falsely be considered too small. 
Focusing on the fundamental mode, we find that biases from unmodeled overtones are less
important overall compared to most, but not all, fundamental modes, if they are injected with the same relative amplitude.

In Sec.~\ref{discussions}, we discussed that quantifying ringdown biases with LSA opens up many possible applications, ranging from analyzing high-precision simulations to practically mitigating theoretical errors in actual ringdown models. 
While some aspects of our work could already be studied using high-precision simulations, extending it to detector-specific studies with realistic noise properties will require some basic extensions.

\acknowledgments

The authors thank Keefe Mitman for useful discussions, and Emanuele Berti, Nicola Franchini, and Saul Teukolsky for valuable comments on the manuscript. 
The authors also thank the anonymous referee for providing feedback that improved the presentation of this work. 
S.\,H.\,V. acknowledges funding from the Deutsche Forschungsgemeinschaft (DFG): Project No. 386119226.

\bibliography{literature}

\begin{thebibliography}{105}%
\makeatletter
\providecommand \@ifxundefined [1]{%
 \@ifx{#1\undefined}
}%
\providecommand \@ifnum [1]{%
 \ifnum #1\expandafter \@firstoftwo
 \else \expandafter \@secondoftwo
 \fi
}%
\providecommand \@ifx [1]{%
 \ifx #1\expandafter \@firstoftwo
 \else \expandafter \@secondoftwo
 \fi
}%
\providecommand \natexlab [1]{#1}%
\providecommand \enquote  [1]{``#1''}%
\providecommand \bibnamefont  [1]{#1}%
\providecommand \bibfnamefont [1]{#1}%
\providecommand \citenamefont [1]{#1}%
\providecommand \href@noop [0]{\@secondoftwo}%
\providecommand \href [0]{\begingroup \@sanitize@url \@href}%
\providecommand \@href[1]{\@@startlink{#1}\@@href}%
\providecommand \@@href[1]{\endgroup#1\@@endlink}%
\providecommand \@sanitize@url [0]{\catcode `\\12\catcode `\$12\catcode `\&12\catcode `\#12\catcode `\^12\catcode `\_12\catcode `\%12\relax}%
\providecommand \@@startlink[1]{}%
\providecommand \@@endlink[0]{}%
\providecommand \url  [0]{\begingroup\@sanitize@url \@url }%
\providecommand \@url [1]{\endgroup\@href {#1}{\urlprefix }}%
\providecommand \urlprefix  [0]{URL }%
\providecommand \Eprint [0]{\href }%
\providecommand \doibase [0]{https://doi.org/}%
\providecommand \selectlanguage [0]{\@gobble}%
\providecommand \bibinfo  [0]{\@secondoftwo}%
\providecommand \bibfield  [0]{\@secondoftwo}%
\providecommand \translation [1]{[#1]}%
\providecommand \BibitemOpen [0]{}%
\providecommand \bibitemStop [0]{}%
\providecommand \bibitemNoStop [0]{.\EOS\space}%
\providecommand \EOS [0]{\spacefactor3000\relax}%
\providecommand \BibitemShut  [1]{\csname bibitem#1\endcsname}%
\let\auto@bib@innerbib\@empty
\bibitem [{\citenamefont {Detweiler}(1980)}]{Detweiler:1980gk}%
  \BibitemOpen
  \bibfield  {author} {\bibinfo {author} {\bibfnamefont {S.~L.}\ \bibnamefont {Detweiler}},\ }\bibfield  {title} {\bibinfo {title} {{ Black holes and gravitational waves. III - The resonant frequencies of rotating holes}},\ }\href {https://doi.org/10.1086/158109} {\bibfield  {journal} {\bibinfo  {journal} {Astrophys. J.}\ }\textbf {\bibinfo {volume} {239}},\ \bibinfo {pages} {292} (\bibinfo {year} {1980})}\BibitemShut {NoStop}%
\bibitem [{\citenamefont {Dreyer}\ \emph {et~al.}(2004)\citenamefont {Dreyer}, \citenamefont {Kelly}, \citenamefont {Krishnan}, \citenamefont {Finn}, \citenamefont {Garrison},\ and\ \citenamefont {Lopez-Aleman}}]{Dreyer:2003bv}%
  \BibitemOpen
  \bibfield  {author} {\bibinfo {author} {\bibfnamefont {O.}~\bibnamefont {Dreyer}}, \bibinfo {author} {\bibfnamefont {B.~J.}\ \bibnamefont {Kelly}}, \bibinfo {author} {\bibfnamefont {B.}~\bibnamefont {Krishnan}}, \bibinfo {author} {\bibfnamefont {L.~S.}\ \bibnamefont {Finn}}, \bibinfo {author} {\bibfnamefont {D.}~\bibnamefont {Garrison}},\ and\ \bibinfo {author} {\bibfnamefont {R.}~\bibnamefont {Lopez-Aleman}},\ }\bibfield  {title} {\bibinfo {title} {{Black hole spectroscopy: Testing general relativity through gravitational wave observations}},\ }\href {https://doi.org/10.1088/0264-9381/21/4/003} {\bibfield  {journal} {\bibinfo  {journal} {Class. Quant. Grav.}\ }\textbf {\bibinfo {volume} {21}},\ \bibinfo {pages} {787} (\bibinfo {year} {2004})},\ \Eprint {https://arxiv.org/abs/gr-qc/0309007} {arXiv:gr-qc/0309007} \BibitemShut {NoStop}%
\bibitem [{\citenamefont {Berti}\ \emph {et~al.}(2006)\citenamefont {Berti}, \citenamefont {Cardoso},\ and\ \citenamefont {Will}}]{Berti:2005ys}%
  \BibitemOpen
  \bibfield  {author} {\bibinfo {author} {\bibfnamefont {E.}~\bibnamefont {Berti}}, \bibinfo {author} {\bibfnamefont {V.}~\bibnamefont {Cardoso}},\ and\ \bibinfo {author} {\bibfnamefont {C.~M.}\ \bibnamefont {Will}},\ }\bibfield  {title} {\bibinfo {title} {{On gravitational-wave spectroscopy of massive black holes with the space interferometer LISA}},\ }\href {https://doi.org/10.1103/PhysRevD.73.064030} {\bibfield  {journal} {\bibinfo  {journal} {Phys. Rev. D}\ }\textbf {\bibinfo {volume} {73}},\ \bibinfo {pages} {064030} (\bibinfo {year} {2006})},\ \Eprint {https://arxiv.org/abs/gr-qc/0512160} {arXiv:gr-qc/0512160} \BibitemShut {NoStop}%
\bibitem [{\citenamefont {Regge}\ and\ \citenamefont {Wheeler}(1957)}]{Regge:1957td}%
  \BibitemOpen
  \bibfield  {author} {\bibinfo {author} {\bibfnamefont {T.}~\bibnamefont {Regge}}\ and\ \bibinfo {author} {\bibfnamefont {J.~A.}\ \bibnamefont {Wheeler}},\ }\bibfield  {title} {\bibinfo {title} {{Stability of a Schwarzschild singularity}},\ }\href {https://doi.org/10.1103/PhysRev.108.1063} {\bibfield  {journal} {\bibinfo  {journal} {Phys. Rev.}\ }\textbf {\bibinfo {volume} {108}},\ \bibinfo {pages} {1063} (\bibinfo {year} {1957})}\BibitemShut {NoStop}%
\bibitem [{\citenamefont {Zerilli}(1970)}]{Zerilli:1970se}%
  \BibitemOpen
  \bibfield  {author} {\bibinfo {author} {\bibfnamefont {F.~J.}\ \bibnamefont {Zerilli}},\ }\bibfield  {title} {\bibinfo {title} {{Effective potential for even parity Regge-Wheeler gravitational perturbation equations}},\ }\href {https://doi.org/10.1103/PhysRevLett.24.737} {\bibfield  {journal} {\bibinfo  {journal} {Phys. Rev. Lett.}\ }\textbf {\bibinfo {volume} {24}},\ \bibinfo {pages} {737} (\bibinfo {year} {1970})}\BibitemShut {NoStop}%
\bibitem [{\citenamefont {Teukolsky}(1973)}]{Teukolsky:1973ha}%
  \BibitemOpen
  \bibfield  {author} {\bibinfo {author} {\bibfnamefont {S.~A.}\ \bibnamefont {Teukolsky}},\ }\bibfield  {title} {\bibinfo {title} {{Perturbations of a rotating black hole. 1. Fundamental equations for gravitational electromagnetic and neutrino field perturbations}},\ }\href {https://doi.org/10.1086/152444} {\bibfield  {journal} {\bibinfo  {journal} {Astrophys. J.}\ }\textbf {\bibinfo {volume} {185}},\ \bibinfo {pages} {635} (\bibinfo {year} {1973})}\BibitemShut {NoStop}%
\bibitem [{\citenamefont {Kokkotas}\ and\ \citenamefont {Schmidt}(1999)}]{Kokkotas:1999bd}%
  \BibitemOpen
  \bibfield  {author} {\bibinfo {author} {\bibfnamefont {K.~D.}\ \bibnamefont {Kokkotas}}\ and\ \bibinfo {author} {\bibfnamefont {B.~G.}\ \bibnamefont {Schmidt}},\ }\bibfield  {title} {\bibinfo {title} {{Quasinormal modes of stars and black holes}},\ }\href {https://doi.org/10.12942/lrr-1999-2} {\bibfield  {journal} {\bibinfo  {journal} {Living Rev. Rel.}\ }\textbf {\bibinfo {volume} {2}},\ \bibinfo {pages} {2} (\bibinfo {year} {1999})},\ \Eprint {https://arxiv.org/abs/gr-qc/9909058} {arXiv:gr-qc/9909058} \BibitemShut {NoStop}%
\bibitem [{\citenamefont {Nollert}(1999)}]{Nollert:1999ji}%
  \BibitemOpen
  \bibfield  {author} {\bibinfo {author} {\bibfnamefont {H.-P.}\ \bibnamefont {Nollert}},\ }\bibfield  {title} {\bibinfo {title} {{Topical Review: Quasinormal modes: the characteristic `sound' of black holes and neutron stars}},\ }\href {https://doi.org/10.1088/0264-9381/16/12/201} {\bibfield  {journal} {\bibinfo  {journal} {Class. Quant. Grav.}\ }\textbf {\bibinfo {volume} {16}},\ \bibinfo {pages} {R159} (\bibinfo {year} {1999})}\BibitemShut {NoStop}%
\bibitem [{\citenamefont {Berti}\ \emph {et~al.}(2009)\citenamefont {Berti}, \citenamefont {Cardoso},\ and\ \citenamefont {Starinets}}]{Berti:2009kk}%
  \BibitemOpen
  \bibfield  {author} {\bibinfo {author} {\bibfnamefont {E.}~\bibnamefont {Berti}}, \bibinfo {author} {\bibfnamefont {V.}~\bibnamefont {Cardoso}},\ and\ \bibinfo {author} {\bibfnamefont {A.~O.}\ \bibnamefont {Starinets}},\ }\bibfield  {title} {\bibinfo {title} {{Quasinormal modes of black holes and black branes}},\ }\href {https://doi.org/10.1088/0264-9381/26/16/163001} {\bibfield  {journal} {\bibinfo  {journal} {Class. Quant. Grav.}\ }\textbf {\bibinfo {volume} {26}},\ \bibinfo {pages} {163001} (\bibinfo {year} {2009})},\ \Eprint {https://arxiv.org/abs/0905.2975} {arXiv:0905.2975 [gr-qc]} \BibitemShut {NoStop}%
\bibitem [{\citenamefont {Konoplya}\ and\ \citenamefont {Zhidenko}(2011)}]{Konoplya:2011qq}%
  \BibitemOpen
  \bibfield  {author} {\bibinfo {author} {\bibfnamefont {R.~A.}\ \bibnamefont {Konoplya}}\ and\ \bibinfo {author} {\bibfnamefont {A.}~\bibnamefont {Zhidenko}},\ }\bibfield  {title} {\bibinfo {title} {{Quasinormal modes of black holes: From astrophysics to string theory}},\ }\href {https://doi.org/10.1103/RevModPhys.83.793} {\bibfield  {journal} {\bibinfo  {journal} {Rev. Mod. Phys.}\ }\textbf {\bibinfo {volume} {83}},\ \bibinfo {pages} {793} (\bibinfo {year} {2011})},\ \Eprint {https://arxiv.org/abs/1102.4014} {arXiv:1102.4014 [gr-qc]} \BibitemShut {NoStop}%
\bibitem [{\citenamefont {Berti}\ \emph {et~al.}(2025)\citenamefont {Berti} \emph {et~al.}}]{Berti:2025hly}%
  \BibitemOpen
  \bibfield  {author} {\bibinfo {author} {\bibfnamefont {E.}~\bibnamefont {Berti}} \emph {et~al.},\ }\bibfield  {title} {\bibinfo {title} {{Black hole spectroscopy: from theory to experiment}},\ }\href@noop {} {\  (\bibinfo {year} {2025})},\ \Eprint {https://arxiv.org/abs/2505.23895} {arXiv:2505.23895 [gr-qc]} \BibitemShut {NoStop}%
\bibitem [{\citenamefont {Israel}(1967)}]{Israel:1967wq}%
  \BibitemOpen
  \bibfield  {author} {\bibinfo {author} {\bibfnamefont {W.}~\bibnamefont {Israel}},\ }\bibfield  {title} {\bibinfo {title} {{Event horizons in static vacuum space-times}},\ }\href {https://doi.org/10.1103/PhysRev.164.1776} {\bibfield  {journal} {\bibinfo  {journal} {Phys. Rev.}\ }\textbf {\bibinfo {volume} {164}},\ \bibinfo {pages} {1776} (\bibinfo {year} {1967})}\BibitemShut {NoStop}%
\bibitem [{\citenamefont {Hawking}(1972)}]{Hawking:1971vc}%
  \BibitemOpen
  \bibfield  {author} {\bibinfo {author} {\bibfnamefont {S.~W.}\ \bibnamefont {Hawking}},\ }\bibfield  {title} {\bibinfo {title} {{Black holes in general relativity}},\ }\href {https://doi.org/10.1007/BF01877517} {\bibfield  {journal} {\bibinfo  {journal} {Commun. Math. Phys.}\ }\textbf {\bibinfo {volume} {25}},\ \bibinfo {pages} {152} (\bibinfo {year} {1972})}\BibitemShut {NoStop}%
\bibitem [{\citenamefont {Carter}(1971)}]{Carter:1971zc}%
  \BibitemOpen
  \bibfield  {author} {\bibinfo {author} {\bibfnamefont {B.}~\bibnamefont {Carter}},\ }\bibfield  {title} {\bibinfo {title} {{Axisymmetric Black Hole Has Only Two Degrees of Freedom}},\ }\href {https://doi.org/10.1103/PhysRevLett.26.331} {\bibfield  {journal} {\bibinfo  {journal} {Phys. Rev. Lett.}\ }\textbf {\bibinfo {volume} {26}},\ \bibinfo {pages} {331} (\bibinfo {year} {1971})}\BibitemShut {NoStop}%
\bibitem [{\citenamefont {Robinson}(1975)}]{Robinson:1975bv}%
  \BibitemOpen
  \bibfield  {author} {\bibinfo {author} {\bibfnamefont {D.~C.}\ \bibnamefont {Robinson}},\ }\bibfield  {title} {\bibinfo {title} {{Uniqueness of the Kerr black hole}},\ }\href {https://doi.org/10.1103/PhysRevLett.34.905} {\bibfield  {journal} {\bibinfo  {journal} {Phys. Rev. Lett.}\ }\textbf {\bibinfo {volume} {34}},\ \bibinfo {pages} {905} (\bibinfo {year} {1975})}\BibitemShut {NoStop}%
\bibitem [{\citenamefont {{Bachelot}}\ and\ \citenamefont {{Motet-Bachelot}}(1993)}]{1993AIHPA..59....3B}%
  \BibitemOpen
  \bibfield  {author} {\bibinfo {author} {\bibfnamefont {A.}~\bibnamefont {{Bachelot}}}\ and\ \bibinfo {author} {\bibfnamefont {A.}~\bibnamefont {{Motet-Bachelot}}},\ }\bibfield  {title} {\bibinfo {title} {{Les r{\'e}sonances d'un trou noir de Schwarzschild.}},\ }\href@noop {} {\bibfield  {journal} {\bibinfo  {journal} {Annales de L'Institut Henri Poincare Section (A) Physique Theorique}\ }\textbf {\bibinfo {volume} {59}},\ \bibinfo {pages} {3} (\bibinfo {year} {1993})}\BibitemShut {NoStop}%
\bibitem [{\citenamefont {Beyer}(1999)}]{Beyer:1998nu}%
  \BibitemOpen
  \bibfield  {author} {\bibinfo {author} {\bibfnamefont {H.~R.}\ \bibnamefont {Beyer}},\ }\bibfield  {title} {\bibinfo {title} {{On the completeness of the quasinormal modes of the Poschl-Teller potential}},\ }\href {https://doi.org/10.1007/s002200050651} {\bibfield  {journal} {\bibinfo  {journal} {Commun. Math. Phys.}\ }\textbf {\bibinfo {volume} {204}},\ \bibinfo {pages} {397} (\bibinfo {year} {1999})},\ \Eprint {https://arxiv.org/abs/gr-qc/9803034} {arXiv:gr-qc/9803034} \BibitemShut {NoStop}%
\bibitem [{\citenamefont {Vishveshwara}(1970)}]{Vishveshwara:1970zz}%
  \BibitemOpen
  \bibfield  {author} {\bibinfo {author} {\bibfnamefont {C.~V.}\ \bibnamefont {Vishveshwara}},\ }\bibfield  {title} {\bibinfo {title} {{Scattering of Gravitational Radiation by a Schwarzschild Black-hole}},\ }\href {https://doi.org/10.1038/227936a0} {\bibfield  {journal} {\bibinfo  {journal} {Nature}\ }\textbf {\bibinfo {volume} {227}},\ \bibinfo {pages} {936} (\bibinfo {year} {1970})}\BibitemShut {NoStop}%
\bibitem [{\citenamefont {Price}(1972)}]{Price:1971fb}%
  \BibitemOpen
  \bibfield  {author} {\bibinfo {author} {\bibfnamefont {R.~H.}\ \bibnamefont {Price}},\ }\bibfield  {title} {\bibinfo {title} {{Nonspherical perturbations of relativistic gravitational collapse. 1. Scalar and gravitational perturbations}},\ }\href {https://doi.org/10.1103/PhysRevD.5.2419} {\bibfield  {journal} {\bibinfo  {journal} {Phys. Rev. D}\ }\textbf {\bibinfo {volume} {5}},\ \bibinfo {pages} {2419} (\bibinfo {year} {1972})}\BibitemShut {NoStop}%
\bibitem [{\citenamefont {Leaver}(1986)}]{Leaver:1986gd}%
  \BibitemOpen
  \bibfield  {author} {\bibinfo {author} {\bibfnamefont {E.~W.}\ \bibnamefont {Leaver}},\ }\bibfield  {title} {\bibinfo {title} {{Spectral decomposition of the perturbation response of the Schwarzschild geometry}},\ }\href {https://doi.org/10.1103/PhysRevD.34.384} {\bibfield  {journal} {\bibinfo  {journal} {Phys. Rev. D}\ }\textbf {\bibinfo {volume} {34}},\ \bibinfo {pages} {384} (\bibinfo {year} {1986})}\BibitemShut {NoStop}%
\bibitem [{\citenamefont {Gundlach}\ \emph {et~al.}(1994{\natexlab{a}})\citenamefont {Gundlach}, \citenamefont {Price},\ and\ \citenamefont {Pullin}}]{Gundlach:1993tp}%
  \BibitemOpen
  \bibfield  {author} {\bibinfo {author} {\bibfnamefont {C.}~\bibnamefont {Gundlach}}, \bibinfo {author} {\bibfnamefont {R.~H.}\ \bibnamefont {Price}},\ and\ \bibinfo {author} {\bibfnamefont {J.}~\bibnamefont {Pullin}},\ }\bibfield  {title} {\bibinfo {title} {{Late time behavior of stellar collapse and explosions: 1. Linearized perturbations}},\ }\href {https://doi.org/10.1103/PhysRevD.49.883} {\bibfield  {journal} {\bibinfo  {journal} {Phys. Rev. D}\ }\textbf {\bibinfo {volume} {49}},\ \bibinfo {pages} {883} (\bibinfo {year} {1994}{\natexlab{a}})},\ \Eprint {https://arxiv.org/abs/gr-qc/9307009} {arXiv:gr-qc/9307009} \BibitemShut {NoStop}%
\bibitem [{\citenamefont {Gundlach}\ \emph {et~al.}(1994{\natexlab{b}})\citenamefont {Gundlach}, \citenamefont {Price},\ and\ \citenamefont {Pullin}}]{Gundlach:1993tn}%
  \BibitemOpen
  \bibfield  {author} {\bibinfo {author} {\bibfnamefont {C.}~\bibnamefont {Gundlach}}, \bibinfo {author} {\bibfnamefont {R.~H.}\ \bibnamefont {Price}},\ and\ \bibinfo {author} {\bibfnamefont {J.}~\bibnamefont {Pullin}},\ }\bibfield  {title} {\bibinfo {title} {{Late time behavior of stellar collapse and explosions: 2. Nonlinear evolution}},\ }\href {https://doi.org/10.1103/PhysRevD.49.890} {\bibfield  {journal} {\bibinfo  {journal} {Phys. Rev. D}\ }\textbf {\bibinfo {volume} {49}},\ \bibinfo {pages} {890} (\bibinfo {year} {1994}{\natexlab{b}})},\ \Eprint {https://arxiv.org/abs/gr-qc/9307010} {arXiv:gr-qc/9307010} \BibitemShut {NoStop}%
\bibitem [{\citenamefont {Barack}(1999)}]{Barack:1998bw}%
  \BibitemOpen
  \bibfield  {author} {\bibinfo {author} {\bibfnamefont {L.}~\bibnamefont {Barack}},\ }\bibfield  {title} {\bibinfo {title} {{Late time dynamics of scalar perturbations outside black holes. 2. Schwarzschild geometry}},\ }\href {https://doi.org/10.1103/PhysRevD.59.044017} {\bibfield  {journal} {\bibinfo  {journal} {Phys. Rev. D}\ }\textbf {\bibinfo {volume} {59}},\ \bibinfo {pages} {044017} (\bibinfo {year} {1999})},\ \Eprint {https://arxiv.org/abs/gr-qc/9811028} {arXiv:gr-qc/9811028} \BibitemShut {NoStop}%
\bibitem [{\citenamefont {Rosato}\ and\ \citenamefont {Pani}(2025)}]{Rosato:2025rtr}%
  \BibitemOpen
  \bibfield  {author} {\bibinfo {author} {\bibfnamefont {R.~F.}\ \bibnamefont {Rosato}}\ and\ \bibinfo {author} {\bibfnamefont {P.}~\bibnamefont {Pani}},\ }\bibfield  {title} {\bibinfo {title} {{On the universality of late-time ringdown tail}},\ }\href@noop {} {\  (\bibinfo {year} {2025})},\ \Eprint {https://arxiv.org/abs/2505.08877} {arXiv:2505.08877 [gr-qc]} \BibitemShut {NoStop}%
\bibitem [{\citenamefont {Nollert}\ and\ \citenamefont {Price}(1999)}]{Nollert:1998ys}%
  \BibitemOpen
  \bibfield  {author} {\bibinfo {author} {\bibfnamefont {H.-P.}\ \bibnamefont {Nollert}}\ and\ \bibinfo {author} {\bibfnamefont {R.~H.}\ \bibnamefont {Price}},\ }\bibfield  {title} {\bibinfo {title} {{Quantifying excitations of quasinormal mode systems}},\ }\href {https://doi.org/10.1063/1.532698} {\bibfield  {journal} {\bibinfo  {journal} {J. Math. Phys.}\ }\textbf {\bibinfo {volume} {40}},\ \bibinfo {pages} {980} (\bibinfo {year} {1999})},\ \Eprint {https://arxiv.org/abs/gr-qc/9810074} {arXiv:gr-qc/9810074} \BibitemShut {NoStop}%
\bibitem [{\citenamefont {Hod}(2000{\natexlab{a}})}]{Hod:1999ci}%
  \BibitemOpen
  \bibfield  {author} {\bibinfo {author} {\bibfnamefont {S.}~\bibnamefont {Hod}},\ }\bibfield  {title} {\bibinfo {title} {{The Radiative tail of realistic gravitational collapse}},\ }\href {https://doi.org/10.1103/PhysRevLett.84.10} {\bibfield  {journal} {\bibinfo  {journal} {Phys. Rev. Lett.}\ }\textbf {\bibinfo {volume} {84}},\ \bibinfo {pages} {10} (\bibinfo {year} {2000}{\natexlab{a}})},\ \Eprint {https://arxiv.org/abs/gr-qc/9907096} {arXiv:gr-qc/9907096} \BibitemShut {NoStop}%
\bibitem [{\citenamefont {Hod}(2000{\natexlab{b}})}]{Hod:2000fh}%
  \BibitemOpen
  \bibfield  {author} {\bibinfo {author} {\bibfnamefont {S.}~\bibnamefont {Hod}},\ }\bibfield  {title} {\bibinfo {title} {{Mode coupling in rotating gravitational collapse: Gravitational and electromagnetic perturbations}},\ }\href {https://doi.org/10.1103/PhysRevD.61.064018} {\bibfield  {journal} {\bibinfo  {journal} {Phys. Rev. D}\ }\textbf {\bibinfo {volume} {61}},\ \bibinfo {pages} {064018} (\bibinfo {year} {2000}{\natexlab{b}})}\BibitemShut {NoStop}%
\bibitem [{\citenamefont {Buonanno}\ \emph {et~al.}(2007)\citenamefont {Buonanno}, \citenamefont {Cook},\ and\ \citenamefont {Pretorius}}]{Buonanno:2006ui}%
  \BibitemOpen
  \bibfield  {author} {\bibinfo {author} {\bibfnamefont {A.}~\bibnamefont {Buonanno}}, \bibinfo {author} {\bibfnamefont {G.~B.}\ \bibnamefont {Cook}},\ and\ \bibinfo {author} {\bibfnamefont {F.}~\bibnamefont {Pretorius}},\ }\bibfield  {title} {\bibinfo {title} {{Inspiral, merger and ring-down of equal-mass black-hole binaries}},\ }\href {https://doi.org/10.1103/PhysRevD.75.124018} {\bibfield  {journal} {\bibinfo  {journal} {Phys. Rev. D}\ }\textbf {\bibinfo {volume} {75}},\ \bibinfo {pages} {124018} (\bibinfo {year} {2007})},\ \Eprint {https://arxiv.org/abs/gr-qc/0610122} {arXiv:gr-qc/0610122} \BibitemShut {NoStop}%
\bibitem [{\citenamefont {Berti}\ \emph {et~al.}(2007{\natexlab{a}})\citenamefont {Berti}, \citenamefont {Cardoso}, \citenamefont {Gonzalez}, \citenamefont {Sperhake}, \citenamefont {Hannam}, \citenamefont {Husa},\ and\ \citenamefont {Bruegmann}}]{Berti:2007fi}%
  \BibitemOpen
  \bibfield  {author} {\bibinfo {author} {\bibfnamefont {E.}~\bibnamefont {Berti}}, \bibinfo {author} {\bibfnamefont {V.}~\bibnamefont {Cardoso}}, \bibinfo {author} {\bibfnamefont {J.~A.}\ \bibnamefont {Gonzalez}}, \bibinfo {author} {\bibfnamefont {U.}~\bibnamefont {Sperhake}}, \bibinfo {author} {\bibfnamefont {M.}~\bibnamefont {Hannam}}, \bibinfo {author} {\bibfnamefont {S.}~\bibnamefont {Husa}},\ and\ \bibinfo {author} {\bibfnamefont {B.}~\bibnamefont {Bruegmann}},\ }\bibfield  {title} {\bibinfo {title} {{Inspiral, merger and ringdown of unequal mass black hole binaries: A Multipolar analysis}},\ }\href {https://doi.org/10.1103/PhysRevD.76.064034} {\bibfield  {journal} {\bibinfo  {journal} {Phys. Rev. D}\ }\textbf {\bibinfo {volume} {76}},\ \bibinfo {pages} {064034} (\bibinfo {year} {2007}{\natexlab{a}})},\ \Eprint {https://arxiv.org/abs/gr-qc/0703053} {arXiv:gr-qc/0703053} \BibitemShut {NoStop}%
\bibitem [{\citenamefont {London}\ \emph {et~al.}(2014)\citenamefont {London}, \citenamefont {Shoemaker},\ and\ \citenamefont {Healy}}]{London:2014cma}%
  \BibitemOpen
  \bibfield  {author} {\bibinfo {author} {\bibfnamefont {L.}~\bibnamefont {London}}, \bibinfo {author} {\bibfnamefont {D.}~\bibnamefont {Shoemaker}},\ and\ \bibinfo {author} {\bibfnamefont {J.}~\bibnamefont {Healy}},\ }\bibfield  {title} {\bibinfo {title} {{Modeling ringdown: Beyond the fundamental quasinormal modes}},\ }\href {https://doi.org/10.1103/PhysRevD.90.124032} {\bibfield  {journal} {\bibinfo  {journal} {Phys. Rev. D}\ }\textbf {\bibinfo {volume} {90}},\ \bibinfo {pages} {124032} (\bibinfo {year} {2014})},\ \bibinfo {note} {[Erratum: Phys.Rev.D 94, 069902 (2016)]},\ \Eprint {https://arxiv.org/abs/1404.3197} {arXiv:1404.3197 [gr-qc]} \BibitemShut {NoStop}%
\bibitem [{\citenamefont {Giesler}\ \emph {et~al.}(2019)\citenamefont {Giesler}, \citenamefont {Isi}, \citenamefont {Scheel},\ and\ \citenamefont {Teukolsky}}]{Giesler:2019uxc}%
  \BibitemOpen
  \bibfield  {author} {\bibinfo {author} {\bibfnamefont {M.}~\bibnamefont {Giesler}}, \bibinfo {author} {\bibfnamefont {M.}~\bibnamefont {Isi}}, \bibinfo {author} {\bibfnamefont {M.~A.}\ \bibnamefont {Scheel}},\ and\ \bibinfo {author} {\bibfnamefont {S.}~\bibnamefont {Teukolsky}},\ }\bibfield  {title} {\bibinfo {title} {{Black Hole Ringdown: The Importance of Overtones}},\ }\href {https://doi.org/10.1103/PhysRevX.9.041060} {\bibfield  {journal} {\bibinfo  {journal} {Phys. Rev. X}\ }\textbf {\bibinfo {volume} {9}},\ \bibinfo {pages} {041060} (\bibinfo {year} {2019})},\ \Eprint {https://arxiv.org/abs/1903.08284} {arXiv:1903.08284 [gr-qc]} \BibitemShut {NoStop}%
\bibitem [{\citenamefont {Finch}\ and\ \citenamefont {Moore}(2022)}]{Finch:2022ynt}%
  \BibitemOpen
  \bibfield  {author} {\bibinfo {author} {\bibfnamefont {E.}~\bibnamefont {Finch}}\ and\ \bibinfo {author} {\bibfnamefont {C.~J.}\ \bibnamefont {Moore}},\ }\bibfield  {title} {\bibinfo {title} {{Searching for a ringdown overtone in GW150914}},\ }\href {https://doi.org/10.1103/PhysRevD.106.043005} {\bibfield  {journal} {\bibinfo  {journal} {Phys. Rev. D}\ }\textbf {\bibinfo {volume} {106}},\ \bibinfo {pages} {043005} (\bibinfo {year} {2022})},\ \Eprint {https://arxiv.org/abs/2205.07809} {arXiv:2205.07809 [gr-qc]} \BibitemShut {NoStop}%
\bibitem [{\citenamefont {Isi}\ and\ \citenamefont {Farr}(2022)}]{Isi:2022mhy}%
  \BibitemOpen
  \bibfield  {author} {\bibinfo {author} {\bibfnamefont {M.}~\bibnamefont {Isi}}\ and\ \bibinfo {author} {\bibfnamefont {W.~M.}\ \bibnamefont {Farr}},\ }\bibfield  {title} {\bibinfo {title} {{Revisiting the ringdown of GW150914}},\ }\href@noop {} {\  (\bibinfo {year} {2022})},\ \Eprint {https://arxiv.org/abs/2202.02941} {arXiv:2202.02941 [gr-qc]} \BibitemShut {NoStop}%
\bibitem [{\citenamefont {Ma}\ \emph {et~al.}(2022)\citenamefont {Ma}, \citenamefont {Mitman}, \citenamefont {Sun}, \citenamefont {Deppe}, \citenamefont {H{\'e}bert}, \citenamefont {Kidder}, \citenamefont {Moxon}, \citenamefont {Throwe}, \citenamefont {Vu},\ and\ \citenamefont {Chen}}]{Ma:2022wpv}%
  \BibitemOpen
  \bibfield  {author} {\bibinfo {author} {\bibfnamefont {S.}~\bibnamefont {Ma}}, \bibinfo {author} {\bibfnamefont {K.}~\bibnamefont {Mitman}}, \bibinfo {author} {\bibfnamefont {L.}~\bibnamefont {Sun}}, \bibinfo {author} {\bibfnamefont {N.}~\bibnamefont {Deppe}}, \bibinfo {author} {\bibfnamefont {F.}~\bibnamefont {H{\'e}bert}}, \bibinfo {author} {\bibfnamefont {L.~E.}\ \bibnamefont {Kidder}}, \bibinfo {author} {\bibfnamefont {J.}~\bibnamefont {Moxon}}, \bibinfo {author} {\bibfnamefont {W.}~\bibnamefont {Throwe}}, \bibinfo {author} {\bibfnamefont {N.~L.}\ \bibnamefont {Vu}},\ and\ \bibinfo {author} {\bibfnamefont {Y.}~\bibnamefont {Chen}},\ }\bibfield  {title} {\bibinfo {title} {{Quasinormal-mode filters: A new approach to analyze the gravitational-wave ringdown of binary black-hole mergers}},\ }\href {https://doi.org/10.1103/PhysRevD.106.084036} {\bibfield  {journal} {\bibinfo  {journal} {Phys. Rev. D}\ }\textbf {\bibinfo {volume} {106}},\ \bibinfo {pages} {084036} (\bibinfo {year} {2022})},\ \Eprint
  {https://arxiv.org/abs/2207.10870} {arXiv:2207.10870 [gr-qc]} \BibitemShut {NoStop}%
\bibitem [{\citenamefont {Ma}\ \emph {et~al.}(2023)\citenamefont {Ma}, \citenamefont {Sun},\ and\ \citenamefont {Chen}}]{Ma:2023cwe}%
  \BibitemOpen
  \bibfield  {author} {\bibinfo {author} {\bibfnamefont {S.}~\bibnamefont {Ma}}, \bibinfo {author} {\bibfnamefont {L.}~\bibnamefont {Sun}},\ and\ \bibinfo {author} {\bibfnamefont {Y.}~\bibnamefont {Chen}},\ }\bibfield  {title} {\bibinfo {title} {{Black Hole Spectroscopy by Mode Cleaning}},\ }\href {https://doi.org/10.1103/PhysRevLett.130.141401} {\bibfield  {journal} {\bibinfo  {journal} {Phys. Rev. Lett.}\ }\textbf {\bibinfo {volume} {130}},\ \bibinfo {pages} {141401} (\bibinfo {year} {2023})},\ \Eprint {https://arxiv.org/abs/2301.06705} {arXiv:2301.06705 [gr-qc]} \BibitemShut {NoStop}%
\bibitem [{\citenamefont {Baibhav}\ \emph {et~al.}(2023)\citenamefont {Baibhav}, \citenamefont {Cheung}, \citenamefont {Berti}, \citenamefont {Cardoso}, \citenamefont {Carullo}, \citenamefont {Cotesta}, \citenamefont {Del~Pozzo},\ and\ \citenamefont {Duque}}]{Baibhav:2023clw}%
  \BibitemOpen
  \bibfield  {author} {\bibinfo {author} {\bibfnamefont {V.}~\bibnamefont {Baibhav}}, \bibinfo {author} {\bibfnamefont {M.~H.-Y.}\ \bibnamefont {Cheung}}, \bibinfo {author} {\bibfnamefont {E.}~\bibnamefont {Berti}}, \bibinfo {author} {\bibfnamefont {V.}~\bibnamefont {Cardoso}}, \bibinfo {author} {\bibfnamefont {G.}~\bibnamefont {Carullo}}, \bibinfo {author} {\bibfnamefont {R.}~\bibnamefont {Cotesta}}, \bibinfo {author} {\bibfnamefont {W.}~\bibnamefont {Del~Pozzo}},\ and\ \bibinfo {author} {\bibfnamefont {F.}~\bibnamefont {Duque}},\ }\bibfield  {title} {\bibinfo {title} {{Agnostic black hole spectroscopy: Quasinormal mode content of numerical relativity waveforms and limits of validity of linear perturbation theory}},\ }\href {https://doi.org/10.1103/PhysRevD.108.104020} {\bibfield  {journal} {\bibinfo  {journal} {Phys. Rev. D}\ }\textbf {\bibinfo {volume} {108}},\ \bibinfo {pages} {104020} (\bibinfo {year} {2023})},\ \Eprint {https://arxiv.org/abs/2302.03050} {arXiv:2302.03050 [gr-qc]} \BibitemShut {NoStop}%
\bibitem [{\citenamefont {Nee}\ \emph {et~al.}(2023)\citenamefont {Nee}, \citenamefont {V\"olkel},\ and\ \citenamefont {Pfeiffer}}]{Nee:2023osy}%
  \BibitemOpen
  \bibfield  {author} {\bibinfo {author} {\bibfnamefont {P.~J.}\ \bibnamefont {Nee}}, \bibinfo {author} {\bibfnamefont {S.~H.}\ \bibnamefont {V\"olkel}},\ and\ \bibinfo {author} {\bibfnamefont {H.~P.}\ \bibnamefont {Pfeiffer}},\ }\bibfield  {title} {\bibinfo {title} {{Role of black hole quasinormal mode overtones for ringdown analysis}},\ }\href {https://doi.org/10.1103/PhysRevD.108.044032} {\bibfield  {journal} {\bibinfo  {journal} {Phys. Rev. D}\ }\textbf {\bibinfo {volume} {108}},\ \bibinfo {pages} {044032} (\bibinfo {year} {2023})},\ \Eprint {https://arxiv.org/abs/2302.06634} {arXiv:2302.06634 [gr-qc]} \BibitemShut {NoStop}%
\bibitem [{\citenamefont {Isi}\ and\ \citenamefont {Farr}(2023)}]{Isi:2023nif}%
  \BibitemOpen
  \bibfield  {author} {\bibinfo {author} {\bibfnamefont {M.}~\bibnamefont {Isi}}\ and\ \bibinfo {author} {\bibfnamefont {W.~M.}\ \bibnamefont {Farr}},\ }\bibfield  {title} {\bibinfo {title} {{Comment on {\textquotedblleft}Analysis of Ringdown Overtones in GW150914{\textquotedblright}}},\ }\href {https://doi.org/10.1103/PhysRevLett.131.169001} {\bibfield  {journal} {\bibinfo  {journal} {Phys. Rev. Lett.}\ }\textbf {\bibinfo {volume} {131}},\ \bibinfo {pages} {169001} (\bibinfo {year} {2023})},\ \Eprint {https://arxiv.org/abs/2310.13869} {arXiv:2310.13869 [astro-ph.HE]} \BibitemShut {NoStop}%
\bibitem [{\citenamefont {Carullo}\ \emph {et~al.}(2023)\citenamefont {Carullo}, \citenamefont {Cotesta}, \citenamefont {Berti},\ and\ \citenamefont {Cardoso}}]{Carullo:2023gtf}%
  \BibitemOpen
  \bibfield  {author} {\bibinfo {author} {\bibfnamefont {G.}~\bibnamefont {Carullo}}, \bibinfo {author} {\bibfnamefont {R.}~\bibnamefont {Cotesta}}, \bibinfo {author} {\bibfnamefont {E.}~\bibnamefont {Berti}},\ and\ \bibinfo {author} {\bibfnamefont {V.}~\bibnamefont {Cardoso}},\ }\bibfield  {title} {\bibinfo {title} {{Reply to Comment on ''Analysis of Ringdown Overtones in GW150914''}},\ }\href {https://doi.org/10.1103/PhysRevLett.131.169002} {\bibfield  {journal} {\bibinfo  {journal} {Phys. Rev. Lett.}\ }\textbf {\bibinfo {volume} {131}},\ \bibinfo {pages} {169002} (\bibinfo {year} {2023})},\ \Eprint {https://arxiv.org/abs/2310.20625} {arXiv:2310.20625 [gr-qc]} \BibitemShut {NoStop}%
\bibitem [{\citenamefont {Cheung}\ \emph {et~al.}(2024)\citenamefont {Cheung}, \citenamefont {Berti}, \citenamefont {Baibhav},\ and\ \citenamefont {Cotesta}}]{Cheung:2023vki}%
  \BibitemOpen
  \bibfield  {author} {\bibinfo {author} {\bibfnamefont {M.~H.-Y.}\ \bibnamefont {Cheung}}, \bibinfo {author} {\bibfnamefont {E.}~\bibnamefont {Berti}}, \bibinfo {author} {\bibfnamefont {V.}~\bibnamefont {Baibhav}},\ and\ \bibinfo {author} {\bibfnamefont {R.}~\bibnamefont {Cotesta}},\ }\bibfield  {title} {\bibinfo {title} {{Extracting linear and nonlinear quasinormal modes from black hole merger simulations}},\ }\href {https://doi.org/10.1103/PhysRevD.109.044069} {\bibfield  {journal} {\bibinfo  {journal} {Phys. Rev. D}\ }\textbf {\bibinfo {volume} {109}},\ \bibinfo {pages} {044069} (\bibinfo {year} {2024})},\ \bibinfo {note} {[Erratum: Phys.Rev.D 110, 049902 (2024)]},\ \Eprint {https://arxiv.org/abs/2310.04489} {arXiv:2310.04489 [gr-qc]} \BibitemShut {NoStop}%
\bibitem [{\citenamefont {Maselli}\ \emph {et~al.}(2024)\citenamefont {Maselli}, \citenamefont {Yi}, \citenamefont {Pierini}, \citenamefont {Vellucci}, \citenamefont {Reali}, \citenamefont {Gualtieri},\ and\ \citenamefont {Berti}}]{Maselli:2023khq}%
  \BibitemOpen
  \bibfield  {author} {\bibinfo {author} {\bibfnamefont {A.}~\bibnamefont {Maselli}}, \bibinfo {author} {\bibfnamefont {S.}~\bibnamefont {Yi}}, \bibinfo {author} {\bibfnamefont {L.}~\bibnamefont {Pierini}}, \bibinfo {author} {\bibfnamefont {V.}~\bibnamefont {Vellucci}}, \bibinfo {author} {\bibfnamefont {L.}~\bibnamefont {Reali}}, \bibinfo {author} {\bibfnamefont {L.}~\bibnamefont {Gualtieri}},\ and\ \bibinfo {author} {\bibfnamefont {E.}~\bibnamefont {Berti}},\ }\bibfield  {title} {\bibinfo {title} {{Black hole spectroscopy beyond Kerr: Agnostic and theory-based tests with next-generation interferometers}},\ }\href {https://doi.org/10.1103/PhysRevD.109.064060} {\bibfield  {journal} {\bibinfo  {journal} {Phys. Rev. D}\ }\textbf {\bibinfo {volume} {109}},\ \bibinfo {pages} {064060} (\bibinfo {year} {2024})},\ \Eprint {https://arxiv.org/abs/2311.14803} {arXiv:2311.14803 [gr-qc]} \BibitemShut {NoStop}%
\bibitem [{\citenamefont {Giesler}\ \emph {et~al.}(2025)\citenamefont {Giesler} \emph {et~al.}}]{Giesler:2024hcr}%
  \BibitemOpen
  \bibfield  {author} {\bibinfo {author} {\bibfnamefont {M.}~\bibnamefont {Giesler}} \emph {et~al.},\ }\bibfield  {title} {\bibinfo {title} {{Overtones and nonlinearities in binary black hole ringdowns}},\ }\href {https://doi.org/10.1103/PhysRevD.111.084041} {\bibfield  {journal} {\bibinfo  {journal} {Phys. Rev. D}\ }\textbf {\bibinfo {volume} {111}},\ \bibinfo {pages} {084041} (\bibinfo {year} {2025})},\ \Eprint {https://arxiv.org/abs/2411.11269} {arXiv:2411.11269 [gr-qc]} \BibitemShut {NoStop}%
\bibitem [{\citenamefont {Mitman}\ \emph {et~al.}(2025)\citenamefont {Mitman} \emph {et~al.}}]{Mitman:2025hgy}%
  \BibitemOpen
  \bibfield  {author} {\bibinfo {author} {\bibfnamefont {K.}~\bibnamefont {Mitman}} \emph {et~al.},\ }\bibfield  {title} {\bibinfo {title} {{Probing the ringdown perturbation in binary black hole coalescences with an improved quasi-normal mode extraction algorithm}},\ }\href@noop {} {\  (\bibinfo {year} {2025})},\ \Eprint {https://arxiv.org/abs/2503.09678} {arXiv:2503.09678 [gr-qc]} \BibitemShut {NoStop}%
\bibitem [{\citenamefont {Nobili}\ \emph {et~al.}(2025)\citenamefont {Nobili}, \citenamefont {Bhagwat}, \citenamefont {Pacilio},\ and\ \citenamefont {Gerosa}}]{Nobili:2025ydt}%
  \BibitemOpen
  \bibfield  {author} {\bibinfo {author} {\bibfnamefont {F.}~\bibnamefont {Nobili}}, \bibinfo {author} {\bibfnamefont {S.}~\bibnamefont {Bhagwat}}, \bibinfo {author} {\bibfnamefont {C.}~\bibnamefont {Pacilio}},\ and\ \bibinfo {author} {\bibfnamefont {D.}~\bibnamefont {Gerosa}},\ }\bibfield  {title} {\bibinfo {title} {{Ringdown mode amplitudes of precessing binary black holes}},\ }\href@noop {} {\  (\bibinfo {year} {2025})},\ \Eprint {https://arxiv.org/abs/2504.17021} {arXiv:2504.17021 [gr-qc]} \BibitemShut {NoStop}%
\bibitem [{\citenamefont {Thomopoulos}\ \emph {et~al.}(2025)\citenamefont {Thomopoulos}, \citenamefont {V{\"o}lkel},\ and\ \citenamefont {Pfeiffer}}]{Thomopoulos:2025nuf}%
  \BibitemOpen
  \bibfield  {author} {\bibinfo {author} {\bibfnamefont {S.}~\bibnamefont {Thomopoulos}}, \bibinfo {author} {\bibfnamefont {S.~H.}\ \bibnamefont {V{\"o}lkel}},\ and\ \bibinfo {author} {\bibfnamefont {H.~P.}\ \bibnamefont {Pfeiffer}},\ }\bibfield  {title} {\bibinfo {title} {{Ringdown spectroscopy of phenomenologically modified black holes}},\ }\href@noop {} {\  (\bibinfo {year} {2025})},\ \Eprint {https://arxiv.org/abs/2504.17848} {arXiv:2504.17848 [gr-qc]} \BibitemShut {NoStop}%
\bibitem [{\citenamefont {Gao}\ \emph {et~al.}(2025)\citenamefont {Gao} \emph {et~al.}}]{Gao:2025zvl}%
  \BibitemOpen
  \bibfield  {author} {\bibinfo {author} {\bibfnamefont {L.}~\bibnamefont {Gao}} \emph {et~al.},\ }\bibfield  {title} {\bibinfo {title} {{Robustness of extracting quasinormal mode information from black hole merger simulations}},\ }\href {https://doi.org/10.1103/3jj6-jc8q} {\bibfield  {journal} {\bibinfo  {journal} {Phys. Rev. D}\ }\textbf {\bibinfo {volume} {112}},\ \bibinfo {pages} {024025} (\bibinfo {year} {2025})},\ \Eprint {https://arxiv.org/abs/2502.15921} {arXiv:2502.15921 [gr-qc]} \BibitemShut {NoStop}%
\bibitem [{\citenamefont {Maga{\~n}a~Zertuche}\ \emph {et~al.}(2024)\citenamefont {Maga{\~n}a~Zertuche} \emph {et~al.}}]{MaganaZertuche:2024ajz}%
  \BibitemOpen
  \bibfield  {author} {\bibinfo {author} {\bibfnamefont {L.}~\bibnamefont {Maga{\~n}a~Zertuche}} \emph {et~al.},\ }\bibfield  {title} {\bibinfo {title} {{High-Precision Ringdown Surrogate Model for Non-Precessing Binary Black Holes}},\ }\href@noop {} {\  (\bibinfo {year} {2024})},\ \Eprint {https://arxiv.org/abs/2408.05300} {arXiv:2408.05300 [gr-qc]} \BibitemShut {NoStop}%
\bibitem [{\citenamefont {Oshita}(2021)}]{Oshita:2021iyn}%
  \BibitemOpen
  \bibfield  {author} {\bibinfo {author} {\bibfnamefont {N.}~\bibnamefont {Oshita}},\ }\bibfield  {title} {\bibinfo {title} {{Ease of excitation of black hole ringing: Quantifying the importance of overtones by the excitation factors}},\ }\href {https://doi.org/10.1103/PhysRevD.104.124032} {\bibfield  {journal} {\bibinfo  {journal} {Phys. Rev. D}\ }\textbf {\bibinfo {volume} {104}},\ \bibinfo {pages} {124032} (\bibinfo {year} {2021})},\ \Eprint {https://arxiv.org/abs/2109.09757} {arXiv:2109.09757 [gr-qc]} \BibitemShut {NoStop}%
\bibitem [{\citenamefont {Motohashi}(2025)}]{Motohashi:2024fwt}%
  \BibitemOpen
  \bibfield  {author} {\bibinfo {author} {\bibfnamefont {H.}~\bibnamefont {Motohashi}},\ }\bibfield  {title} {\bibinfo {title} {{Resonant Excitation of Quasinormal Modes of Black Holes}},\ }\href {https://doi.org/10.1103/PhysRevLett.134.141401} {\bibfield  {journal} {\bibinfo  {journal} {Phys. Rev. Lett.}\ }\textbf {\bibinfo {volume} {134}},\ \bibinfo {pages} {141401} (\bibinfo {year} {2025})},\ \Eprint {https://arxiv.org/abs/2407.15191} {arXiv:2407.15191 [gr-qc]} \BibitemShut {NoStop}%
\bibitem [{\citenamefont {Oshita}\ and\ \citenamefont {Cardoso}(2025)}]{Oshita:2024wgt}%
  \BibitemOpen
  \bibfield  {author} {\bibinfo {author} {\bibfnamefont {N.}~\bibnamefont {Oshita}}\ and\ \bibinfo {author} {\bibfnamefont {V.}~\bibnamefont {Cardoso}},\ }\bibfield  {title} {\bibinfo {title} {{Reconstruction of ringdown with excitation factors}},\ }\href {https://doi.org/10.1103/PhysRevD.111.104043} {\bibfield  {journal} {\bibinfo  {journal} {Phys. Rev. D}\ }\textbf {\bibinfo {volume} {111}},\ \bibinfo {pages} {104043} (\bibinfo {year} {2025})},\ \Eprint {https://arxiv.org/abs/2407.02563} {arXiv:2407.02563 [gr-qc]} \BibitemShut {NoStop}%
\bibitem [{\citenamefont {Oshita}\ \emph {et~al.}(2025)\citenamefont {Oshita}, \citenamefont {Berti},\ and\ \citenamefont {Cardoso}}]{Oshita:2025ibu}%
  \BibitemOpen
  \bibfield  {author} {\bibinfo {author} {\bibfnamefont {N.}~\bibnamefont {Oshita}}, \bibinfo {author} {\bibfnamefont {E.}~\bibnamefont {Berti}},\ and\ \bibinfo {author} {\bibfnamefont {V.}~\bibnamefont {Cardoso}},\ }\bibfield  {title} {\bibinfo {title} {{Unstable Chords and Destructive Resonant Excitation of Black Hole Quasinormal Modes}},\ }\href {https://doi.org/10.1103/ht2n-vvvh} {\bibfield  {journal} {\bibinfo  {journal} {Phys. Rev. Lett.}\ }\textbf {\bibinfo {volume} {135}},\ \bibinfo {pages} {031401} (\bibinfo {year} {2025})},\ \Eprint {https://arxiv.org/abs/2503.21276} {arXiv:2503.21276 [gr-qc]} \BibitemShut {NoStop}%
\bibitem [{\citenamefont {Lo}\ \emph {et~al.}(2025)\citenamefont {Lo}, \citenamefont {Sabani},\ and\ \citenamefont {Cardoso}}]{Lo:2025njp}%
  \BibitemOpen
  \bibfield  {author} {\bibinfo {author} {\bibfnamefont {R.~K.~L.}\ \bibnamefont {Lo}}, \bibinfo {author} {\bibfnamefont {L.}~\bibnamefont {Sabani}},\ and\ \bibinfo {author} {\bibfnamefont {V.}~\bibnamefont {Cardoso}},\ }\bibfield  {title} {\bibinfo {title} {{Quasinormal modes and excitation factors of Kerr black holes}},\ }\href {https://doi.org/10.1103/PhysRevD.111.124002} {\bibfield  {journal} {\bibinfo  {journal} {Phys. Rev. D}\ }\textbf {\bibinfo {volume} {111}},\ \bibinfo {pages} {124002} (\bibinfo {year} {2025})},\ \Eprint {https://arxiv.org/abs/2504.00084} {arXiv:2504.00084 [gr-qc]} \BibitemShut {NoStop}%
\bibitem [{\citenamefont {Berti}\ and\ \citenamefont {Cardoso}(2006)}]{Berti:2006wq}%
  \BibitemOpen
  \bibfield  {author} {\bibinfo {author} {\bibfnamefont {E.}~\bibnamefont {Berti}}\ and\ \bibinfo {author} {\bibfnamefont {V.}~\bibnamefont {Cardoso}},\ }\bibfield  {title} {\bibinfo {title} {{Quasinormal ringing of Kerr black holes. I. The Excitation factors}},\ }\href {https://doi.org/10.1103/PhysRevD.74.104020} {\bibfield  {journal} {\bibinfo  {journal} {Phys. Rev. D}\ }\textbf {\bibinfo {volume} {74}},\ \bibinfo {pages} {104020} (\bibinfo {year} {2006})},\ \Eprint {https://arxiv.org/abs/gr-qc/0605118} {arXiv:gr-qc/0605118} \BibitemShut {NoStop}%
\bibitem [{\citenamefont {Zhang}\ \emph {et~al.}(2013)\citenamefont {Zhang}, \citenamefont {Berti},\ and\ \citenamefont {Cardoso}}]{Zhang:2013ksa}%
  \BibitemOpen
  \bibfield  {author} {\bibinfo {author} {\bibfnamefont {Z.}~\bibnamefont {Zhang}}, \bibinfo {author} {\bibfnamefont {E.}~\bibnamefont {Berti}},\ and\ \bibinfo {author} {\bibfnamefont {V.}~\bibnamefont {Cardoso}},\ }\bibfield  {title} {\bibinfo {title} {{Quasinormal ringing of Kerr black holes. II. Excitation by particles falling radially with arbitrary energy}},\ }\href {https://doi.org/10.1103/PhysRevD.88.044018} {\bibfield  {journal} {\bibinfo  {journal} {Phys. Rev. D}\ }\textbf {\bibinfo {volume} {88}},\ \bibinfo {pages} {044018} (\bibinfo {year} {2013})},\ \Eprint {https://arxiv.org/abs/1305.4306} {arXiv:1305.4306 [gr-qc]} \BibitemShut {NoStop}%
\bibitem [{\citenamefont {Dhani}(2021)}]{Dhani:2020nik}%
  \BibitemOpen
  \bibfield  {author} {\bibinfo {author} {\bibfnamefont {A.}~\bibnamefont {Dhani}},\ }\bibfield  {title} {\bibinfo {title} {{Importance of mirror modes in binary black hole ringdown waveform}},\ }\href {https://doi.org/10.1103/PhysRevD.103.104048} {\bibfield  {journal} {\bibinfo  {journal} {Phys. Rev. D}\ }\textbf {\bibinfo {volume} {103}},\ \bibinfo {pages} {104048} (\bibinfo {year} {2021})},\ \Eprint {https://arxiv.org/abs/2010.08602} {arXiv:2010.08602 [gr-qc]} \BibitemShut {NoStop}%
\bibitem [{\citenamefont {Dhani}\ and\ \citenamefont {Sathyaprakash}(2021)}]{Dhani:2021vac}%
  \BibitemOpen
  \bibfield  {author} {\bibinfo {author} {\bibfnamefont {A.}~\bibnamefont {Dhani}}\ and\ \bibinfo {author} {\bibfnamefont {B.~S.}\ \bibnamefont {Sathyaprakash}},\ }\bibfield  {title} {\bibinfo {title} {{Overtones, mirror modes, and mode-mixing in binary black hole mergers}},\ }\href@noop {} {\  (\bibinfo {year} {2021})},\ \Eprint {https://arxiv.org/abs/2107.14195} {arXiv:2107.14195 [gr-qc]} \BibitemShut {NoStop}%
\bibitem [{\citenamefont {Forteza}\ and\ \citenamefont {Mourier}(2021)}]{Forteza:2021wfq}%
  \BibitemOpen
  \bibfield  {author} {\bibinfo {author} {\bibfnamefont {X.~J.}\ \bibnamefont {Forteza}}\ and\ \bibinfo {author} {\bibfnamefont {P.}~\bibnamefont {Mourier}},\ }\bibfield  {title} {\bibinfo {title} {{High-overtone fits to numerical relativity ringdowns: Beyond the dismissed n=8 special tone}},\ }\href {https://doi.org/10.1103/PhysRevD.104.124072} {\bibfield  {journal} {\bibinfo  {journal} {Phys. Rev. D}\ }\textbf {\bibinfo {volume} {104}},\ \bibinfo {pages} {124072} (\bibinfo {year} {2021})},\ \Eprint {https://arxiv.org/abs/2107.11829} {arXiv:2107.11829 [gr-qc]} \BibitemShut {NoStop}%
\bibitem [{\citenamefont {Mitman}\ \emph {et~al.}(2023)\citenamefont {Mitman} \emph {et~al.}}]{Mitman:2022qdl}%
  \BibitemOpen
  \bibfield  {author} {\bibinfo {author} {\bibfnamefont {K.}~\bibnamefont {Mitman}} \emph {et~al.},\ }\bibfield  {title} {\bibinfo {title} {{Nonlinearities in Black Hole Ringdowns}},\ }\href {https://doi.org/10.1103/PhysRevLett.130.081402} {\bibfield  {journal} {\bibinfo  {journal} {Phys. Rev. Lett.}\ }\textbf {\bibinfo {volume} {130}},\ \bibinfo {pages} {081402} (\bibinfo {year} {2023})},\ \Eprint {https://arxiv.org/abs/2208.07380} {arXiv:2208.07380 [gr-qc]} \BibitemShut {NoStop}%
\bibitem [{\citenamefont {Cook}(2020)}]{Cook:2020otn}%
  \BibitemOpen
  \bibfield  {author} {\bibinfo {author} {\bibfnamefont {G.~B.}\ \bibnamefont {Cook}},\ }\bibfield  {title} {\bibinfo {title} {{Aspects of multimode Kerr ringdown fitting}},\ }\href {https://doi.org/10.1103/PhysRevD.102.024027} {\bibfield  {journal} {\bibinfo  {journal} {Phys. Rev. D}\ }\textbf {\bibinfo {volume} {102}},\ \bibinfo {pages} {024027} (\bibinfo {year} {2020})},\ \Eprint {https://arxiv.org/abs/2004.08347} {arXiv:2004.08347 [gr-qc]} \BibitemShut {NoStop}%
\bibitem [{\citenamefont {Carullo}\ and\ \citenamefont {De~Amicis}(2023)}]{Carullo:2023tff}%
  \BibitemOpen
  \bibfield  {author} {\bibinfo {author} {\bibfnamefont {G.}~\bibnamefont {Carullo}}\ and\ \bibinfo {author} {\bibfnamefont {M.}~\bibnamefont {De~Amicis}},\ }\bibfield  {title} {\bibinfo {title} {{Late-time tails in nonlinear evolutions of merging black hole binaries}},\ }\href@noop {} {\  (\bibinfo {year} {2023})},\ \Eprint {https://arxiv.org/abs/2310.12968} {arXiv:2310.12968 [gr-qc]} \BibitemShut {NoStop}%
\bibitem [{\citenamefont {Cardoso}\ \emph {et~al.}(2024)\citenamefont {Cardoso}, \citenamefont {Carullo}, \citenamefont {De~Amicis}, \citenamefont {Duque}, \citenamefont {Katagiri}, \citenamefont {Pereniguez}, \citenamefont {Redondo-Yuste}, \citenamefont {Spieksma},\ and\ \citenamefont {Zhong}}]{Cardoso:2024jme}%
  \BibitemOpen
  \bibfield  {author} {\bibinfo {author} {\bibfnamefont {V.}~\bibnamefont {Cardoso}}, \bibinfo {author} {\bibfnamefont {G.}~\bibnamefont {Carullo}}, \bibinfo {author} {\bibfnamefont {M.}~\bibnamefont {De~Amicis}}, \bibinfo {author} {\bibfnamefont {F.}~\bibnamefont {Duque}}, \bibinfo {author} {\bibfnamefont {T.}~\bibnamefont {Katagiri}}, \bibinfo {author} {\bibfnamefont {D.}~\bibnamefont {Pereniguez}}, \bibinfo {author} {\bibfnamefont {J.}~\bibnamefont {Redondo-Yuste}}, \bibinfo {author} {\bibfnamefont {T.~F.~M.}\ \bibnamefont {Spieksma}},\ and\ \bibinfo {author} {\bibfnamefont {Z.}~\bibnamefont {Zhong}},\ }\bibfield  {title} {\bibinfo {title} {{Hushing black holes: Tails in dynamical spacetimes}},\ }\href {https://doi.org/10.1103/PhysRevD.109.L121502} {\bibfield  {journal} {\bibinfo  {journal} {Phys. Rev. D}\ }\textbf {\bibinfo {volume} {109}},\ \bibinfo {pages} {L121502} (\bibinfo {year} {2024})},\ \Eprint {https://arxiv.org/abs/2405.12290} {arXiv:2405.12290 [gr-qc]} \BibitemShut {NoStop}%
\bibitem [{\citenamefont {De~Amicis}\ \emph {et~al.}(2024{\natexlab{a}})\citenamefont {De~Amicis}, \citenamefont {Albanesi},\ and\ \citenamefont {Carullo}}]{DeAmicis:2024not}%
  \BibitemOpen
  \bibfield  {author} {\bibinfo {author} {\bibfnamefont {M.}~\bibnamefont {De~Amicis}}, \bibinfo {author} {\bibfnamefont {S.}~\bibnamefont {Albanesi}},\ and\ \bibinfo {author} {\bibfnamefont {G.}~\bibnamefont {Carullo}},\ }\bibfield  {title} {\bibinfo {title} {{Inspiral-inherited ringdown tails}},\ }\href {https://doi.org/10.1103/PhysRevD.110.104005} {\bibfield  {journal} {\bibinfo  {journal} {Phys. Rev. D}\ }\textbf {\bibinfo {volume} {110}},\ \bibinfo {pages} {104005} (\bibinfo {year} {2024}{\natexlab{a}})},\ \Eprint {https://arxiv.org/abs/2406.17018} {arXiv:2406.17018 [gr-qc]} \BibitemShut {NoStop}%
\bibitem [{\citenamefont {Islam}\ \emph {et~al.}(2024)\citenamefont {Islam}, \citenamefont {Faggioli}, \citenamefont {Khanna}, \citenamefont {Field}, \citenamefont {van~de Meent},\ and\ \citenamefont {Buonanno}}]{Islam:2024vro}%
  \BibitemOpen
  \bibfield  {author} {\bibinfo {author} {\bibfnamefont {T.}~\bibnamefont {Islam}}, \bibinfo {author} {\bibfnamefont {G.}~\bibnamefont {Faggioli}}, \bibinfo {author} {\bibfnamefont {G.}~\bibnamefont {Khanna}}, \bibinfo {author} {\bibfnamefont {S.~E.}\ \bibnamefont {Field}}, \bibinfo {author} {\bibfnamefont {M.}~\bibnamefont {van~de Meent}},\ and\ \bibinfo {author} {\bibfnamefont {A.}~\bibnamefont {Buonanno}},\ }\bibfield  {title} {\bibinfo {title} {{Phenomenology and origin of late-time tails in eccentric binary black hole mergers}},\ }\href@noop {} {\  (\bibinfo {year} {2024})},\ \Eprint {https://arxiv.org/abs/2407.04682} {arXiv:2407.04682 [gr-qc]} \BibitemShut {NoStop}%
\bibitem [{\citenamefont {Ma}\ \emph {et~al.}(2025{\natexlab{a}})\citenamefont {Ma}, \citenamefont {Nelli}, \citenamefont {Moxon}, \citenamefont {Scheel}, \citenamefont {Deppe}, \citenamefont {Kidder}, \citenamefont {Throwe},\ and\ \citenamefont {Vu}}]{Ma:2024bed}%
  \BibitemOpen
  \bibfield  {author} {\bibinfo {author} {\bibfnamefont {S.}~\bibnamefont {Ma}}, \bibinfo {author} {\bibfnamefont {K.~C.}\ \bibnamefont {Nelli}}, \bibinfo {author} {\bibfnamefont {J.}~\bibnamefont {Moxon}}, \bibinfo {author} {\bibfnamefont {M.~A.}\ \bibnamefont {Scheel}}, \bibinfo {author} {\bibfnamefont {N.}~\bibnamefont {Deppe}}, \bibinfo {author} {\bibfnamefont {L.~E.}\ \bibnamefont {Kidder}}, \bibinfo {author} {\bibfnamefont {W.}~\bibnamefont {Throwe}},\ and\ \bibinfo {author} {\bibfnamefont {N.~L.}\ \bibnamefont {Vu}},\ }\bibfield  {title} {\bibinfo {title} {{Einstein{\textendash}Klein{\textendash}Gordon system via Cauchy-characteristic evolution: computation of memory and ringdown tail}},\ }\href {https://doi.org/10.1088/1361-6382/adaf6f} {\bibfield  {journal} {\bibinfo  {journal} {Class. Quant. Grav.}\ }\textbf {\bibinfo {volume} {42}},\ \bibinfo {pages} {055006} (\bibinfo {year} {2025}{\natexlab{a}})},\ \Eprint {https://arxiv.org/abs/2409.06141} {arXiv:2409.06141 [gr-qc]} \BibitemShut {NoStop}%
\bibitem [{\citenamefont {De~Amicis}\ \emph {et~al.}(2024{\natexlab{b}})\citenamefont {De~Amicis} \emph {et~al.}}]{DeAmicis:2024eoy}%
  \BibitemOpen
  \bibfield  {author} {\bibinfo {author} {\bibfnamefont {M.}~\bibnamefont {De~Amicis}} \emph {et~al.},\ }\bibfield  {title} {\bibinfo {title} {{Late-time tails in nonlinear evolutions of merging black holes}},\ }\href@noop {} {\  (\bibinfo {year} {2024}{\natexlab{b}})},\ \Eprint {https://arxiv.org/abs/2412.06887} {arXiv:2412.06887 [gr-qc]} \BibitemShut {NoStop}%
\bibitem [{\citenamefont {Ma}\ \emph {et~al.}(2025{\natexlab{b}})\citenamefont {Ma}, \citenamefont {Scheel}, \citenamefont {Moxon}, \citenamefont {Nelli}, \citenamefont {Deppe}, \citenamefont {Kidder}, \citenamefont {Throwe},\ and\ \citenamefont {Vu}}]{Ma:2024hzq}%
  \BibitemOpen
  \bibfield  {author} {\bibinfo {author} {\bibfnamefont {S.}~\bibnamefont {Ma}}, \bibinfo {author} {\bibfnamefont {M.~A.}\ \bibnamefont {Scheel}}, \bibinfo {author} {\bibfnamefont {J.}~\bibnamefont {Moxon}}, \bibinfo {author} {\bibfnamefont {K.~C.}\ \bibnamefont {Nelli}}, \bibinfo {author} {\bibfnamefont {N.}~\bibnamefont {Deppe}}, \bibinfo {author} {\bibfnamefont {L.~E.}\ \bibnamefont {Kidder}}, \bibinfo {author} {\bibfnamefont {W.}~\bibnamefont {Throwe}},\ and\ \bibinfo {author} {\bibfnamefont {N.~L.}\ \bibnamefont {Vu}},\ }\bibfield  {title} {\bibinfo {title} {{Merging black holes with Cauchy-characteristic matching: Computation of late-time tails}},\ }\href {https://doi.org/10.1103/jd26-8q5w} {\bibfield  {journal} {\bibinfo  {journal} {Phys. Rev. D}\ }\textbf {\bibinfo {volume} {112}},\ \bibinfo {pages} {024003} (\bibinfo {year} {2025}{\natexlab{b}})},\ \Eprint {https://arxiv.org/abs/2412.06906} {arXiv:2412.06906 [gr-qc]} \BibitemShut {NoStop}%
\bibitem [{\citenamefont {O'Shaughnessy}\ \emph {et~al.}(2013)\citenamefont {O'Shaughnessy}, \citenamefont {London}, \citenamefont {Healy},\ and\ \citenamefont {Shoemaker}}]{OShaughnessy:2012iol}%
  \BibitemOpen
  \bibfield  {author} {\bibinfo {author} {\bibfnamefont {R.}~\bibnamefont {O'Shaughnessy}}, \bibinfo {author} {\bibfnamefont {L.}~\bibnamefont {London}}, \bibinfo {author} {\bibfnamefont {J.}~\bibnamefont {Healy}},\ and\ \bibinfo {author} {\bibfnamefont {D.}~\bibnamefont {Shoemaker}},\ }\bibfield  {title} {\bibinfo {title} {{Precession during merger: Strong polarization changes are observationally accessible features of strong-field gravity during binary black hole merger}},\ }\href {https://doi.org/10.1103/PhysRevD.87.044038} {\bibfield  {journal} {\bibinfo  {journal} {Phys. Rev. D}\ }\textbf {\bibinfo {volume} {87}},\ \bibinfo {pages} {044038} (\bibinfo {year} {2013})},\ \Eprint {https://arxiv.org/abs/1209.3712} {arXiv:1209.3712 [gr-qc]} \BibitemShut {NoStop}%
\bibitem [{\citenamefont {Zhu}\ \emph {et~al.}(2025)\citenamefont {Zhu} \emph {et~al.}}]{Zhu:2023fnf}%
  \BibitemOpen
  \bibfield  {author} {\bibinfo {author} {\bibfnamefont {H.}~\bibnamefont {Zhu}} \emph {et~al.},\ }\bibfield  {title} {\bibinfo {title} {{Black hole spectroscopy for precessing binary black hole coalescences}},\ }\href {https://doi.org/10.1103/PhysRevD.111.064052} {\bibfield  {journal} {\bibinfo  {journal} {Phys. Rev. D}\ }\textbf {\bibinfo {volume} {111}},\ \bibinfo {pages} {064052} (\bibinfo {year} {2025})},\ \Eprint {https://arxiv.org/abs/2312.08588} {arXiv:2312.08588 [gr-qc]} \BibitemShut {NoStop}%
\bibitem [{\citenamefont {Sinha}\ \emph {et~al.}(2025)\citenamefont {Sinha}, \citenamefont {Sun},\ and\ \citenamefont {Ma}}]{Sinha:2025snr}%
  \BibitemOpen
  \bibfield  {author} {\bibinfo {author} {\bibfnamefont {M.~R.}\ \bibnamefont {Sinha}}, \bibinfo {author} {\bibfnamefont {L.}~\bibnamefont {Sun}},\ and\ \bibinfo {author} {\bibfnamefont {S.}~\bibnamefont {Ma}},\ }\bibfield  {title} {\bibinfo {title} {{Impact of Detector Calibration Accuracy on Black Hole Spectroscopy}},\ }\href@noop {} {\  (\bibinfo {year} {2025})},\ \Eprint {https://arxiv.org/abs/2506.15979} {arXiv:2506.15979 [gr-qc]} \BibitemShut {NoStop}%
\bibitem [{\citenamefont {Cheung}\ \emph {et~al.}(2023)\citenamefont {Cheung} \emph {et~al.}}]{Cheung:2022rbm}%
  \BibitemOpen
  \bibfield  {author} {\bibinfo {author} {\bibfnamefont {M.~H.-Y.}\ \bibnamefont {Cheung}} \emph {et~al.},\ }\bibfield  {title} {\bibinfo {title} {{Nonlinear Effects in Black Hole Ringdown}},\ }\href {https://doi.org/10.1103/PhysRevLett.130.081401} {\bibfield  {journal} {\bibinfo  {journal} {Phys. Rev. Lett.}\ }\textbf {\bibinfo {volume} {130}},\ \bibinfo {pages} {081401} (\bibinfo {year} {2023})},\ \Eprint {https://arxiv.org/abs/2208.07374} {arXiv:2208.07374 [gr-qc]} \BibitemShut {NoStop}%
\bibitem [{\citenamefont {Redondo-Yuste}\ \emph {et~al.}(2024)\citenamefont {Redondo-Yuste}, \citenamefont {Carullo}, \citenamefont {Ripley}, \citenamefont {Berti},\ and\ \citenamefont {Cardoso}}]{Redondo-Yuste:2023seq}%
  \BibitemOpen
  \bibfield  {author} {\bibinfo {author} {\bibfnamefont {J.}~\bibnamefont {Redondo-Yuste}}, \bibinfo {author} {\bibfnamefont {G.}~\bibnamefont {Carullo}}, \bibinfo {author} {\bibfnamefont {J.~L.}\ \bibnamefont {Ripley}}, \bibinfo {author} {\bibfnamefont {E.}~\bibnamefont {Berti}},\ and\ \bibinfo {author} {\bibfnamefont {V.}~\bibnamefont {Cardoso}},\ }\bibfield  {title} {\bibinfo {title} {{Spin dependence of black hole ringdown nonlinearities}},\ }\href {https://doi.org/10.1103/PhysRevD.109.L101503} {\bibfield  {journal} {\bibinfo  {journal} {Phys. Rev. D}\ }\textbf {\bibinfo {volume} {109}},\ \bibinfo {pages} {L101503} (\bibinfo {year} {2024})},\ \Eprint {https://arxiv.org/abs/2308.14796} {arXiv:2308.14796 [gr-qc]} \BibitemShut {NoStop}%
\bibitem [{\citenamefont {Sberna}\ \emph {et~al.}(2022)\citenamefont {Sberna}, \citenamefont {Bosch}, \citenamefont {East}, \citenamefont {Green},\ and\ \citenamefont {Lehner}}]{Sberna:2021eui}%
  \BibitemOpen
  \bibfield  {author} {\bibinfo {author} {\bibfnamefont {L.}~\bibnamefont {Sberna}}, \bibinfo {author} {\bibfnamefont {P.}~\bibnamefont {Bosch}}, \bibinfo {author} {\bibfnamefont {W.~E.}\ \bibnamefont {East}}, \bibinfo {author} {\bibfnamefont {S.~R.}\ \bibnamefont {Green}},\ and\ \bibinfo {author} {\bibfnamefont {L.}~\bibnamefont {Lehner}},\ }\bibfield  {title} {\bibinfo {title} {{Nonlinear effects in the black hole ringdown: Absorption-induced mode excitation}},\ }\href {https://doi.org/10.1103/PhysRevD.105.064046} {\bibfield  {journal} {\bibinfo  {journal} {Phys. Rev. D}\ }\textbf {\bibinfo {volume} {105}},\ \bibinfo {pages} {064046} (\bibinfo {year} {2022})},\ \Eprint {https://arxiv.org/abs/2112.11168} {arXiv:2112.11168 [gr-qc]} \BibitemShut {NoStop}%
\bibitem [{\citenamefont {Yi}\ \emph {et~al.}(2024)\citenamefont {Yi}, \citenamefont {Kuntz}, \citenamefont {Barausse}, \citenamefont {Berti}, \citenamefont {Cheung}, \citenamefont {Kritos},\ and\ \citenamefont {Maselli}}]{Yi:2024elj}%
  \BibitemOpen
  \bibfield  {author} {\bibinfo {author} {\bibfnamefont {S.}~\bibnamefont {Yi}}, \bibinfo {author} {\bibfnamefont {A.}~\bibnamefont {Kuntz}}, \bibinfo {author} {\bibfnamefont {E.}~\bibnamefont {Barausse}}, \bibinfo {author} {\bibfnamefont {E.}~\bibnamefont {Berti}}, \bibinfo {author} {\bibfnamefont {M.~H.-Y.}\ \bibnamefont {Cheung}}, \bibinfo {author} {\bibfnamefont {K.}~\bibnamefont {Kritos}},\ and\ \bibinfo {author} {\bibfnamefont {A.}~\bibnamefont {Maselli}},\ }\bibfield  {title} {\bibinfo {title} {{Nonlinear quasinormal mode detectability with next-generation gravitational wave detectors}},\ }\href {https://doi.org/10.1103/PhysRevD.109.124029} {\bibfield  {journal} {\bibinfo  {journal} {Phys. Rev. D}\ }\textbf {\bibinfo {volume} {109}},\ \bibinfo {pages} {124029} (\bibinfo {year} {2024})},\ \Eprint {https://arxiv.org/abs/2403.09767} {arXiv:2403.09767 [gr-qc]} \BibitemShut {NoStop}%
\bibitem [{\citenamefont {Khera}\ \emph {et~al.}(2025)\citenamefont {Khera}, \citenamefont {Ma},\ and\ \citenamefont {Yang}}]{Khera:2024bjs}%
  \BibitemOpen
  \bibfield  {author} {\bibinfo {author} {\bibfnamefont {N.}~\bibnamefont {Khera}}, \bibinfo {author} {\bibfnamefont {S.}~\bibnamefont {Ma}},\ and\ \bibinfo {author} {\bibfnamefont {H.}~\bibnamefont {Yang}},\ }\bibfield  {title} {\bibinfo {title} {{Quadratic Mode Couplings in Rotating Black Holes and Their Detectability}},\ }\href {https://doi.org/10.1103/PhysRevLett.134.211404} {\bibfield  {journal} {\bibinfo  {journal} {Phys. Rev. Lett.}\ }\textbf {\bibinfo {volume} {134}},\ \bibinfo {pages} {211404} (\bibinfo {year} {2025})},\ \Eprint {https://arxiv.org/abs/2410.14529} {arXiv:2410.14529 [gr-qc]} \BibitemShut {NoStop}%
\bibitem [{\citenamefont {Steppohn}\ \emph {et~al.}(2025)\citenamefont {Steppohn}, \citenamefont {V{\"o}lkel},\ and\ \citenamefont {Dietrich}}]{Steppohn:2025kbh}%
  \BibitemOpen
  \bibfield  {author} {\bibinfo {author} {\bibfnamefont {O.}~\bibnamefont {Steppohn}}, \bibinfo {author} {\bibfnamefont {S.~H.}\ \bibnamefont {V{\"o}lkel}},\ and\ \bibinfo {author} {\bibfnamefont {T.}~\bibnamefont {Dietrich}},\ }\bibfield  {title} {\bibinfo {title} {{Black hole spectroscopy of collapsing and merging neutron stars}},\ }\href@noop {} {\  (\bibinfo {year} {2025})},\ \Eprint {https://arxiv.org/abs/2508.15534} {arXiv:2508.15534 [gr-qc]} \BibitemShut {NoStop}%
\bibitem [{\citenamefont {Crisostomi}\ \emph {et~al.}(2023)\citenamefont {Crisostomi}, \citenamefont {Dey}, \citenamefont {Barausse},\ and\ \citenamefont {Trotta}}]{Crisostomi:2023tle}%
  \BibitemOpen
  \bibfield  {author} {\bibinfo {author} {\bibfnamefont {M.}~\bibnamefont {Crisostomi}}, \bibinfo {author} {\bibfnamefont {K.}~\bibnamefont {Dey}}, \bibinfo {author} {\bibfnamefont {E.}~\bibnamefont {Barausse}},\ and\ \bibinfo {author} {\bibfnamefont {R.}~\bibnamefont {Trotta}},\ }\bibfield  {title} {\bibinfo {title} {{Neural posterior estimation with guaranteed exact coverage: The ringdown of GW150914}},\ }\href {https://doi.org/10.1103/PhysRevD.108.044029} {\bibfield  {journal} {\bibinfo  {journal} {Phys. Rev. D}\ }\textbf {\bibinfo {volume} {108}},\ \bibinfo {pages} {044029} (\bibinfo {year} {2023})},\ \Eprint {https://arxiv.org/abs/2305.18528} {arXiv:2305.18528 [gr-qc]} \BibitemShut {NoStop}%
\bibitem [{\citenamefont {Pacilio}\ \emph {et~al.}(2024)\citenamefont {Pacilio}, \citenamefont {Bhagwat},\ and\ \citenamefont {Cotesta}}]{Pacilio:2024qcq}%
  \BibitemOpen
  \bibfield  {author} {\bibinfo {author} {\bibfnamefont {C.}~\bibnamefont {Pacilio}}, \bibinfo {author} {\bibfnamefont {S.}~\bibnamefont {Bhagwat}},\ and\ \bibinfo {author} {\bibfnamefont {R.}~\bibnamefont {Cotesta}},\ }\bibfield  {title} {\bibinfo {title} {{Simulation-based inference of black hole ringdowns in the time domain}},\ }\href {https://doi.org/10.1103/PhysRevD.110.083010} {\bibfield  {journal} {\bibinfo  {journal} {Phys. Rev. D}\ }\textbf {\bibinfo {volume} {110}},\ \bibinfo {pages} {083010} (\bibinfo {year} {2024})},\ \Eprint {https://arxiv.org/abs/2404.11373} {arXiv:2404.11373 [gr-qc]} \BibitemShut {NoStop}%
\bibitem [{\citenamefont {Finn}\ and\ \citenamefont {Chernoff}(1993)}]{Finn:1992xs}%
  \BibitemOpen
  \bibfield  {author} {\bibinfo {author} {\bibfnamefont {L.~S.}\ \bibnamefont {Finn}}\ and\ \bibinfo {author} {\bibfnamefont {D.~F.}\ \bibnamefont {Chernoff}},\ }\bibfield  {title} {\bibinfo {title} {{Observing binary inspiral in gravitational radiation: One interferometer}},\ }\href {https://doi.org/10.1103/PhysRevD.47.2198} {\bibfield  {journal} {\bibinfo  {journal} {Phys. Rev. D}\ }\textbf {\bibinfo {volume} {47}},\ \bibinfo {pages} {2198} (\bibinfo {year} {1993})},\ \Eprint {https://arxiv.org/abs/gr-qc/9301003} {arXiv:gr-qc/9301003} \BibitemShut {NoStop}%
\bibitem [{\citenamefont {Krolak}\ \emph {et~al.}(1993)\citenamefont {Krolak}, \citenamefont {Lobo},\ and\ \citenamefont {Meers}}]{Krolak:1993zy}%
  \BibitemOpen
  \bibfield  {author} {\bibinfo {author} {\bibfnamefont {A.}~\bibnamefont {Krolak}}, \bibinfo {author} {\bibfnamefont {J.~A.}\ \bibnamefont {Lobo}},\ and\ \bibinfo {author} {\bibfnamefont {B.~J.}\ \bibnamefont {Meers}},\ }\bibfield  {title} {\bibinfo {title} {{Estimation of the parameters of the gravitational wave signal of a coalescing binary system}},\ }\href {https://doi.org/10.1103/PhysRevD.48.3451} {\bibfield  {journal} {\bibinfo  {journal} {Phys. Rev. D}\ }\textbf {\bibinfo {volume} {48}},\ \bibinfo {pages} {3451} (\bibinfo {year} {1993})}\BibitemShut {NoStop}%
\bibitem [{\citenamefont {Kokkotas}\ \emph {et~al.}(1994)\citenamefont {Kokkotas}, \citenamefont {Tsegas},\ and\ \citenamefont {Krolak}}]{Kokkotas:1994ef}%
  \BibitemOpen
  \bibfield  {author} {\bibinfo {author} {\bibfnamefont {K.~D.}\ \bibnamefont {Kokkotas}}, \bibinfo {author} {\bibfnamefont {G.}~\bibnamefont {Tsegas}},\ and\ \bibinfo {author} {\bibfnamefont {A.}~\bibnamefont {Krolak}},\ }\bibfield  {title} {\bibinfo {title} {{Statistical analysis of the estimators of the parameters of the gravitational wave signal from a coalescing binary}},\ }\href {https://doi.org/10.1088/0264-9381/11/7/023} {\bibfield  {journal} {\bibinfo  {journal} {Class. Quant. Grav.}\ }\textbf {\bibinfo {volume} {11}},\ \bibinfo {pages} {1901} (\bibinfo {year} {1994})}\BibitemShut {NoStop}%
\bibitem [{\citenamefont {Flanagan}\ and\ \citenamefont {Hughes}(1998)}]{Flanagan:1997kp}%
  \BibitemOpen
  \bibfield  {author} {\bibinfo {author} {\bibfnamefont {E.~E.}\ \bibnamefont {Flanagan}}\ and\ \bibinfo {author} {\bibfnamefont {S.~A.}\ \bibnamefont {Hughes}},\ }\bibfield  {title} {\bibinfo {title} {{Measuring gravitational waves from binary black hole coalescences: 2. The Waves' information and its extraction, with and without templates}},\ }\href {https://doi.org/10.1103/PhysRevD.57.4566} {\bibfield  {journal} {\bibinfo  {journal} {Phys. Rev. D}\ }\textbf {\bibinfo {volume} {57}},\ \bibinfo {pages} {4566} (\bibinfo {year} {1998})},\ \Eprint {https://arxiv.org/abs/gr-qc/9710129} {arXiv:gr-qc/9710129} \BibitemShut {NoStop}%
\bibitem [{\citenamefont {Cutler}\ and\ \citenamefont {Vallisneri}(2007)}]{Cutler:2007mi}%
  \BibitemOpen
  \bibfield  {author} {\bibinfo {author} {\bibfnamefont {C.}~\bibnamefont {Cutler}}\ and\ \bibinfo {author} {\bibfnamefont {M.}~\bibnamefont {Vallisneri}},\ }\bibfield  {title} {\bibinfo {title} {{LISA detections of massive black hole inspirals: Parameter extraction errors due to inaccurate template waveforms}},\ }\href {https://doi.org/10.1103/PhysRevD.76.104018} {\bibfield  {journal} {\bibinfo  {journal} {Phys. Rev. D}\ }\textbf {\bibinfo {volume} {76}},\ \bibinfo {pages} {104018} (\bibinfo {year} {2007})},\ \Eprint {https://arxiv.org/abs/0707.2982} {arXiv:0707.2982 [gr-qc]} \BibitemShut {NoStop}%
\bibitem [{\citenamefont {Gupta}\ \emph {et~al.}(2024)\citenamefont {Gupta} \emph {et~al.}}]{Gupta:2024gun}%
  \BibitemOpen
  \bibfield  {author} {\bibinfo {author} {\bibfnamefont {A.}~\bibnamefont {Gupta}} \emph {et~al.},\ }\bibfield  {title} {\bibinfo {title} {{Possible causes of false general relativity violations in gravitational wave observations}}\ }\href {https://doi.org/10.21468/SciPostPhysCommRep.5} {10.21468/SciPostPhysCommRep.5} (\bibinfo {year} {2024}),\ \Eprint {https://arxiv.org/abs/2405.02197} {arXiv:2405.02197 [gr-qc]} \BibitemShut {NoStop}%
\bibitem [{\citenamefont {Chandramouli}\ \emph {et~al.}(2025)\citenamefont {Chandramouli}, \citenamefont {Prokup}, \citenamefont {Berti},\ and\ \citenamefont {Yunes}}]{Chandramouli:2024vhw}%
  \BibitemOpen
  \bibfield  {author} {\bibinfo {author} {\bibfnamefont {R.~S.}\ \bibnamefont {Chandramouli}}, \bibinfo {author} {\bibfnamefont {K.}~\bibnamefont {Prokup}}, \bibinfo {author} {\bibfnamefont {E.}~\bibnamefont {Berti}},\ and\ \bibinfo {author} {\bibfnamefont {N.}~\bibnamefont {Yunes}},\ }\bibfield  {title} {\bibinfo {title} {{Systematic biases due to waveform mismodeling in parametrized post-Einsteinian tests of general relativity: The impact of neglecting spin precession and higher modes}},\ }\href {https://doi.org/10.1103/PhysRevD.111.044026} {\bibfield  {journal} {\bibinfo  {journal} {Phys. Rev. D}\ }\textbf {\bibinfo {volume} {111}},\ \bibinfo {pages} {044026} (\bibinfo {year} {2025})},\ \Eprint {https://arxiv.org/abs/2410.06254} {arXiv:2410.06254 [gr-qc]} \BibitemShut {NoStop}%
\bibitem [{\citenamefont {Cano}\ \emph {et~al.}(2024)\citenamefont {Cano}, \citenamefont {Capuano}, \citenamefont {Franchini}, \citenamefont {Maenaut},\ and\ \citenamefont {V{\"o}lkel}}]{Cano:2024jkd}%
  \BibitemOpen
  \bibfield  {author} {\bibinfo {author} {\bibfnamefont {P.~A.}\ \bibnamefont {Cano}}, \bibinfo {author} {\bibfnamefont {L.}~\bibnamefont {Capuano}}, \bibinfo {author} {\bibfnamefont {N.}~\bibnamefont {Franchini}}, \bibinfo {author} {\bibfnamefont {S.}~\bibnamefont {Maenaut}},\ and\ \bibinfo {author} {\bibfnamefont {S.~H.}\ \bibnamefont {V{\"o}lkel}},\ }\bibfield  {title} {\bibinfo {title} {{Parametrized quasinormal mode framework for modified Teukolsky equations}},\ }\href {https://doi.org/10.1103/PhysRevD.110.104007} {\bibfield  {journal} {\bibinfo  {journal} {Phys. Rev. D}\ }\textbf {\bibinfo {volume} {110}},\ \bibinfo {pages} {104007} (\bibinfo {year} {2024})},\ \Eprint {https://arxiv.org/abs/2407.15947} {arXiv:2407.15947 [gr-qc]} \BibitemShut {NoStop}%
\bibitem [{\citenamefont {Völkel}\ and\ \citenamefont {Franchini}(2024)}]{sebastian_volkel_2024_14001739}%
  \BibitemOpen
  \bibfield  {author} {\bibinfo {author} {\bibfnamefont {S.}~\bibnamefont {Völkel}}\ and\ \bibinfo {author} {\bibfnamefont {N.}~\bibnamefont {Franchini}},\ }\href {https://doi.org/10.5281/zenodo.14001739} {\bibinfo {title} {sebastianvoelkel/parametrized\_qnm\_framework: Parametrized qnm framework}} (\bibinfo {year} {2024})\BibitemShut {NoStop}%
\bibitem [{\citenamefont {Carullo}\ \emph {et~al.}(2019)\citenamefont {Carullo}, \citenamefont {Del~Pozzo},\ and\ \citenamefont {Veitch}}]{Carullo:2019flw}%
  \BibitemOpen
  \bibfield  {author} {\bibinfo {author} {\bibfnamefont {G.}~\bibnamefont {Carullo}}, \bibinfo {author} {\bibfnamefont {W.}~\bibnamefont {Del~Pozzo}},\ and\ \bibinfo {author} {\bibfnamefont {J.}~\bibnamefont {Veitch}},\ }\bibfield  {title} {\bibinfo {title} {{Observational Black Hole Spectroscopy: A time-domain multimode analysis of GW150914}},\ }\href {https://doi.org/10.1103/PhysRevD.99.123029} {\bibfield  {journal} {\bibinfo  {journal} {Phys. Rev. D}\ }\textbf {\bibinfo {volume} {99}},\ \bibinfo {pages} {123029} (\bibinfo {year} {2019})},\ \bibinfo {note} {[Erratum: Phys.Rev.D 100, 089903 (2019)]},\ \Eprint {https://arxiv.org/abs/1902.07527} {arXiv:1902.07527 [gr-qc]} \BibitemShut {NoStop}%
\bibitem [{\citenamefont {Isi}\ and\ \citenamefont {Farr}(2021)}]{Isi:2021iql}%
  \BibitemOpen
  \bibfield  {author} {\bibinfo {author} {\bibfnamefont {M.}~\bibnamefont {Isi}}\ and\ \bibinfo {author} {\bibfnamefont {W.~M.}\ \bibnamefont {Farr}},\ }\bibfield  {title} {\bibinfo {title} {{Analyzing black-hole ringdowns}},\ }\href@noop {} {\  (\bibinfo {year} {2021})},\ \Eprint {https://arxiv.org/abs/2107.05609} {arXiv:2107.05609 [gr-qc]} \BibitemShut {NoStop}%
\bibitem [{\citenamefont {Foreman-Mackey}\ \emph {et~al.}(2013)\citenamefont {Foreman-Mackey}, \citenamefont {Hogg}, \citenamefont {Lang},\ and\ \citenamefont {Goodman}}]{Foreman-Mackey:2012any}%
  \BibitemOpen
  \bibfield  {author} {\bibinfo {author} {\bibfnamefont {D.}~\bibnamefont {Foreman-Mackey}}, \bibinfo {author} {\bibfnamefont {D.~W.}\ \bibnamefont {Hogg}}, \bibinfo {author} {\bibfnamefont {D.}~\bibnamefont {Lang}},\ and\ \bibinfo {author} {\bibfnamefont {J.}~\bibnamefont {Goodman}},\ }\bibfield  {title} {\bibinfo {title} {{emcee: The MCMC Hammer}},\ }\href {https://doi.org/10.1086/670067} {\bibfield  {journal} {\bibinfo  {journal} {Publ. Astron. Soc. Pac.}\ }\textbf {\bibinfo {volume} {125}},\ \bibinfo {pages} {306} (\bibinfo {year} {2013})},\ \Eprint {https://arxiv.org/abs/1202.3665} {arXiv:1202.3665 [astro-ph.IM]} \BibitemShut {NoStop}%
\bibitem [{\citenamefont {Dhani}\ \emph {et~al.}(2025)\citenamefont {Dhani}, \citenamefont {V{\"o}lkel}, \citenamefont {Buonanno}, \citenamefont {Estelles}, \citenamefont {Gair}, \citenamefont {Pfeiffer}, \citenamefont {Pompili},\ and\ \citenamefont {Toubiana}}]{Dhani:2024jja}%
  \BibitemOpen
  \bibfield  {author} {\bibinfo {author} {\bibfnamefont {A.}~\bibnamefont {Dhani}}, \bibinfo {author} {\bibfnamefont {S.~H.}\ \bibnamefont {V{\"o}lkel}}, \bibinfo {author} {\bibfnamefont {A.}~\bibnamefont {Buonanno}}, \bibinfo {author} {\bibfnamefont {H.}~\bibnamefont {Estelles}}, \bibinfo {author} {\bibfnamefont {J.}~\bibnamefont {Gair}}, \bibinfo {author} {\bibfnamefont {H.~P.}\ \bibnamefont {Pfeiffer}}, \bibinfo {author} {\bibfnamefont {L.}~\bibnamefont {Pompili}},\ and\ \bibinfo {author} {\bibfnamefont {A.}~\bibnamefont {Toubiana}},\ }\bibfield  {title} {\bibinfo {title} {{Systematic Biases in Estimating the Properties of Black Holes Due to Inaccurate Gravitational-Wave Models}},\ }\href {https://doi.org/10.1103/5pks-qz6b} {\bibfield  {journal} {\bibinfo  {journal} {Phys. Rev. X}\ }\textbf {\bibinfo {volume} {15}},\ \bibinfo {pages} {031036} (\bibinfo {year} {2025})},\ \Eprint {https://arxiv.org/abs/2404.05811} {arXiv:2404.05811 [gr-qc]} \BibitemShut {NoStop}%
\bibitem [{\citenamefont {Kapil}\ \emph {et~al.}(2024)\citenamefont {Kapil}, \citenamefont {Reali}, \citenamefont {Cotesta},\ and\ \citenamefont {Berti}}]{Kapil:2024zdn}%
  \BibitemOpen
  \bibfield  {author} {\bibinfo {author} {\bibfnamefont {V.}~\bibnamefont {Kapil}}, \bibinfo {author} {\bibfnamefont {L.}~\bibnamefont {Reali}}, \bibinfo {author} {\bibfnamefont {R.}~\bibnamefont {Cotesta}},\ and\ \bibinfo {author} {\bibfnamefont {E.}~\bibnamefont {Berti}},\ }\bibfield  {title} {\bibinfo {title} {{Systematic bias from waveform modeling for binary black hole populations in next-generation gravitational wave detectors}},\ }\href {https://doi.org/10.1103/PhysRevD.109.104043} {\bibfield  {journal} {\bibinfo  {journal} {Phys. Rev. D}\ }\textbf {\bibinfo {volume} {109}},\ \bibinfo {pages} {104043} (\bibinfo {year} {2024})},\ \Eprint {https://arxiv.org/abs/2404.00090} {arXiv:2404.00090 [gr-qc]} \BibitemShut {NoStop}%
\bibitem [{\citenamefont {Yi}\ \emph {et~al.}(2025)\citenamefont {Yi}, \citenamefont {Iacovelli}, \citenamefont {Marsat}, \citenamefont {Wadekar},\ and\ \citenamefont {Berti}}]{Yi:2025pxe}%
  \BibitemOpen
  \bibfield  {author} {\bibinfo {author} {\bibfnamefont {S.}~\bibnamefont {Yi}}, \bibinfo {author} {\bibfnamefont {F.}~\bibnamefont {Iacovelli}}, \bibinfo {author} {\bibfnamefont {S.}~\bibnamefont {Marsat}}, \bibinfo {author} {\bibfnamefont {D.}~\bibnamefont {Wadekar}},\ and\ \bibinfo {author} {\bibfnamefont {E.}~\bibnamefont {Berti}},\ }\bibfield  {title} {\bibinfo {title} {{Systematic biases from the exclusion of higher harmonics in parameter estimation on LISA binaries}},\ }\href@noop {} {\  (\bibinfo {year} {2025})},\ \Eprint {https://arxiv.org/abs/2502.12237} {arXiv:2502.12237 [gr-qc]} \BibitemShut {NoStop}%
\bibitem [{\citenamefont {Capuano}\ \emph {et~al.}(2025)\citenamefont {Capuano}, \citenamefont {Vaglio}, \citenamefont {Chandramouli}, \citenamefont {Pitte}, \citenamefont {Kuntz},\ and\ \citenamefont {Barausse}}]{Capuano:2025kkl}%
  \BibitemOpen
  \bibfield  {author} {\bibinfo {author} {\bibfnamefont {L.}~\bibnamefont {Capuano}}, \bibinfo {author} {\bibfnamefont {M.}~\bibnamefont {Vaglio}}, \bibinfo {author} {\bibfnamefont {R.~S.}\ \bibnamefont {Chandramouli}}, \bibinfo {author} {\bibfnamefont {C.~L.}\ \bibnamefont {Pitte}}, \bibinfo {author} {\bibfnamefont {A.}~\bibnamefont {Kuntz}},\ and\ \bibinfo {author} {\bibfnamefont {E.}~\bibnamefont {Barausse}},\ }\bibfield  {title} {\bibinfo {title} {{Systematic bias in LISA ringdown analysis due to waveform inaccuracy}},\ }\href@noop {} {\  (\bibinfo {year} {2025})},\ \Eprint {https://arxiv.org/abs/2506.21181} {arXiv:2506.21181 [gr-qc]} \BibitemShut {NoStop}%
\bibitem [{\citenamefont {Forteza}\ \emph {et~al.}(2023)\citenamefont {Forteza}, \citenamefont {Bhagwat}, \citenamefont {Kumar},\ and\ \citenamefont {Pani}}]{Forteza:2022tgq}%
  \BibitemOpen
  \bibfield  {author} {\bibinfo {author} {\bibfnamefont {X.~J.}\ \bibnamefont {Forteza}}, \bibinfo {author} {\bibfnamefont {S.}~\bibnamefont {Bhagwat}}, \bibinfo {author} {\bibfnamefont {S.}~\bibnamefont {Kumar}},\ and\ \bibinfo {author} {\bibfnamefont {P.}~\bibnamefont {Pani}},\ }\bibfield  {title} {\bibinfo {title} {{Novel Ringdown Amplitude-Phase Consistency Test}},\ }\href {https://doi.org/10.1103/PhysRevLett.130.021001} {\bibfield  {journal} {\bibinfo  {journal} {Phys. Rev. Lett.}\ }\textbf {\bibinfo {volume} {130}},\ \bibinfo {pages} {021001} (\bibinfo {year} {2023})},\ \Eprint {https://arxiv.org/abs/2205.14910} {arXiv:2205.14910 [gr-qc]} \BibitemShut {NoStop}%
\bibitem [{\citenamefont {Berti}\ \emph {et~al.}(2007{\natexlab{b}})\citenamefont {Berti}, \citenamefont {Cardoso}, \citenamefont {Cardoso},\ and\ \citenamefont {Cavaglia}}]{Berti:2007zu}%
  \BibitemOpen
  \bibfield  {author} {\bibinfo {author} {\bibfnamefont {E.}~\bibnamefont {Berti}}, \bibinfo {author} {\bibfnamefont {J.}~\bibnamefont {Cardoso}}, \bibinfo {author} {\bibfnamefont {V.}~\bibnamefont {Cardoso}},\ and\ \bibinfo {author} {\bibfnamefont {M.}~\bibnamefont {Cavaglia}},\ }\bibfield  {title} {\bibinfo {title} {{Matched-filtering and parameter estimation of ringdown waveforms}},\ }\href {https://doi.org/10.1103/PhysRevD.76.104044} {\bibfield  {journal} {\bibinfo  {journal} {Phys. Rev. D}\ }\textbf {\bibinfo {volume} {76}},\ \bibinfo {pages} {104044} (\bibinfo {year} {2007}{\natexlab{b}})},\ \Eprint {https://arxiv.org/abs/0707.1202} {arXiv:0707.1202 [gr-qc]} \BibitemShut {NoStop}%
\bibitem [{\citenamefont {Isi}\ \emph {et~al.}(2019)\citenamefont {Isi}, \citenamefont {Giesler}, \citenamefont {Farr}, \citenamefont {Scheel},\ and\ \citenamefont {Teukolsky}}]{Isi:2019aib}%
  \BibitemOpen
  \bibfield  {author} {\bibinfo {author} {\bibfnamefont {M.}~\bibnamefont {Isi}}, \bibinfo {author} {\bibfnamefont {M.}~\bibnamefont {Giesler}}, \bibinfo {author} {\bibfnamefont {W.~M.}\ \bibnamefont {Farr}}, \bibinfo {author} {\bibfnamefont {M.~A.}\ \bibnamefont {Scheel}},\ and\ \bibinfo {author} {\bibfnamefont {S.~A.}\ \bibnamefont {Teukolsky}},\ }\bibfield  {title} {\bibinfo {title} {{Testing the no-hair theorem with GW150914}},\ }\href {https://doi.org/10.1103/PhysRevLett.123.111102} {\bibfield  {journal} {\bibinfo  {journal} {Phys. Rev. Lett.}\ }\textbf {\bibinfo {volume} {123}},\ \bibinfo {pages} {111102} (\bibinfo {year} {2019})},\ \Eprint {https://arxiv.org/abs/1905.00869} {arXiv:1905.00869 [gr-qc]} \BibitemShut {NoStop}%
\bibitem [{\citenamefont {Cotesta}\ \emph {et~al.}(2022)\citenamefont {Cotesta}, \citenamefont {Carullo}, \citenamefont {Berti},\ and\ \citenamefont {Cardoso}}]{Cotesta:2022pci}%
  \BibitemOpen
  \bibfield  {author} {\bibinfo {author} {\bibfnamefont {R.}~\bibnamefont {Cotesta}}, \bibinfo {author} {\bibfnamefont {G.}~\bibnamefont {Carullo}}, \bibinfo {author} {\bibfnamefont {E.}~\bibnamefont {Berti}},\ and\ \bibinfo {author} {\bibfnamefont {V.}~\bibnamefont {Cardoso}},\ }\bibfield  {title} {\bibinfo {title} {{Analysis of Ringdown Overtones in GW150914}},\ }\href {https://doi.org/10.1103/PhysRevLett.129.111102} {\bibfield  {journal} {\bibinfo  {journal} {Phys. Rev. Lett.}\ }\textbf {\bibinfo {volume} {129}},\ \bibinfo {pages} {111102} (\bibinfo {year} {2022})},\ \Eprint {https://arxiv.org/abs/2201.00822} {arXiv:2201.00822 [gr-qc]} \BibitemShut {NoStop}%
\bibitem [{\citenamefont {Wang}\ \emph {et~al.}(2023{\natexlab{a}})\citenamefont {Wang}, \citenamefont {Capano}, \citenamefont {Abedi}, \citenamefont {Kastha}, \citenamefont {Krishnan}, \citenamefont {Nielsen}, \citenamefont {Nitz},\ and\ \citenamefont {Westerweck}}]{Wang:2023xsy}%
  \BibitemOpen
  \bibfield  {author} {\bibinfo {author} {\bibfnamefont {Y.-F.}\ \bibnamefont {Wang}}, \bibinfo {author} {\bibfnamefont {C.~D.}\ \bibnamefont {Capano}}, \bibinfo {author} {\bibfnamefont {J.}~\bibnamefont {Abedi}}, \bibinfo {author} {\bibfnamefont {S.}~\bibnamefont {Kastha}}, \bibinfo {author} {\bibfnamefont {B.}~\bibnamefont {Krishnan}}, \bibinfo {author} {\bibfnamefont {A.~B.}\ \bibnamefont {Nielsen}}, \bibinfo {author} {\bibfnamefont {A.~H.}\ \bibnamefont {Nitz}},\ and\ \bibinfo {author} {\bibfnamefont {J.}~\bibnamefont {Westerweck}},\ }\bibfield  {title} {\bibinfo {title} {{A gating-and-inpainting perspective on GW150914 ringdown overtone: understanding the data analysis systematics}},\ }\href@noop {} {\  (\bibinfo {year} {2023}{\natexlab{a}})},\ \Eprint {https://arxiv.org/abs/2310.19645} {arXiv:2310.19645 [gr-qc]} \BibitemShut {NoStop}%
\bibitem [{\citenamefont {Toubiana}\ \emph {et~al.}(2024)\citenamefont {Toubiana}, \citenamefont {Pompili}, \citenamefont {Buonanno}, \citenamefont {Gair},\ and\ \citenamefont {Katz}}]{Toubiana:2023cwr}%
  \BibitemOpen
  \bibfield  {author} {\bibinfo {author} {\bibfnamefont {A.}~\bibnamefont {Toubiana}}, \bibinfo {author} {\bibfnamefont {L.}~\bibnamefont {Pompili}}, \bibinfo {author} {\bibfnamefont {A.}~\bibnamefont {Buonanno}}, \bibinfo {author} {\bibfnamefont {J.~R.}\ \bibnamefont {Gair}},\ and\ \bibinfo {author} {\bibfnamefont {M.~L.}\ \bibnamefont {Katz}},\ }\bibfield  {title} {\bibinfo {title} {{Measuring source properties and quasinormal mode frequencies of heavy massive black-hole binaries with LISA}},\ }\href {https://doi.org/10.1103/PhysRevD.109.104019} {\bibfield  {journal} {\bibinfo  {journal} {Phys. Rev. D}\ }\textbf {\bibinfo {volume} {109}},\ \bibinfo {pages} {104019} (\bibinfo {year} {2024})},\ \Eprint {https://arxiv.org/abs/2307.15086} {arXiv:2307.15086 [gr-qc]} \BibitemShut {NoStop}%
\bibitem [{LIG(2025)}]{LIGOScientific:2025obp}%
  \BibitemOpen
  \bibfield  {title} {\bibinfo {title} {{Black Hole Spectroscopy and Tests of General Relativity with GW250114}},\ }\href@noop {} {\  (\bibinfo {year} {2025})},\ \Eprint {https://arxiv.org/abs/2509.08099} {arXiv:2509.08099 [gr-qc]} \BibitemShut {NoStop}%
\bibitem [{\citenamefont {Correia}\ \emph {et~al.}(2024)\citenamefont {Correia}, \citenamefont {Wang}, \citenamefont {Westerweck},\ and\ \citenamefont {Capano}}]{Correia:2023bfn}%
  \BibitemOpen
  \bibfield  {author} {\bibinfo {author} {\bibfnamefont {A.}~\bibnamefont {Correia}}, \bibinfo {author} {\bibfnamefont {Y.-F.}\ \bibnamefont {Wang}}, \bibinfo {author} {\bibfnamefont {J.}~\bibnamefont {Westerweck}},\ and\ \bibinfo {author} {\bibfnamefont {C.~D.}\ \bibnamefont {Capano}},\ }\bibfield  {title} {\bibinfo {title} {{Low evidence for ringdown overtone in GW150914 when marginalizing over time and sky location uncertainty}},\ }\href {https://doi.org/10.1103/PhysRevD.110.L041501} {\bibfield  {journal} {\bibinfo  {journal} {Phys. Rev. D}\ }\textbf {\bibinfo {volume} {110}},\ \bibinfo {pages} {L041501} (\bibinfo {year} {2024})},\ \Eprint {https://arxiv.org/abs/2312.14118} {arXiv:2312.14118 [gr-qc]} \BibitemShut {NoStop}%
\bibitem [{\citenamefont {Wang}\ \emph {et~al.}(2023{\natexlab{b}})\citenamefont {Wang}, \citenamefont {Capano}, \citenamefont {Abedi}, \citenamefont {Kastha}, \citenamefont {Krishnan}, \citenamefont {Nielsen}, \citenamefont {Nitz},\ and\ \citenamefont {Westerweck}}]{Wang:2023ljx}%
  \BibitemOpen
  \bibfield  {author} {\bibinfo {author} {\bibfnamefont {Y.-F.}\ \bibnamefont {Wang}}, \bibinfo {author} {\bibfnamefont {C.~D.}\ \bibnamefont {Capano}}, \bibinfo {author} {\bibfnamefont {J.}~\bibnamefont {Abedi}}, \bibinfo {author} {\bibfnamefont {S.}~\bibnamefont {Kastha}}, \bibinfo {author} {\bibfnamefont {B.}~\bibnamefont {Krishnan}}, \bibinfo {author} {\bibfnamefont {A.~B.}\ \bibnamefont {Nielsen}}, \bibinfo {author} {\bibfnamefont {A.~H.}\ \bibnamefont {Nitz}},\ and\ \bibinfo {author} {\bibfnamefont {J.}~\bibnamefont {Westerweck}},\ }\bibfield  {title} {\bibinfo {title} {{A gating-and-inpainting perspective on GW150914 ringdown overtone: understanding the data analysis systematics}},\ }\href@noop {} {\  (\bibinfo {year} {2023}{\natexlab{b}})},\ \Eprint {https://arxiv.org/abs/2310.19645} {arXiv:2310.19645 [gr-qc]} \BibitemShut {NoStop}%
\bibitem [{\citenamefont {Others}(2025)}]{Others:2025nbi}%
  \BibitemOpen
  \bibfield  {author} {\bibinfo {author} {\bibnamefont {Others}} (\bibinfo {collaboration} {LIGO Scientific, VIRGO, Kagra}),\ }\bibfield  {title} {\bibinfo {title} {{GW230814: investigation of a loud gravitational-wave signal observed with a single detector}},\ }\href@noop {} {\  (\bibinfo {year} {2025})},\ \Eprint {https://arxiv.org/abs/2509.07348} {arXiv:2509.07348 [gr-qc]} \BibitemShut {NoStop}%
\bibitem [{\citenamefont {Maggiore}\ \emph {et~al.}(2020)\citenamefont {Maggiore} \emph {et~al.}}]{ET:2019dnz}%
  \BibitemOpen
  \bibfield  {author} {\bibinfo {author} {\bibfnamefont {M.}~\bibnamefont {Maggiore}} \emph {et~al.} (\bibinfo {collaboration} {ET}),\ }\bibfield  {title} {\bibinfo {title} {{Science Case for the Einstein Telescope}},\ }\href {https://doi.org/10.1088/1475-7516/2020/03/050} {\bibfield  {journal} {\bibinfo  {journal} {JCAP}\ }\textbf {\bibinfo {volume} {03}},\ \bibinfo {pages} {050}},\ \Eprint {https://arxiv.org/abs/1912.02622} {arXiv:1912.02622 [astro-ph.CO]} \BibitemShut {NoStop}%
\bibitem [{\citenamefont {Bhagwat}\ \emph {et~al.}(2020)\citenamefont {Bhagwat}, \citenamefont {Forteza}, \citenamefont {Pani},\ and\ \citenamefont {Ferrari}}]{Bhagwat:2019dtm}%
  \BibitemOpen
  \bibfield  {author} {\bibinfo {author} {\bibfnamefont {S.}~\bibnamefont {Bhagwat}}, \bibinfo {author} {\bibfnamefont {X.~J.}\ \bibnamefont {Forteza}}, \bibinfo {author} {\bibfnamefont {P.}~\bibnamefont {Pani}},\ and\ \bibinfo {author} {\bibfnamefont {V.}~\bibnamefont {Ferrari}},\ }\bibfield  {title} {\bibinfo {title} {{Ringdown overtones, black hole spectroscopy, and no-hair theorem tests}},\ }\href {https://doi.org/10.1103/PhysRevD.101.044033} {\bibfield  {journal} {\bibinfo  {journal} {Phys. Rev. D}\ }\textbf {\bibinfo {volume} {101}},\ \bibinfo {pages} {044033} (\bibinfo {year} {2020})},\ \Eprint {https://arxiv.org/abs/1910.08708} {arXiv:1910.08708 [gr-qc]} \BibitemShut {NoStop}%
\end{thebibliography}%

\appendix 

\section{Additional Results}\label{app_1}

\begin{figure*}[t]
\includegraphics[width=0.49\linewidth]{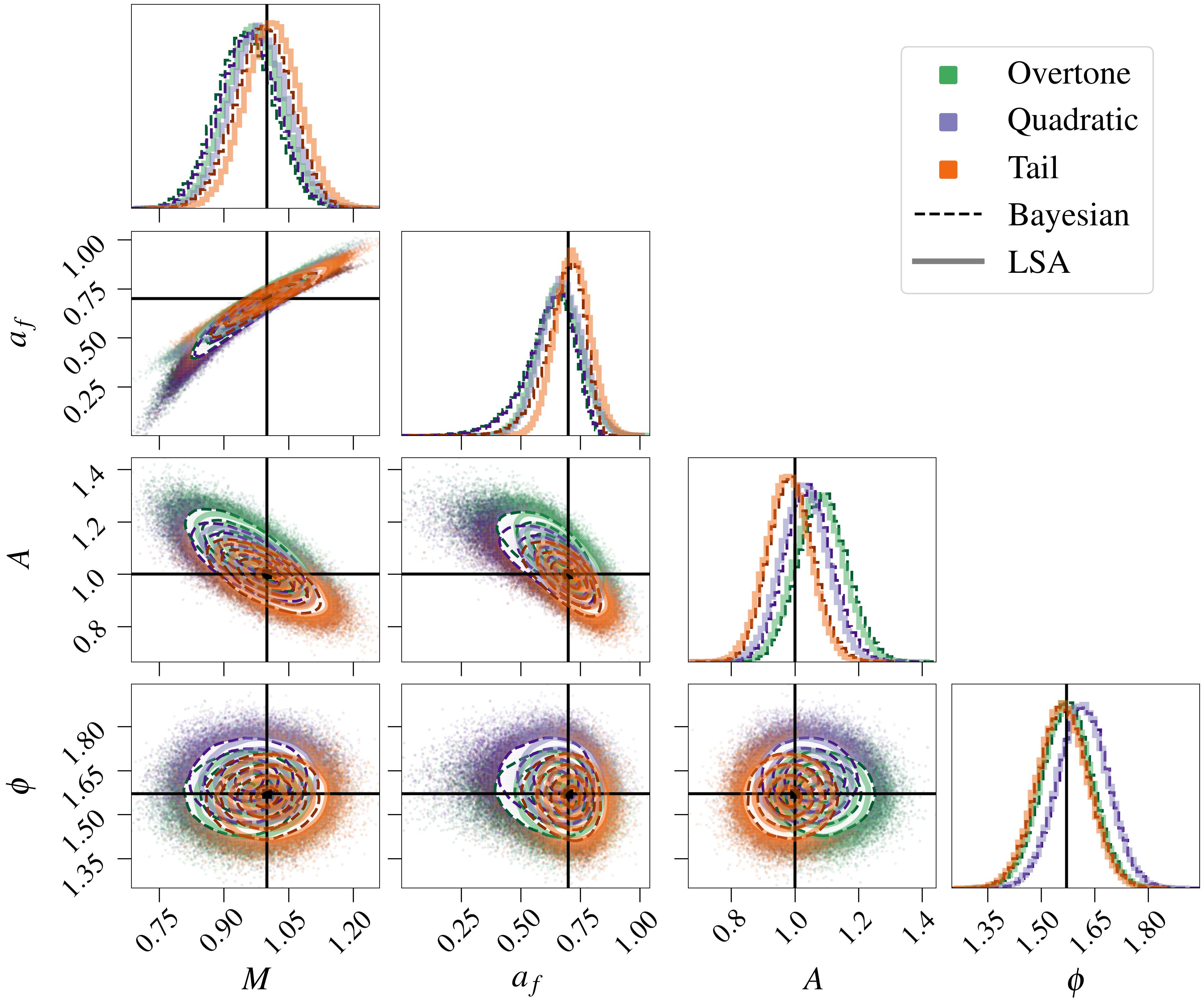}
\hfill
\includegraphics[width=0.49\linewidth]{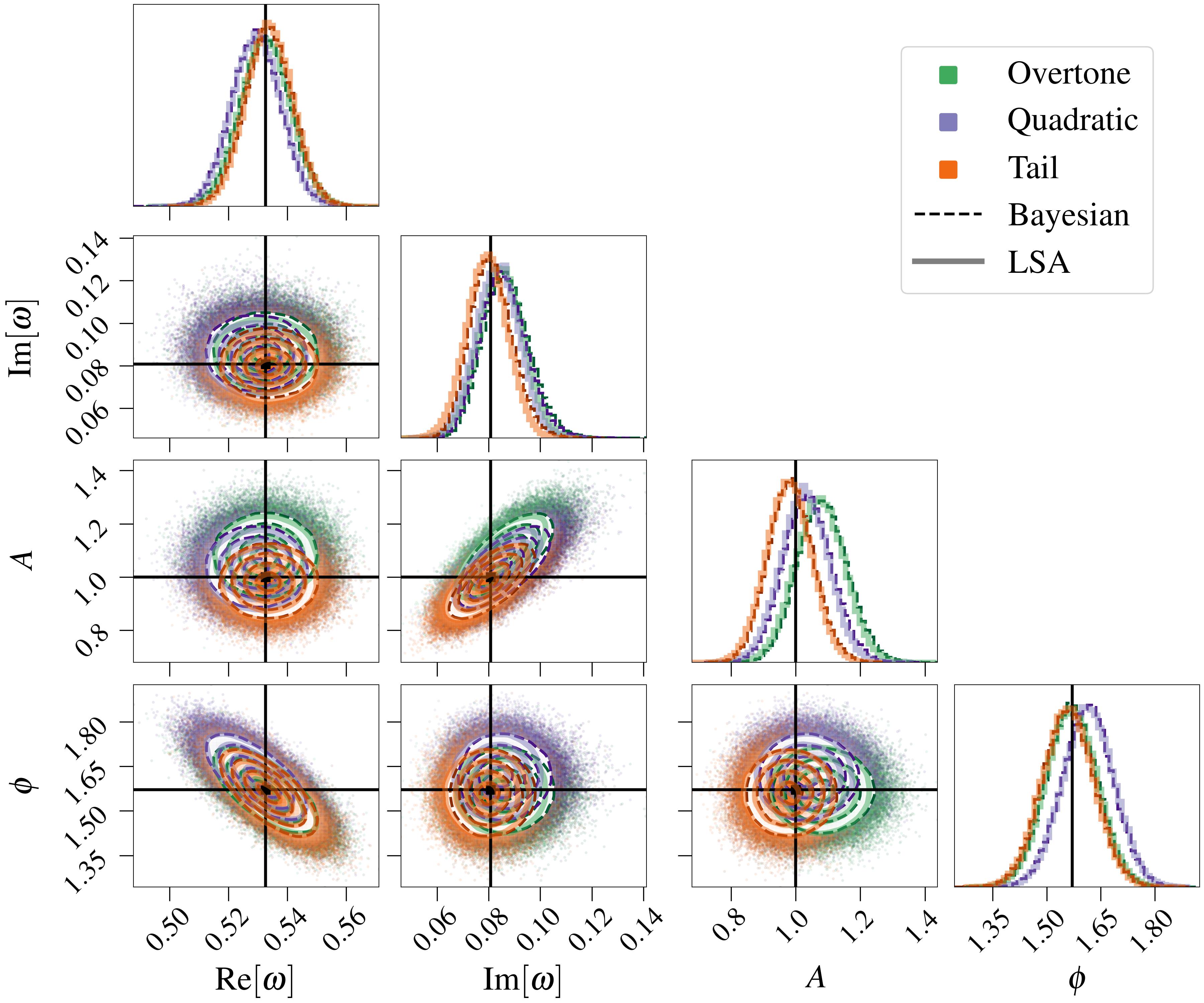}
\\
\vspace{0.3cm}
\includegraphics[width=0.49\linewidth]{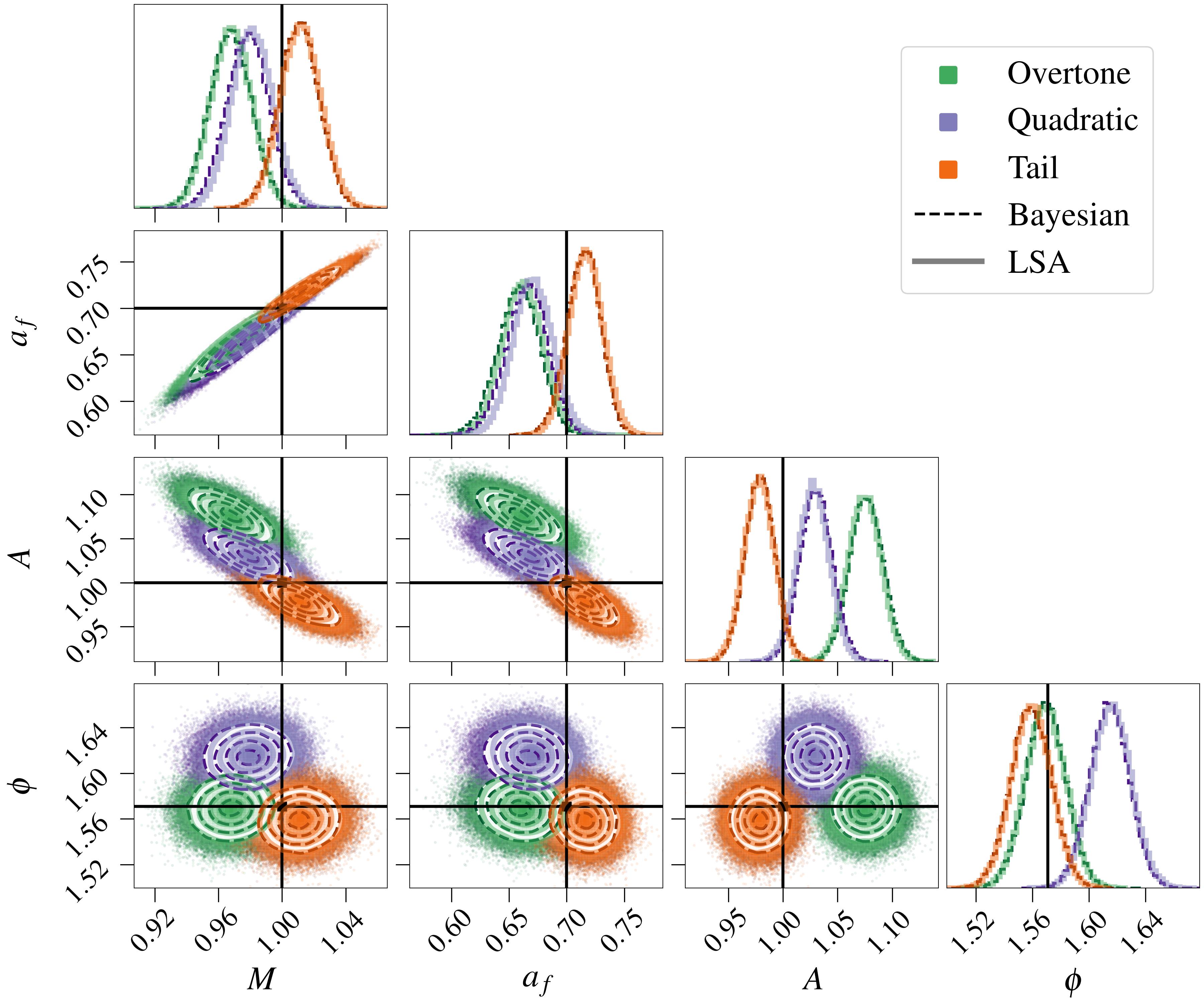}
\hfill
\includegraphics[width=0.49\linewidth]{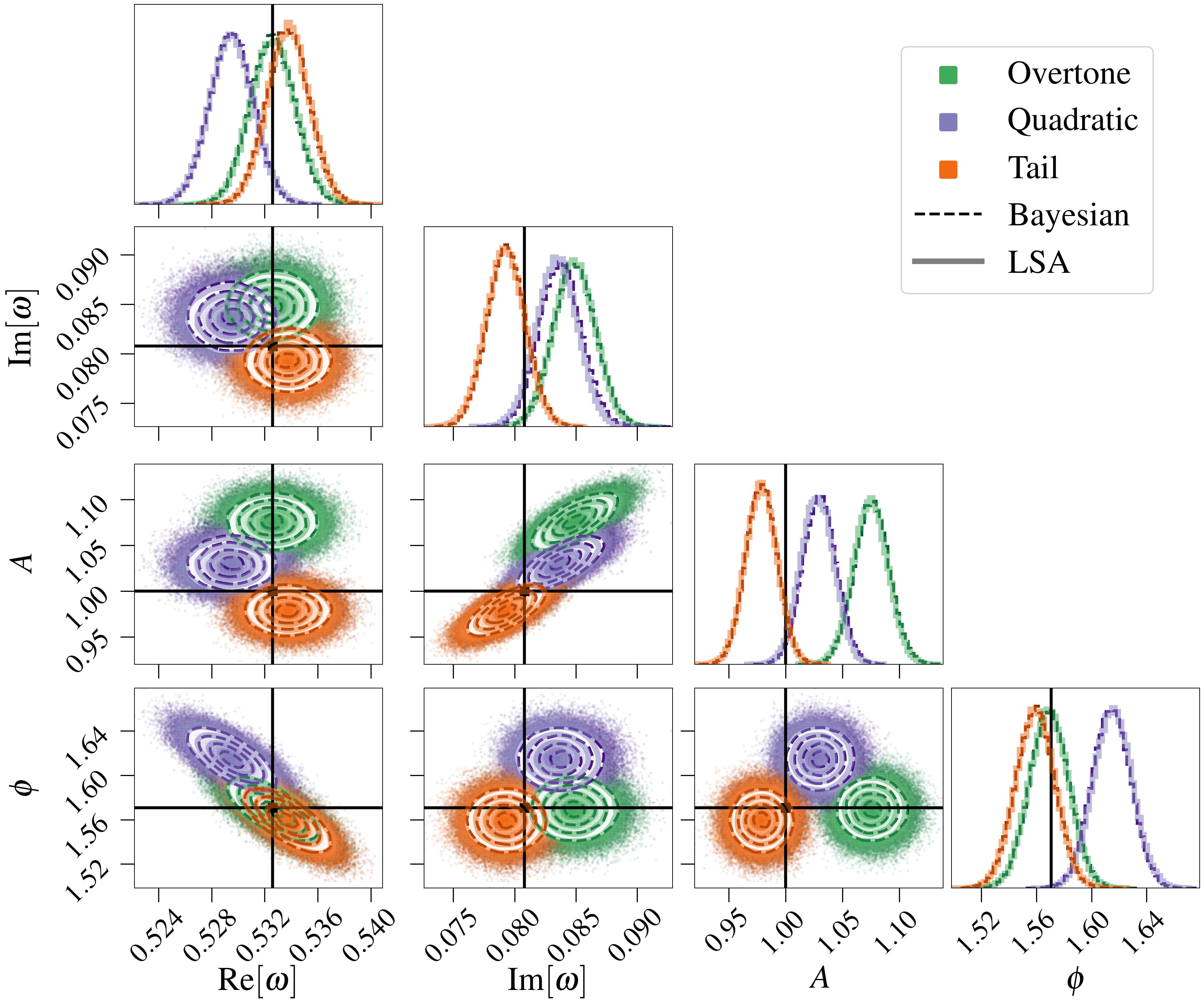}
\caption{\textit{Comparison of LSA versus Bayesian analysis:}
Same cases of unmodeled effects as studied in Fig.~\ref{fig_mcmc_fisher_50}, but for lower SNR=20 and higher SNR=100.  
} 
\label{fig_mcmc_fisher_20_100}
\end{figure*}

\begin{figure*}[t]
\includegraphics[width=1.0\linewidth]{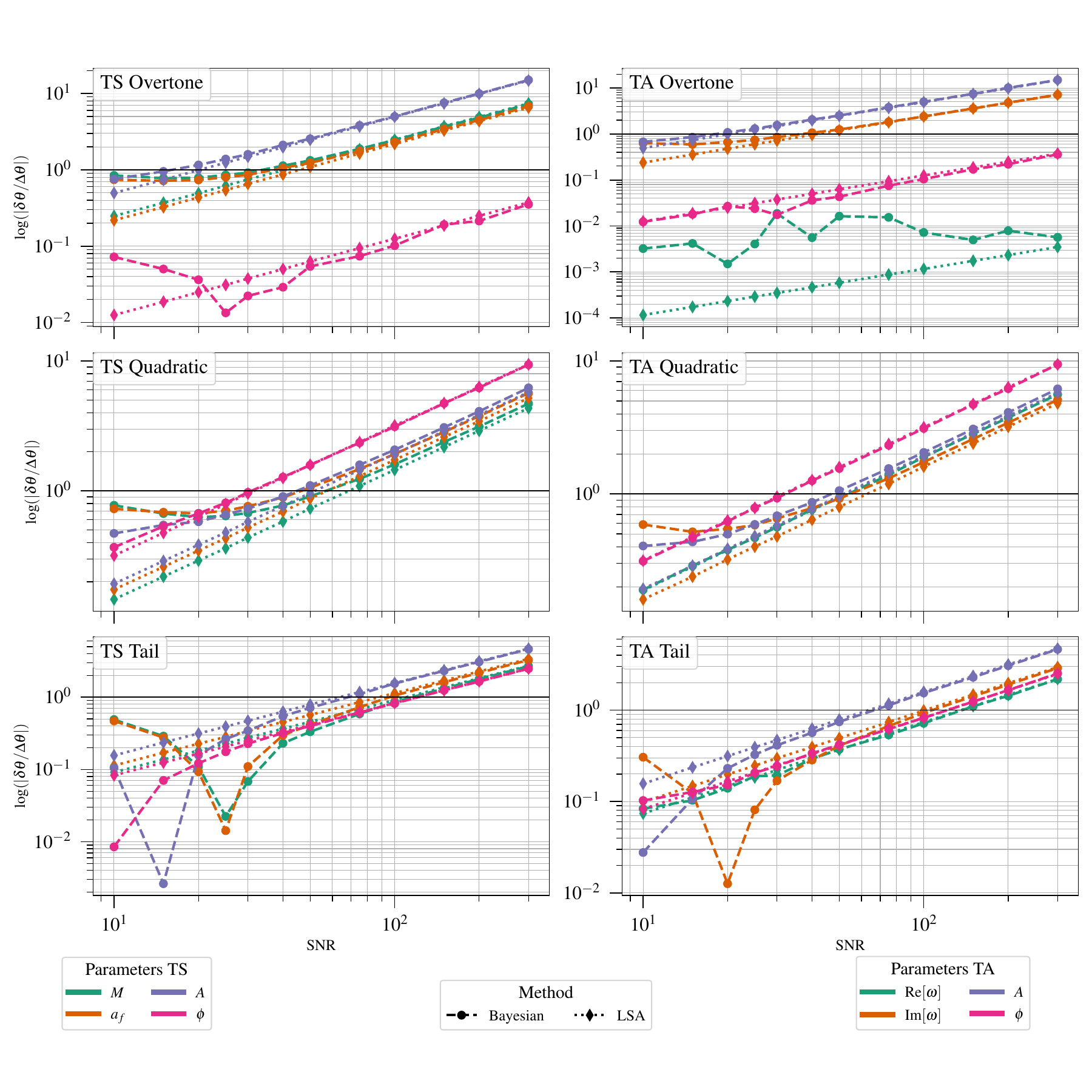}
\caption{\textit{Bias ratios obtained with Bayesian analysis and LSA as a function of SNR:} 
Here we show the bias ratio for the theory-specific (left column) and theory-agnostic (right column) approach as a function of SNR. 
\label{fig_snr_grid}}
\end{figure*}

\begin{figure}[t]
\includegraphics[width=1.0\linewidth]{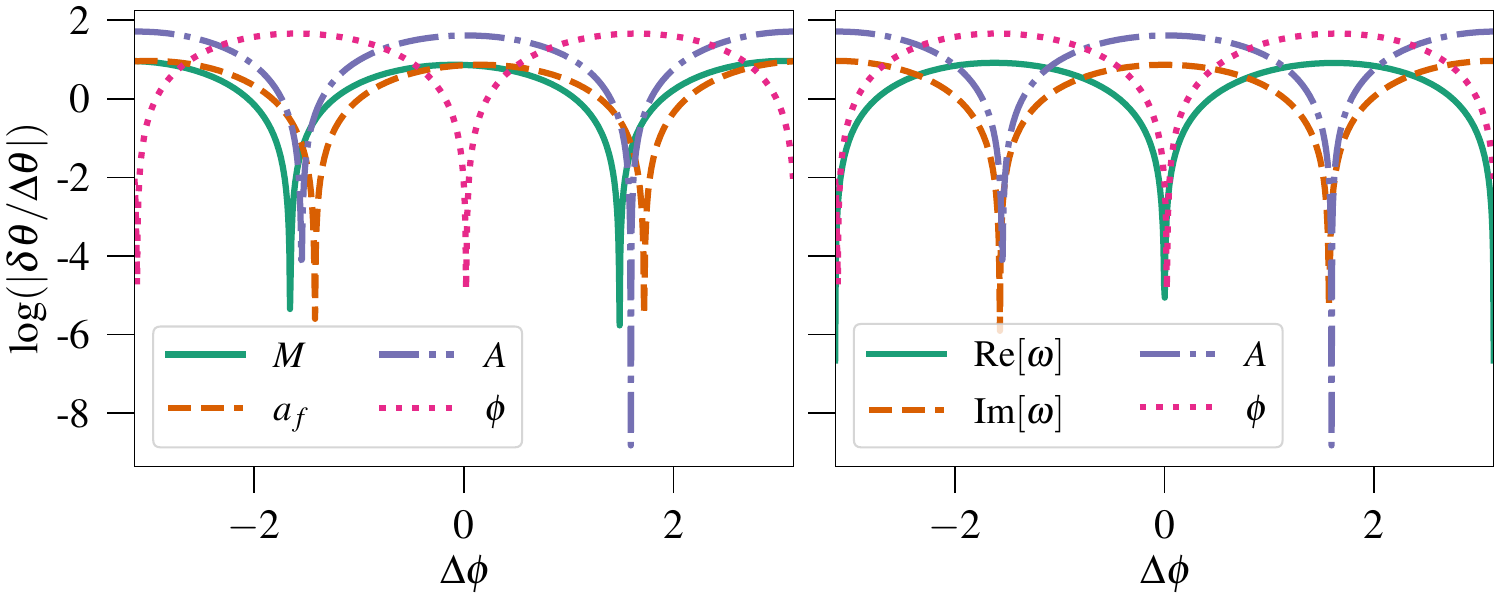}
\includegraphics[width=1.0\linewidth]{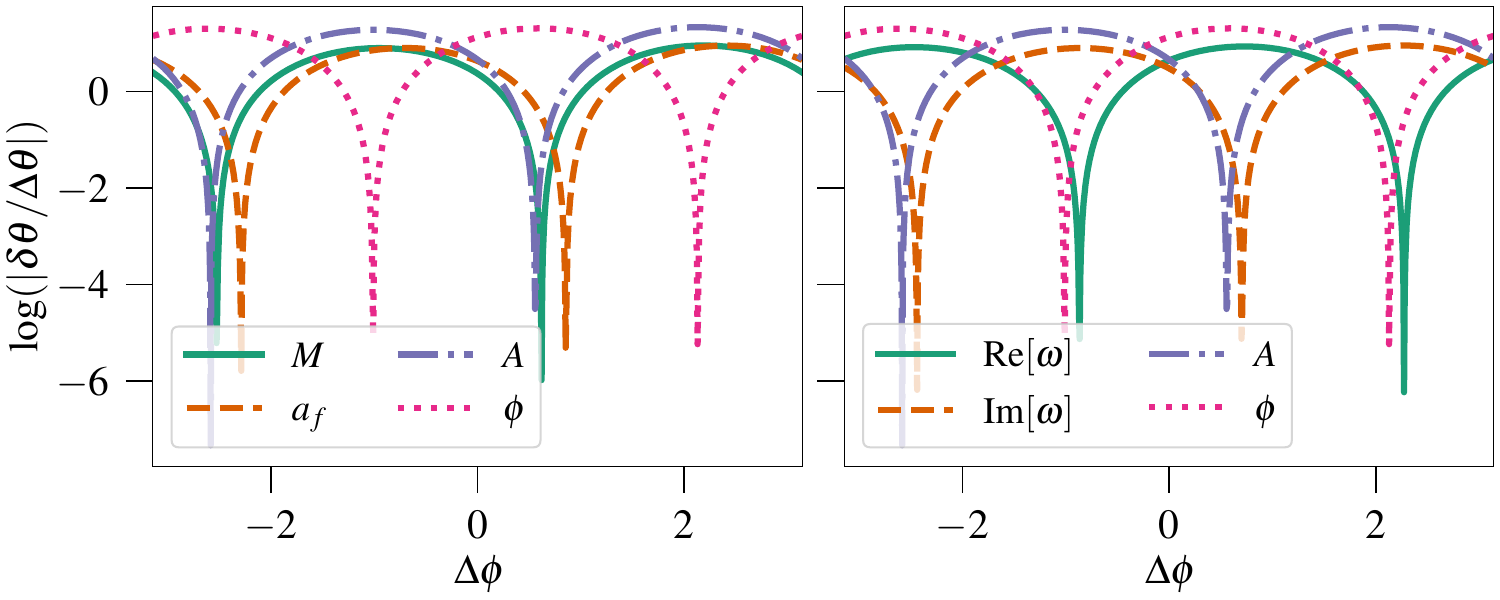}
\caption{\textit{Bias ratios as a function of QNM phase differences:} 
Complementary results to those presented in Fig.~\ref{fig_phase_difference_1}, but for SNR=100. 
\label{fig_phase_difference_2}
}
\end{figure}

\begin{figure}[t]
\includegraphics[width=1.0\linewidth]{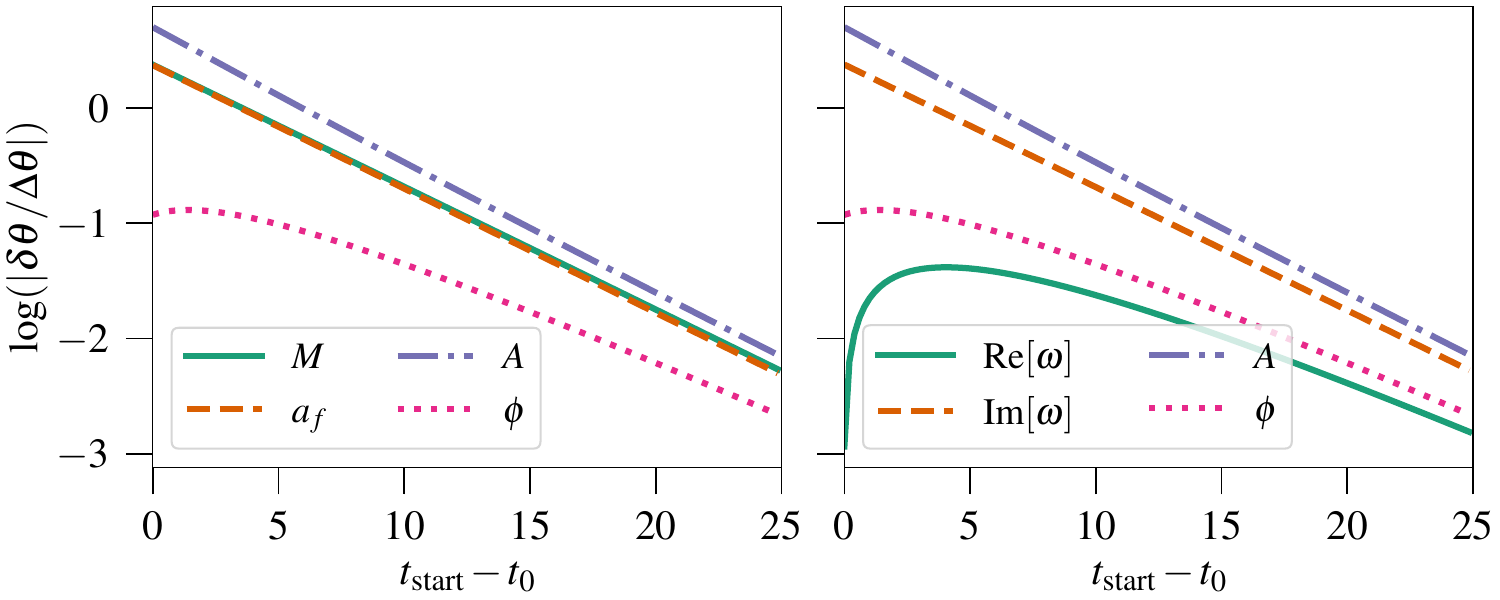}
\includegraphics[width=1.0\linewidth]{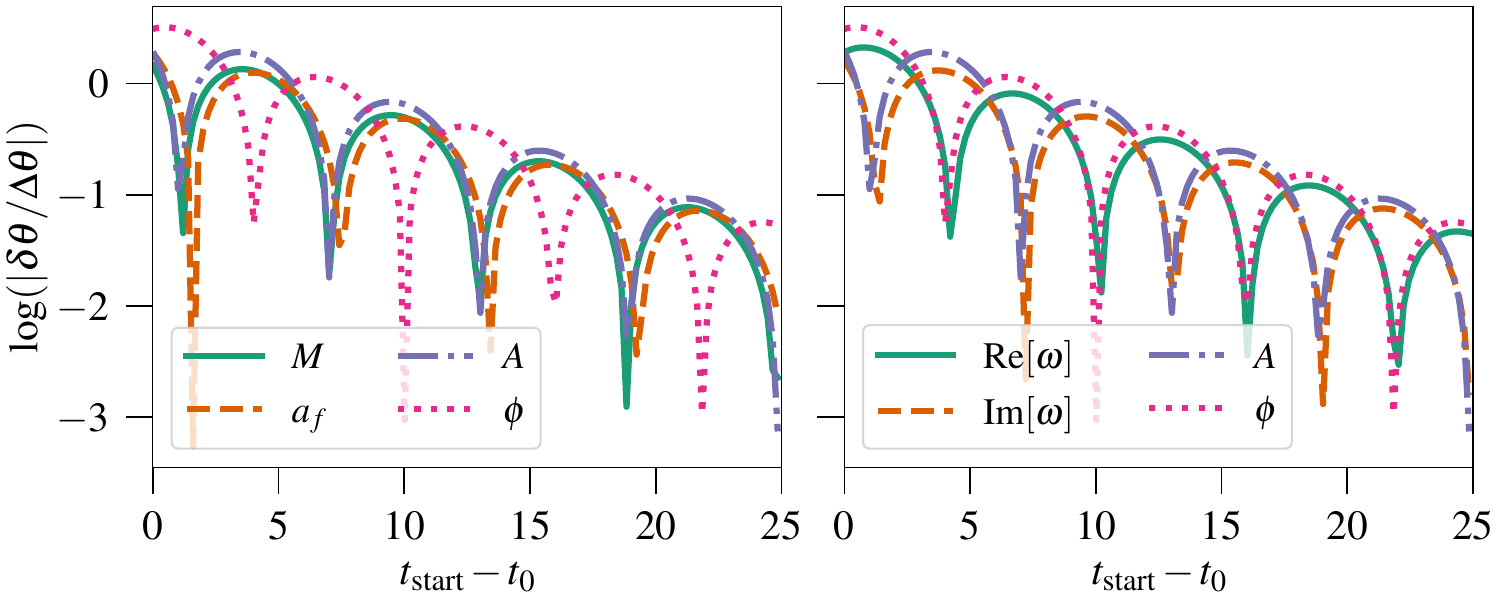}
\includegraphics[width=1.0\linewidth]{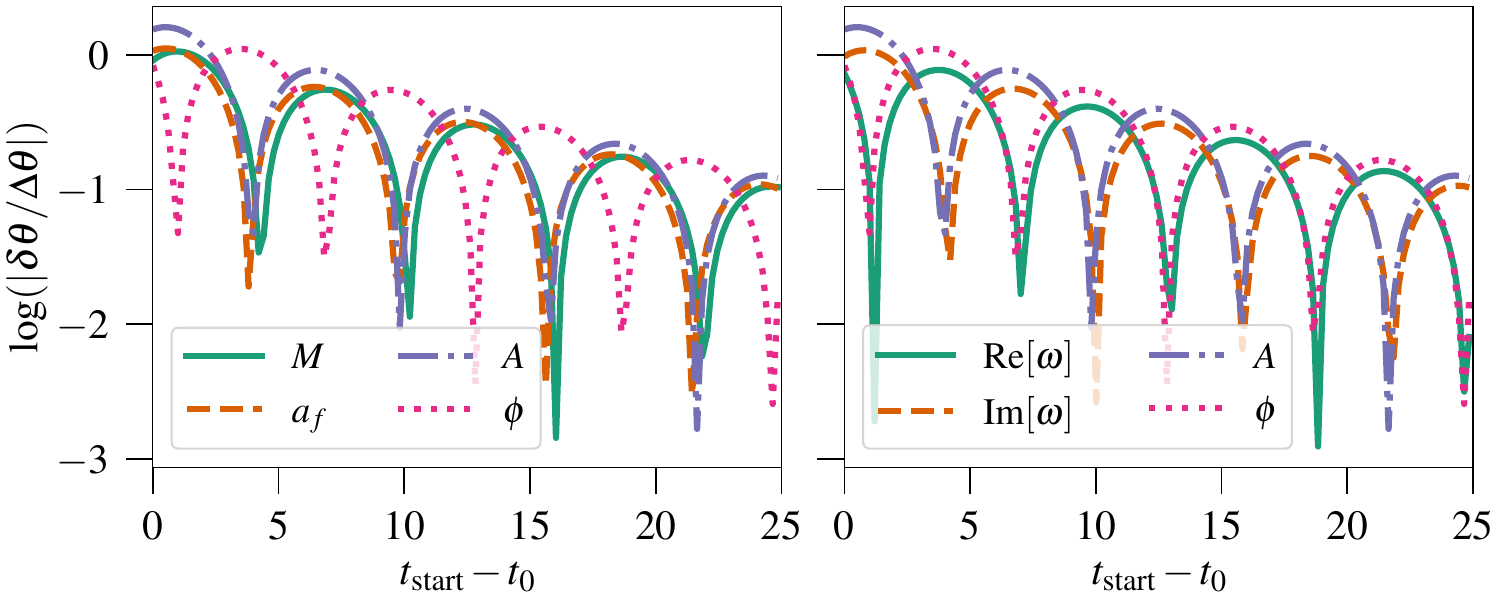}
\caption{\textit{Bias ratios as a function of stating time:} 
Complementary results to those presented in Fig.~\ref{fig_starting_time_1}, but for SNR=100. 
\label{fig_starting_time_2}}
\end{figure}

\begin{figure}[t]
\includegraphics[width=1.0\linewidth]{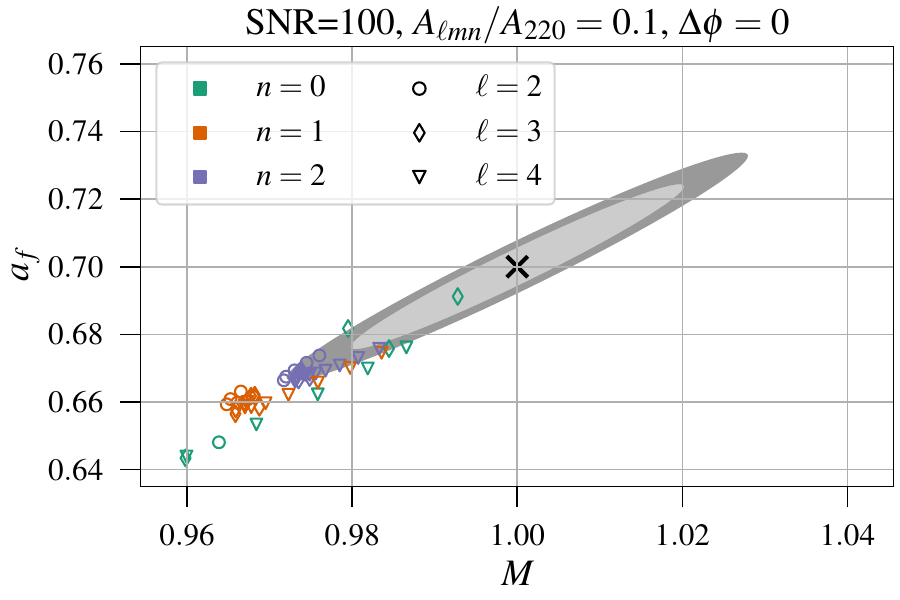}
\includegraphics[width=1.0\linewidth]{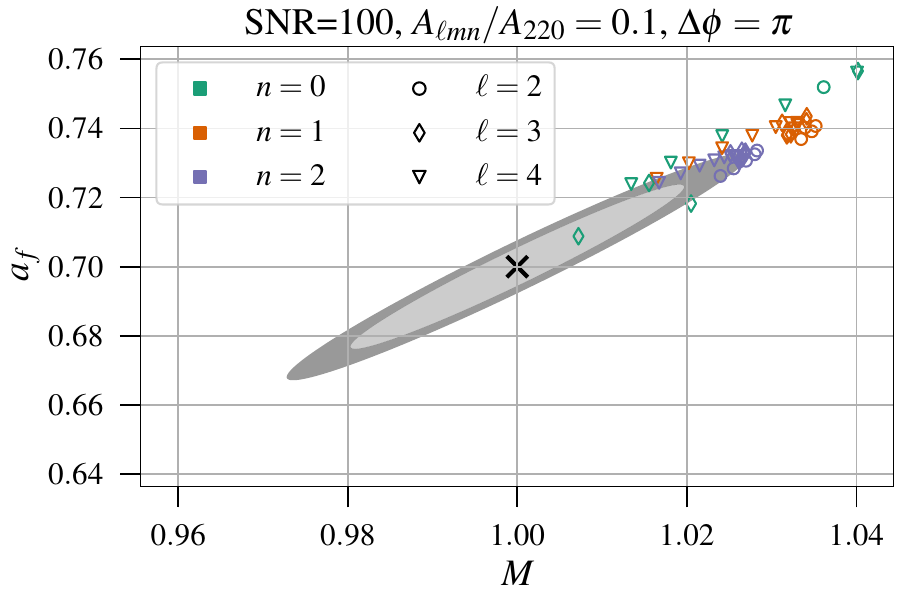}
\caption{\textit{Exploring biases from unmodeled QNMs:} 
Here we show biases of various unmodeled linear QNMs with the same properties as those presented in Fig.~\ref{fig_scatter_v1} but for SNR=100. 
\label{fig_scatter_v2}}
\end{figure}

In the following, we provide additional results to those presented in the main text. 
In Fig.~\ref{fig_mcmc_fisher_20_100}, we compare LSA with Bayesian analysis for the same cases as in the main text, but for SNR=20 and SNR=100. 
In Fig.~\ref{fig_snr_grid}, we provide a detailed comparison of the bias ratio predicted from LSA and Bayesian analysis for selected SNRs. 
We define the Bayesian bias as the difference between the true parameter and the median of the MCMC samples. 
The statistical error is defined as half of the 68\,\% highest density interval, which for a Gaussian distribution (valid at high SNR) corresponds to one sigma. In Fig.~\ref{fig_phase_difference_2}, we show the bias ratio as a function of phase difference for SNR=100. In Fig.~\ref{fig_starting_time_2}, we show the bias ratio as a function of starting time for SNR=100. In Fig.~\ref{fig_scatter_v2}, we provide the biases from various unmodeled QNMs at SNR=100.

\end{document}